\documentclass[useAMS,usenatbib]{mn2e}
\usepackage{url,times,graphicx,
    amsmath,amsfonts,amssymb,aas_macros,color,epsfig,varioref,subfigure,comment,caption,float}
\usepackage[normalem]{ulem}

\usepackage{color}

\def\bepsilon{\;\pmb{\epsilon}}

\def\bsigma{\;\pmb{\sigma}}

\def\lesssim{{^<_\sim}}
\newcommand{\kms}{{\,\rm km \ s^{-1}}}
\newcommand{\kmsMpc}{{\,\rm km \ s^{-1}Mpc^{-1}}}
\newcommand{\hMpc}{{\ifmmode{h^{-1}{\rm Mpc}}\else{$h^{-1}$Mpc}\fi}}
\newcommand{\hgpc}{{\ifmmode{h^{-1}{\rm Gpc}}\else{$h^{-1}$Gpc}\fi}}
\newcommand{\hmpc}{{\ifmmode{h^{-1}{\rm Mpc}}\else{$h^{-1}$Mpc}\fi}}
\newcommand{\hkpc}{{\ifmmode{h^{-1}{\rm kpc}}\else{$h^{-1}$kpc}\fi}}
\newcommand{\Mpc}{{\ifmmode{{\rm Mpc}}\else{Mpc}\fi}}
\newcommand{\kpc}{{\ifmmode{{\rm kpc}}\else{kpc}\fi}}
\newcommand{\hMsun}{{\ifmmode{h^{-1}{\rm {M_{\odot}}}}\else{$h^{-1}{\rm{M_{\odot}}}$}\fi}}
\newcommand{\hmsun}{{\ifmmode{h^{-1}{\rm {M_{\odot}}}}\else{$h^{-1}{\rm{M_{\odot}}}$}\fi}}
\newcommand{\Msun}{{\ifmmode{{\rm {M_{\odot}}}}\else{${\rm{M_{\odot}}}$}\fi}}
\newcommand{\msun}{{\ifmmode{{\rm {M_{\odot}}}}\else{${\rm{M_{\odot}}}$}\fi}}
\newcommand{\LCDM}{{$\Lambda$CDM}}
\newcommand{\Vbulk}{{$V_{\mathrm{bulk}}$}}
\newcommand{\VbulkR}{{$V_{\mathrm{bulk}}(R)$}}
\newcommand{\DeltaLR}{{$\Delta_{\mathrm L}(R)$}}

\def \apjl   {{\it Ap. J. Lett.}}

\def \apjs {{\it Ap. J. Suppl.}}

\def\aap{A\&A}

\def\apjs{ApJS}
\def\apjl{ApJL}

\definecolor{ashgrey}{rgb}{0.7, 0.75, 0.71}
\definecolor{ao(english)}{rgb}{0.0, 0.5, 0.0}

\title[Velocity field from   Cosmicflows-4]
{The large scale velocity field from the Cosmicflows-4 data}

\author[Hoffman et al.]{\newauthor
Yehuda Hoffman,$^{1}$\thanks{E-mail: Hoffman@huji.ac.il}
Aur\`elien Valade,$^{2}$
Noam I. Libeskind,$^{2}$
Jenny G. Sorce$^{3,4,2}$,\newauthor 
R. Brent Tully,$^{5}$
Simon Pfeifer,$^{2}$
Stefan Gottlöber,$^{2}$
and Daniel Pomar\`ede$^{6}$
\\
$^{1}$Racah Institute of Physics, Hebrew University, Jerusalem 91904, Israel\\
$^{2}$Leibniz Institut f\"ur Astrophysik Potsdam (AIP), An der Sternwarte 16, D-14482 Potsdam, Germany\\
$^3$ Univ. Lille, CNRS, Centrale Lille, UMR 9189 CRIStAL, F-59000 Lille, France\\
$^4$Universit\'e Paris-Saclay, CNRS, Institut d'Astrophysique Spatiale, 91405, Orsay, France\\
$^{5}$Institute for Astronomy, University of Hawaii, Honolulu HI 96822, USA\\
$^{6}$Institut de Recherche sur les Lois Fondamentales de l’Univers, CEA, Universit\'e Paris-Saclay, 91191 Gif-sur-Yvette, France
}

\begin{document}

\date{Submitted XXXX XXX XXXX}

\pagerange{\pageref{firstpage}--\pageref{lastpage}} \pubyear{xxxx}

\maketitle

\label{firstpage}


\begin{abstract}
The reconstruction of the large scale velocity field from the grouped Cosmicflows-4 (CF4) database is presented. The lognormal bias of the inferred distances and velocities   data is corrected by   the Bias Gaussianization correction  (BGc) scheme, and the linear density and velocity fields are reconstructed by means of the Wiener filter (WF) and constrained realizations (CRs) algorithm. These   tools are tested against a suite of random  and constrained Cosmicflows-3-like mock data. The CF4 data consists of 3 main subsamples - the 6dFGS and the SDSS data   - and the `others'. The individual contributions of the subsamples have been studied.   The quantitative analysis of the velocity field is done mostly by the mean overdensity ($\Delta_L(R)$) and the  bulk velocity (\VbulkR) profiles of the velocity field out to  $300\,\hmpc$.

The \VbulkR\ and \DeltaLR\ profiles of the CF4 data without its 6dFGS component are consistent with the cosmic variance to within $1\sigma$. The 6dFGS sample dominates the \Vbulk\ ($\Delta_{\mathrm L}$) profile beyond $\sim120\,\hmpc$, and drives it to roughly a $3.4\sigma$  ($-1.9\sigma$) excess (deficiency) relative to the cosmic variance at $R\sim250\ (190)\ \,\hmpc$. The excess in the amplitude of \Vbulk\ is dominated by its Supergalactic X component, roughly in the direction of the Shapley Concentration. The amplitude and alignment of the inferred velocity field  from the CF4 data is at $\sim(2\,-\,3)\,\sigma$  discrepancy with respect to  the \LCDM\ model. Namely, it is somewhat atypical but yet there is no compelling tension with the   model.


\end{abstract}

\begin{keywords}
cosmology: large-scale structure of Universe -- methods: data analysis -- techniques: radial velocities
\end{keywords}

\section{Introduction}

In the standard model of cosmology departures from uniform density and from a pure Hubble flow are strongly coupled - density irregularities induce peculiar velocities on top of the Hubble flow; peculiar  velocities   drive  the matter away from uniform distribution. The equation of continuity tells it all  \citep{1980lssu.book.....P,2008cosm.book.....W}.  This is why surveys of peculiar velocities of galaxies  play such  an important  role in unveiling   the  underlying - luminous and dark - mass  distribution  in  the  nearby universe  
\citep[][is only a partial list]{1986ApJ...307...91L,1988ApJ...326...19L,1990ApJ...364..349D,1996ApJ...468L...5D,2006ApJ...653..861M}. 
Peculiar velocity surveys have also been used to constrain   cosmological parameters 
\citep[e.g.][]{1993ApJ...412....1D,2001MNRAS.326..375Z,2011ApJ...736...93N,2017MNRAS.470..445N,2018MNRAS.481.1368P}. In fact, local surveys of peculiar velocities - extending out to a cosmological redshift of $\sim0.1$ - are the only tracers that map the local (total) mass distribution directly.
Yet, velocity surveys due to their large errors and sparse sampling are less effective in constraining  the values of cosmological parameters compared with other probes - CMB  anisotropies in particular  \citep[e.g.][]{Planck:2013}.

Of particular interest   is the Cosmicflows project\footnote{https://www.ip2i.in2p3.fr/projet/cosmicflows} of measuring and compiling distances and redshifts of galaxies,  and thereby estimating their peculiar velocities. Four generations of  data  have been  released so far:  Cosmicflows-1   \citep{2008ApJ...676..184T}, Cosmicflows-2  \citep[CF2;][]{2013AJ....146...86T}, Cosmicflows-3 \citep[CF3;][]{2016AJ....152...50T} and Cosmicflow-4 \citep[CF4;][]{2023ApJ...944...94T}. The Constrained Local UniversE Simulation's (CLUES) collaboration\footnote{https://www.clues-project.org/cms/} primary focus  is on   the reconstruction of the present epoch  density  and velocity  fields 
\citep[e.g.][]{2012ApJ...744...43C,2013AJ....146...69C,
2014Natur.513...71T,
2017NatAs...1E..36H,
2017ApJ...845...55P,
2020ApJ...897..133P}
and on setting  initial conditions for constrained simulations of the local universe
\citep[e.g.][]{2008MNRAS.386..390H,
2011MNRAS.417.1434F,
2014NewAR..58....1Y,
2014MNRAS.437.3586S,
2016MNRAS.455.2078S,
2018NatAs...2..680H,2018MNRAS.478.5199S,
2020MNRAS.496.4087O,
2020MNRAS.498.2968L, 2021MNRAS.504.2998S,2023arXiv230101305S,   
2023arXiv230210960D,  
2023MNRAS.tmp.1862P}   
from the Cosmicflows data. 

Theorists do like peculiar velocities  -  their emergence in the standard  cosmological model, the \LCDM\ model, is well understood and in the linear regime the velocity and the density fields are related by a simple linear relation. This stands in sharp contrast to the difficulties arising in estimating  velocities  from observations.  Velocity surveys are actually galaxy distance moduli  and redshift surveys, from which the distances and peculiar radial velocities of the galaxies are derived. 

The Cosmicflows data exemplifies the complexities and intricacies of velocity surveys in general. The data is not homogeneously nor isotropically  sampled. It is not sampled in a rigorous manner, namely it is not subjected to a given selection function. It is made of various subsamples of data, assembled by different observational groups, based on different methods and applying these to different type of galaxies \citep[see][for a detailed description of the composition of the CF4 data] {2023ApJ...944...94T}. 
Velocity surveys are sparse, very inhomogenous covering one part of the sky within a given redshift range and another part in another range, and in particular are very noisy. For the majority of Cosmicflows data the typical distance error is of the order of 20\%. At a distance of, say, $100\,\hmpc$ (where $h$ is Hubble's constant ($H_0$) measured in units of $100\kmsMpc$), such an error in distance implies a $2000\,\kms$ error in velocity. The standard model of cosmology - the \LCDM\ model - predicts that r.m.s. value of the radial velocities of galaxies is somewhat smaller than  $400\,\kms$. Namely the signal to noise ratio of galaxies $100\,\hmpc$ away is smaller than  20\%. 

The analysis of such a database, and in particular the reconstruction of the entirety of the large scale structure (LSS), i.e. the density and three-dimensional  velocity fields on a regular grid, out to distances of hundreds of megaparsecs away is challenging. Yet, Bayesian algorithms, relying on the long range correlations of velocities and assuming the \LCDM\ model, have stood up to the challenge and have reconstructed the LSS remarkably well. This is the case of the   linear reconstruction  by means of the Wiener filter (WF) and constrained realizations (CRs) of Gaussian random fields  \citep[WF/CRs;][]{1991ApJ...380L...5H,1995ApJ...449..446Z,1999ApJ...520..413Z}, and by Markov chain Monte Carlo \citep[MCMC;][]{2016MNRAS.457..172L,2019MNRAS.488.5438G,2023A&A...670L..15C} and by Hamiltonian Monte Carlo \citep[HMC;][]{2022MNRAS.513.5148V,2023MNRAS.519.2981V,2022MNRAS.517.4529B,2023MNRAS.518.4191P} methods. 
\cite{2021ApJ...913...76H} presented a new approach to the problem, applying a deep supervised machine learning algorithm to the Cosmicflows-3 data to reconstruct the very local LSS. The method is still at its infancy stage but it will certainly play a key role in future studies.

Surveys of peculiar velocities like Cosmicflows pose further challenges to their analysis. Peculiar velocities are not directly observed but are rather inferred physical variables and as such inherit their errors from the distance   and the (much smaller) redshift errors. The estimation of extragalactic distances suffer from a variety of Malmquist-like biases \citep[see][for a detailed discussion]{1995PhR...261..271S}. In the Cosmicflows data, and other similar data bases, the distance errors are derived from the normally distributed errors on the distance moduli. It follows that the distance errors are lognormal distributed. In the context of the WF/CRs reconstruction the bias in the distance-velocity distribution stems from that so-called lognormal bias \citep{2021MNRAS.505.3380H}.

The paper starts with a brief description of the CF4 grouped data (\S \ref{sec:CF4}),  
followed by a review of the tools of reconstruction of the LSS from the Cosmicflows data (\S \ref{sec:WF_CRs_BGc}). 
A description of the tools of the analysis of the recovered LSS is presented in \S\ref{sec:tools_analysis}.
A detailed description and analysis of the WF/CRs reconstructed nearby LSS is described in \S\ref{sec:reconstruction-CF4}. 
A comparison with other recent reconstructions of the nearby LSS is given in \S \ref{sec:compare} 
and a final summary and discussion are presented in \S \ref{sec:summary}. 
Appendices  \ref{appdx:random_mocks} and \ref{appdx:constrained_mocks}   describe the random  and constrained mock CF3-like data, respectively, and their analysis. Appendix \ref{appdx:CF4_CF2} presents a comparison of the present results with the WF/CRs application to the CF2 data.

\section{Cosmicflows-4 data}
\label{sec:CF4}

\begin{figure}
    \centering,
\includegraphics[width=1.\linewidth]{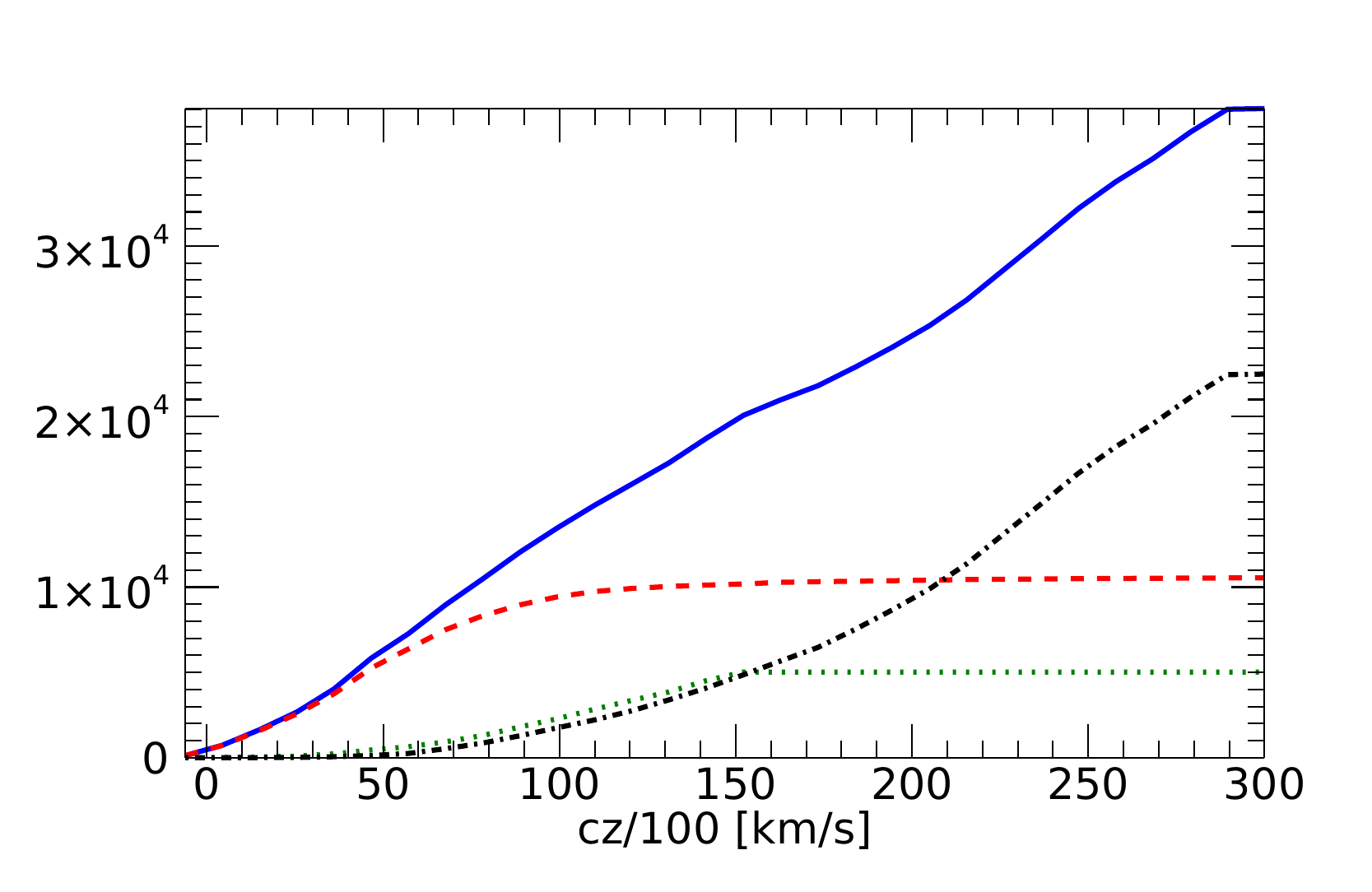}
\caption{The cumulative number of data points within a given redshift is plotted against $cz/100$. The curves correspond to all the data points (blue, solid line), the 6dFGS data (green, dotted), the SDSS galaxies (black, dot-dashed line) and all the 'others' namely non 6dFGS and non SDSS (red, dashed line).  }
\label{fig:CDF}
\end{figure}

\begin{figure}
    \centering
        \includegraphics[width=1.\linewidth]{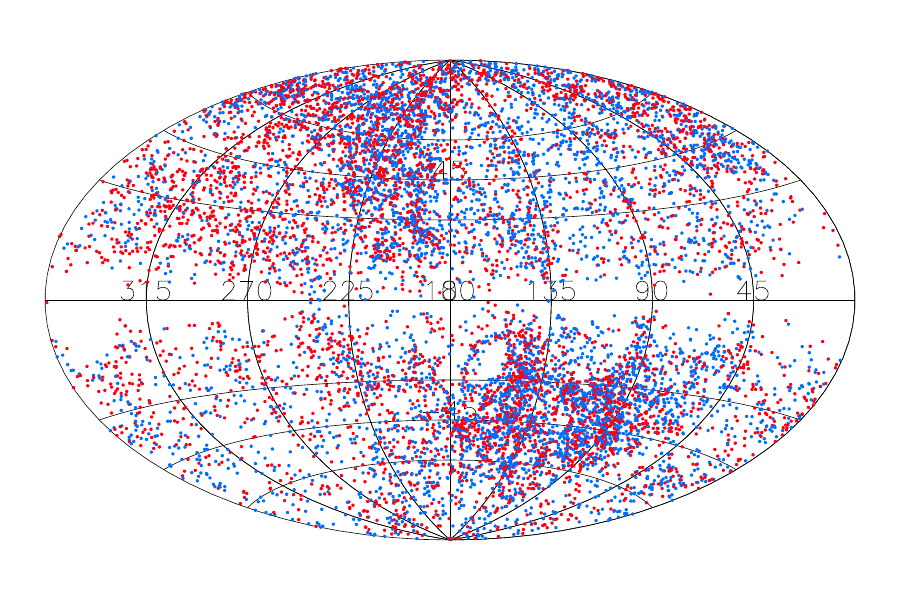}
        \includegraphics[width=1.\linewidth]{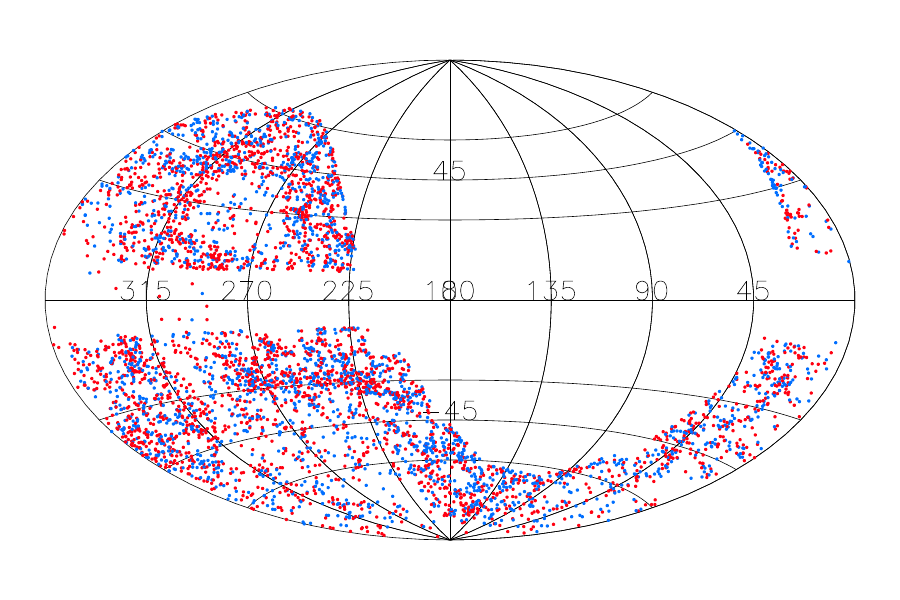}
        \includegraphics[width=1.\linewidth]{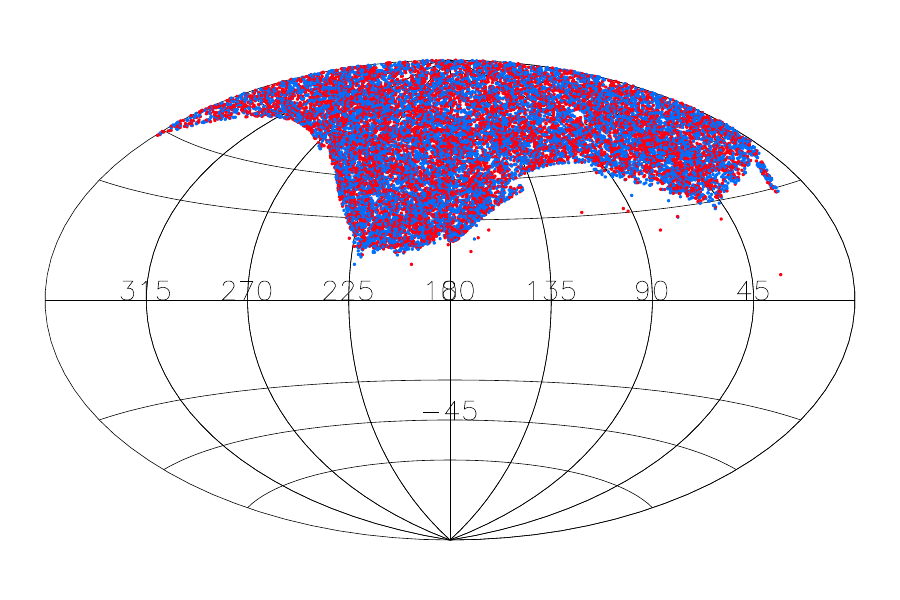}
        \caption{Aitoff projection in Galactic ($l,b$) coordinates of the  distribution  of the sub-samples of the Cosmicflows-4 data: SDSS (bottom), 6dFGS (middle) and all the others (upper panel). Positive peculiar velocities are marked in red and negative ones in blue. 
        }
    \label{fig:aitoffl}
\end{figure}

A detailed description of the Cosmicflows database, and in particular the Cosmicflows-4 (CF4) data, is presented in  \cite{2023ApJ...944...94T}\footnote{The CF4 data used here is the   May 17, 2023, version.}. The CF4 data consists of distance moduli and redshifts of 
roughly $56,000$ galaxies, gathered into $\sim38,000$ groups. Eight different distance measurement methodologies have been employed - the largest numbers coming from the correlation between the photometric and kinematic properties of spiral galaxies (Tully-Fisher; TF) and elliptical galaxies (fundamental flane; FP). The CF4 data consists of three major sub-samples: the 6dFGS, the SDSS  and the `others' samples   \citep[see][for details]{2023ApJ...944...94T}. The redshift distributions of these samples are shown in Fig. \ref{fig:CDF} and their angular distribution in Fig. \ref{fig:aitoffl}. 

\begin{figure}
    \centering
\includegraphics[width=1.\linewidth]{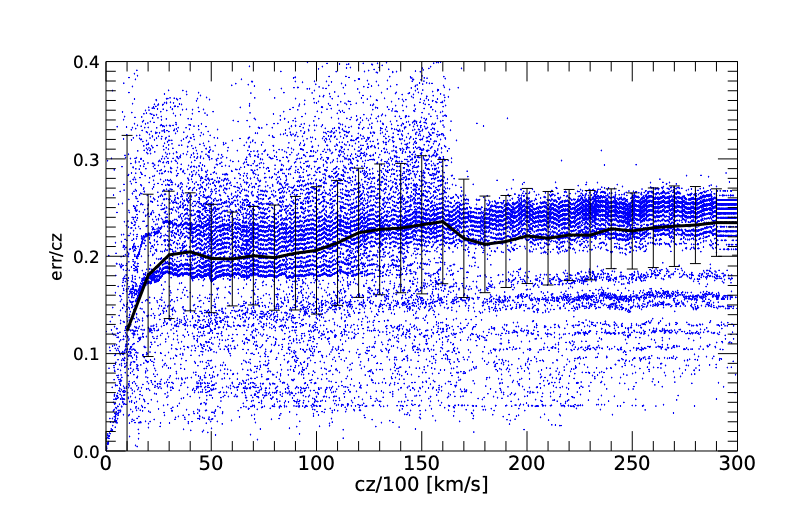}
\caption{A scatter plot of the velocity errors, normalized by $c\,z$, against their redshift distance ($c\,z\,/\,100$). The solid line and the error bars show the mean and standard deviation of the normalized errors.    }
\label{fig:err}
\end{figure}

The reconstruction of the LSS of our local patch of the Universe from the CF4 data is challenging. The three main sub-samples have different radial and angular distributions, with a significant anisotropy that varies with depth. The 'others', 6dFGS and SDSS  components range mostly within the redshift intervals of $cz/100 \sim[0 - 120],\ \sim[60 - 160]\ {\rm and}\ \sim[60 - 300]\,\kms$, respectively.  The 'others' are distributed roughly isotropically, outside the Zone of Avoidance, compared with the 6dFGS that is distributed mostly in the Southern Galactic hemisphere and the SDSS that lies entirely in the Northern Galactic hemisphere. Fig. \ref{fig:err} shows the redshift distribution of the velocity errors, normalized by $c\,z$ (where $c$ is the speed of light and $z$ is the redshift). For the vast majority of the data points the typical fractional error is $\sim20\%$. It follows that the typical uncertainties in the CF4 inferred velocities amount to $\sim2,000\,\kms$ at a redshift of $c\,z\,=\,10000\,\kms$ and  that for the \LCDM\ model the typical signal-to-noise ratio of inferred velocities is $0.15$ at that redshift and about $0.05$ at the edge of the data ($cz/100\,=\,300\,\kms$).

\begin{figure}
    \centering
        \includegraphics[width=1.\linewidth]{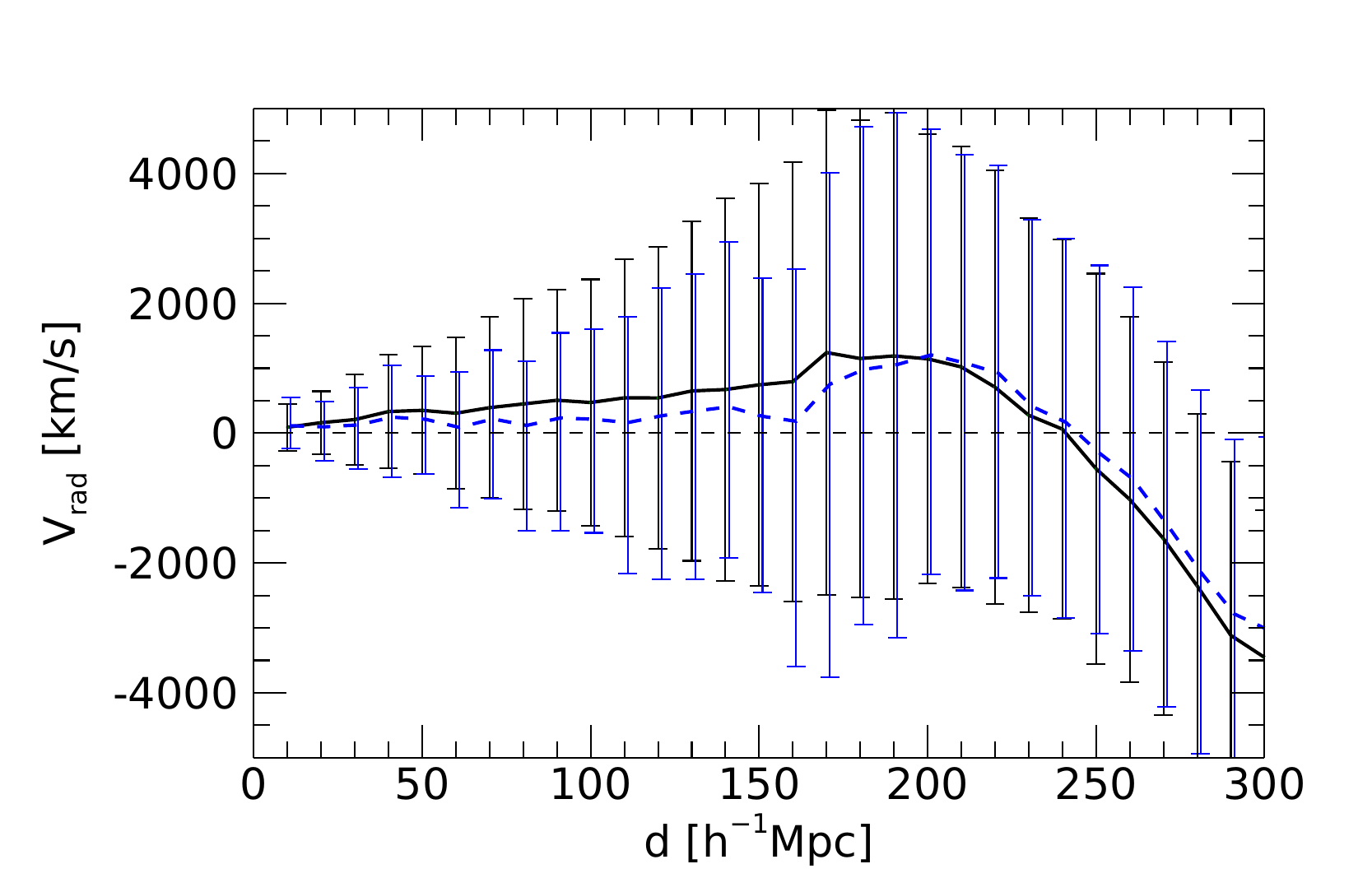}
        \includegraphics[width=1.\linewidth]{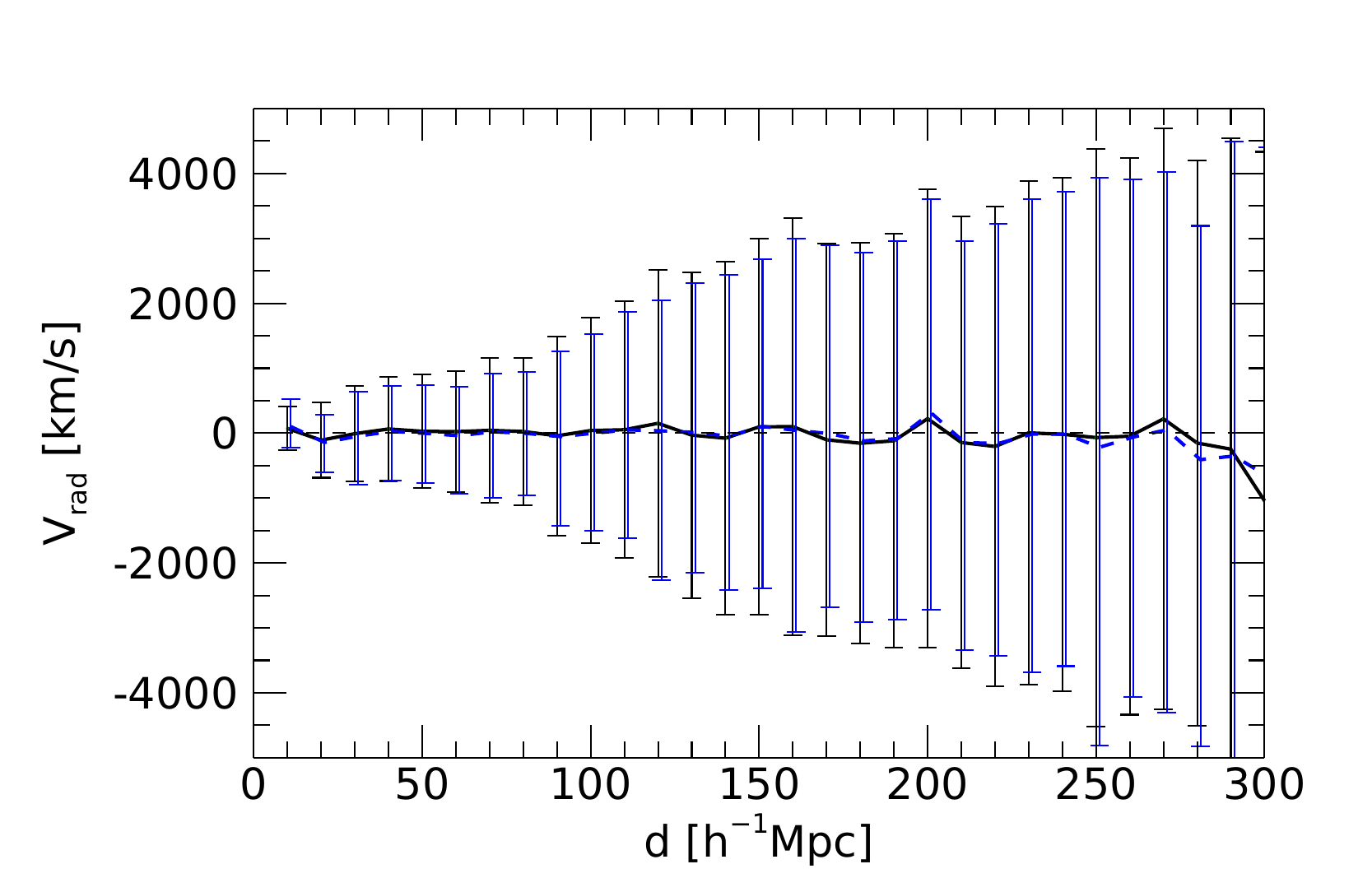}
        \caption{CF4 data: Mean (black, solid line) and median (blue, dashed line) of the peculiar radial velocities vs. their distances.   Black error bars correspond to one standard deviation around the mean value and the blue ones to the 1st and 3rd quartiles around the median, rescaled to correspond to the standard deviation of a normal distribution. The blue error bars are slightly shifted horizontally. The upper panel presents the uncorrected values of the radial velocities and distances, as derived from the measure distances moduli and redshifts. The lower panel presents the BGc corrected distances and velocities.}
    \label{fig:bias}
\end{figure}

The lognormal bias  is clearly manifested by the upper panel of 
Fig.\ref{fig:bias}, which presents the mean and the median of the distribution of the CF4 uncorrected velocities vs their uncorrected distances. It consists of an excess of positive peculiar velocities, namely an outflow, out to roughly $200\,\hmpc$ that is followed by a strong inflow all the way to the edge of the data - in a strong disagreement with the \LCDM\ model. 

\section{Tools of reconstruction}
\label{sec:WF_CRs_BGc}

\subsection{Wiener filter and constrained realizations: reconstruction}
\label{sec:WF/CRs}

In the framework of the standard model of cosmology the LSS of the Universe has emerged out of a primordial Gaussian random perturbations field. The \LCDM\ model predicts that small scales have more power than the large ones hence they collapse and virialize first, while the long waves still remain in the linear regime. Hence, grouping the data acts as a non-linear filter, where members of a collapsed group, e.g. galaxies belonging to  a given cluster, are replaced by one single data point whose dynamics are well approximated by the linear theory.

The Bayesian framework employed here is the linear theory of the \LCDM\ standard model of cosmology. The algorithm of the WF and CRs is the optimal tool for dealing with data within that framework \citep{1991ApJ...380L...5H,1995ApJ...449..446Z}. The essence of the algorithm is that in the case where the underlying field is assumed to be Gaussian the following linear estimators, the minimal variance solution, the conditional mean field given the data, the most probable field and the maximum entropy, are all identical. The WF provides an analytical expression for that estimator. The CRs constitutes random realizations designed to obey the imposed constraints. In the case of `strong' data the WF estimators and the CRs follow closely the constraining data. Where the data is `weak' the null field prevails and the CRs are essentially random realizations.

\subsection{Lognormal bias correction}
\label{sec:BGc}
Peculiar velocities of surveys like the Cosmicflows are not directly observed but are rather inferred physical variables from surveys of galaxy distances. The Cosmicflows database, to be specific, is composed of the angular positions, redshifts and distances of galaxies and their associated uncertainties. The major source of the uncertainties of the derived peculiar radial velocities (and from here on they are referred to as 'velocities') are the distance errors.  Additionally, the estimation of extragalactic distances suffers from a variety of Malmquist-like biases \citep[see][for a detailed discussion]{1995PhR...261..271S}. In the Cosmicflows data, and other similar data bases, the estimated luminosity distance ($d_L$) is derived from the 'observed' distance modulus ($\mu$) via
\begin{equation}
    \mu=5\log_{10}\left({d_L\over 10\,{\mathrm pc}}\right).
\end{equation}
Consequently  the normally distributed observational uncertainties on $\mu$ are transformed into lognormal errors on $d_L$ and thereby also on  proper distance and velocities.   This so-called lognormal  bias is analyzed in detail in 
\citet{2021MNRAS.505.3380H}. In the context of the WF/CRs reconstruction the lognormal bias leads to a nearby (faraway) spurious outflow (inflow) of the inferred velocities of the data points. 
It follows that the reconstruction of the LSS from velocity surveys cannot be done without accounting for that bias. Two approaches to the bias correction has been pursued. In the WF/CRs framework the correction of the bias is done ahead of and independently of the application of the WF/CRs algorithm \citep{2015MNRAS.450.2644S,2021MNRAS.505.3380H,2023arXiv230903945S}. In the MCMC and HMC approach the undoing of the bias is done in conjunction with the estimation of the LSS, within a unified Bayesian approach.

The Bias Gaussianization correction (BGc) algorithm \citep{2021MNRAS.505.3380H} is applied to the one-point probability distribution function of the inferred distances and velocities  and is designed to transform these distributions from lognormal to normal ones. The BGc is applied here to the grouped CF4 data and this bias corrected data is then used as an input for the WF/CRs reconstruction of the LSS of the nearby universe.

\subsection{Random and constrained mock Cosmicflows-like data}
\label{sec:mocks}

Given the complex nature of the CF4 data and the approximate nature of the bias correction scheme and the WF/CRs reconstruction it is essential to examine these tools and test them against mock Cosmicflows-like data. The basic logic followed here is to use halos drawn from cosmological DM-only \LCDM\ simulations for generating such mock catalogs. The simulation from which the mocks are drawn are the target against which the reconstructed LSS is to be compared and its merits are to be judged. Two kinds of mock data are constructed here - one is   of random mock data where the mock data points are drawn from  random \LCDM\  simulations. The other kind is constrained mock data, where the data is drawn from constrained simulations. 
Such mock data sets are `random' in the sense that they are drawn from random realizations of the LSS that are consistent with the \LCDM\ model and without imposing any constraints related to the particularities of our local realization of the Universe.  
Random mock data have been recently used to test the BGc and HMC lognormal bias correction and reconstruction schemes     
\citep{2021MNRAS.505.3380H,2023MNRAS.519.2981V}   One of the main virtues of using random mock data is that it enables the probing of the cosmic variance, i.e. the variance introduced by moving the observer randomly in the universe, together with the errors and sampling variance. 

The random mock CF3 data described by \citet{2021MNRAS.505.3380H} are used here.  Ten different random observers are drawn from a random \LCDM\ simulation, and ten different errors realizations are constructed for each mock observer, resulting in an ensemble of 100 mock data sets. A detailed presentation and analysis of the monopole and dipole moments of the WF/CRs reconstruction from these mocks is given in Appendix \ref{appdx:random_mocks}.

 The reconstruction of the LSS from galaxy peculiar velocity surveys is very appealing - as the velocities constitute an unbiased probe of the underlying matter density field. Yet, the fact that galaxies, and their groups, are used as tracers of the velocity may lead to bias. Consider the case of voids in the galaxy distribution. The contribution of underdense regions to the velocity field is as important as the overdense ones \citep{2017NatAs...1E..36H,2017ApJ...847L...6C}, and  galaxies in voids partake  in the flow field. Yet, galaxies are under-abundant  in voids hence also in galaxy velocities surveys. This can lead to a bias in the reconstruction from such surveys. A way to address such a sampling bias   is to construct mock Cosmicflows-like data from constrained simulations of the local universe 
 (Doumler et al. 2013a,b,c).
  Here we use the constrained CF3 mock data that has been drawn from the constrained DM-only simulation from the   data \citep{2023arXiv230101305S}. That simulation was constrained by the CF2 data and it recovers all the prominent structures within the nearby $\sim100\,\hmpc$ around the LG. Detailed testing of the WF/CRs reconstruction from the constrained mock CF3 data  is presented in Appendix \ref{appdx:constrained_mocks}.

\nocite{2013MNRAS.430..912D,2013MNRAS.430..902D,2013MNRAS.430..888D} 

A few words are due on why the mocks used here are CF3-like while the data is the CF4 catalog. We wanted to test the algorithm based on random and constrained mocks. So far only CF3 based constrained simulations are available and by construction they are not deep enough to construct realistic constrained mocks for CF4. Therefore, we decided to test the algorithm with CF3 mocks. 
Yet, Fig. \ref{fig:wf_CF4_CF3_CF2_dipole_divv} shows that the SDSS component of the CF4 data, the main component of CF4 which distinguishes it from CF3, hardly affects the monopole and dipole moments of the velocity fields. Hence, it seems very likely that the mock CF3 data is adequate for testing   the reconstruction from the CF4 data by applying its methodology to the mock CF3 data.

 The main conclusion one may draw from the comparison with the mock is that for the case studied here the WF/CRs estimated profiles faithfully reconstruct the bulk velocity and the mean overdensity profiles   of the target simulations for spherical volumes of $R\gtrsim40\,\hmpc$.

\section{Tools of analysis}
\label{sec:tools_analysis}

{\bf Cosmic and constrained variance:}
The primordial density and velocity fields are random Gaussian fields. Furthermore, the standard cosmological model dictates that these fields are statistically homogeneous and isotropic. The variance of the  fields,  taken over large enough volume and at a given resolution, has therefore a universal value, also called the cosmic variance. The variance of the possible realizations of the $\delta$ and of the velocity fields constrained by the Cosmicflows data, say, is smaller than the cosmic variance, as it samples a sub-volume of all possible realizations of random realizations of the model. This so-call constrained variance depends on the strength and quality of the constraining data and on the nature of the assumed cosmological model. For the Cosmicflows data one expects the constrained variance to converge to the cosmic variance away from the data zone - in configuration as well as in Fourier space.   In the \LCDM\ model the high wavenumber ($k$) modes are dominated by non-linear dynamics and hence are very poorly constrained by the data, hence one expects that at the high $k$ limit, namely at high resolution, the constrained variance converges to the cosmic one.

{\bf Ensembles of constrained and random realizations:}
The constrained and cosmological unconstrained variance are calculated here over an ensemble of 60 constrained  and 3,000 random realizations, respectively. Random realizations, namely unconstrained ones,  are much `cheaper' to construct, hence the imbalance  in the number of realizations of the different kinds. Two sets of such realizations have been prepared, in boxes of side length of $L=600$ and $=1000\,\hmpc$. In the presentation of the different results the size of the box is clearly stated. The constrained and random realizations are calculated by means of an FFT algorithm, where periodic boundary conditions are assumed. The practice followed here for accounting for the missing power inherent to such an approach is to perform all the FFT calculations in computational boxes that are $4^3$ times larger in volume   and then crop  the resulting fields to the desired box.

{\bf The linear density field:} The WF/CRs methodology provides an estimation of the density and velocity fields within the framework of the linear regime, using the \LCDM\ standard model of cosmology as the Bayesian prior model. Within the linear theory the fractional overdensity ($\delta=\rho/\bar{\rho}-1$), where $\rho$ is the local density and $\bar{\rho}$ is its mean cosmological value) and the velocity fields are related by,
\begin{equation}
    \delta_L = -{1\over H_0 f(\Omega)} \nabla\cdot{\bf v},
    \label{eq:delta_L}
\end{equation}
where $f(\Omega)$  is the linear growth factor and the subscript L denotes that it is the linear $\delta$. A comparison of the WF/CRs estimated density field from mock data should be made against $\delta_L$ inferred from the velocity field of the target simulation.

{\bf Visualization of the velocity field by streamlines:} 
 Streamlines are a graphical visualization of a velocity field. The equation of 'motion' of the line element of a given stream line ${\bf s}({\bf l})$, where ${\bf l}$ is the line parameter, is:
\begin{equation}
    {d\,{\bf s}\over d\,{\bf l}}={\bf v}\left({\bf s}\left({\bf l}\right)\right)
    \label{eq:line}
\end{equation}
The choice of the seeding points of the streamlines is a matter of convention, randomly or uniformly distributed on a grid. Particles move along streamlines at a given moment, yet they do not follow the entirety of given streamlines. The flow field is represented here by colored streamlines whose color reflects the local amplitude of the velocity and the   tangent of the line is in the direction of the velocity vector.

{\bf Multipole expansion:} A common presentation of the velocity field is by means of the monopole and dipole moments of the velocity field. Given a velocity field evaluated over a regular grid these are defined as the volume average over a sphere of radius $R$ of the monopole moment either as,
\begin{equation}
    \nabla\cdot{\bf v}(R) =  {1\over 4\pi R^3/3} \int_{<R} \nabla\cdot{\bf v}\,d^3r,
    \label{eq:monopole}
\end{equation}
or as  
$\Delta_{\mathrm L}(R)=-\nabla\cdot{\bf v}(R)/\left(H_0\,f\left(\Omega\right)\right)$.
 For the \LCDM\ standard model $f(\Omega)  \sim 0.52$.  We use  the term 'monopole' in a somewhat loose sense - it stands here for the isotropic component of the linear expansion of the velocity field, namely the `breathing' mode of the velocity field.  
 The dipole moment, which is also the bulk velocity of the sphere, is given by:
\begin{equation}
    {\bf V}_{\mathrm bulk} (R) =  {1\over 4\pi R^3/3} \int_{<R} {\bf v}\,d^3r.
    \label{eq:dipole}
\end{equation}

{\bf $\chi^2$ statistics:} 
The  Cartesian components of the bulk velocity vector of   given spheres of radius $R$s constitute  a set of correlated Gaussian variables, and as such are best suited for a $\chi^2$ statistics to measure their likelihood given the \LCDM\ model. Following   \cite{2023MNRAS.524.1885W} we consider here the bulk velocity of a given sphere and vary the radius of that sphere.  The analysis consists of calculating the $\chi^2(R)$ of the three Cartesian components of the mean of the ensemble of the bulk velocity of an ensemble of CRs, $\mathbf{V}_{\mathrm bulk}^{\mathrm CRs}(R)$,
\begin{equation}
    \chi^2(R) = V_{\mathrm{bulk},\alpha}^{\mathrm CRs} 
    \large< V_{\mathrm{bulk},\,\alpha}V_{\mathrm{bulk},\,\beta} \large >^{-1}
    V_{\mathrm{bulk},\beta}^{\mathrm CRs},
    \label{eq:chisq_R} 
\end{equation}
where $\large< V_{\mathrm{bulk},\,\alpha}V_{\mathrm{bulk},\,\beta} \large >$ is the covariance matrix of the bulk velocity and the angular brackets denote an ensemble average. In principle this covariance can be easily calculated  analytically, yet given the fact that it corresponds to the velocity field calculated over a finite grid we have chosen to calculate it numerically by calculating it as an average evaluated over an ensemble of the bulk velocity of 3,000 random realizations of the \LCDM\ model.

{\bf Tidal decomposition:}
An inherent feature of the present implementation of the WF/CRs algorithm is the recovery, in principle,  of the full (linear) velocity field, i.e. the velocity field induced by all the matter  in the whole Universe.   Practical considerations limit the evaluation of the velocity field to a given finite computational, that often encompasses the entire data zone. That recovered velocity field can be decomposed into two components - the one introduced by the matter distribution within the box, or a sub-volume of that box and the one induced by the matter outside that volume. These are the 'local' (divergent) and the tidal components, respectively. By construction, the tidal field within that volume is divergence-less.   The technique employed here for the tidal decomposition uses the WF reconstructed density and velocity fields within a box  \citep[see][]{1999ApJ...520..413Z,2001astro.ph..2190H}. Fourier/FFT decomposition is applied to the  density  field within the box, and employing the assumption of the linear theory, the divergent velocity field and the density are related by,
\begin{equation}
    {\bf v}^{\mathrm div}_k = -{1\over H_0 f(\Omega)}\, {{\bf k}\over k^2}\,\delta_k,
    \label{eq:v_k}
\end{equation}
where $\textbf{v}^{\mathrm div}_k$ and $\delta_k$ are the Fourier transformed divergent velocity and density fields, respectively.  The volume  considered here for the tidal decomposition is a sphere of radius $R=300\,\hmpc$.

\section{Reconstruction from  Cosmicflows-4 data}
\label{sec:reconstruction-CF4}

\subsection{Large scale structure}
\label{sec:LSS}

\begin{figure*}
\centerline{
\includegraphics[width=0.5\linewidth]     {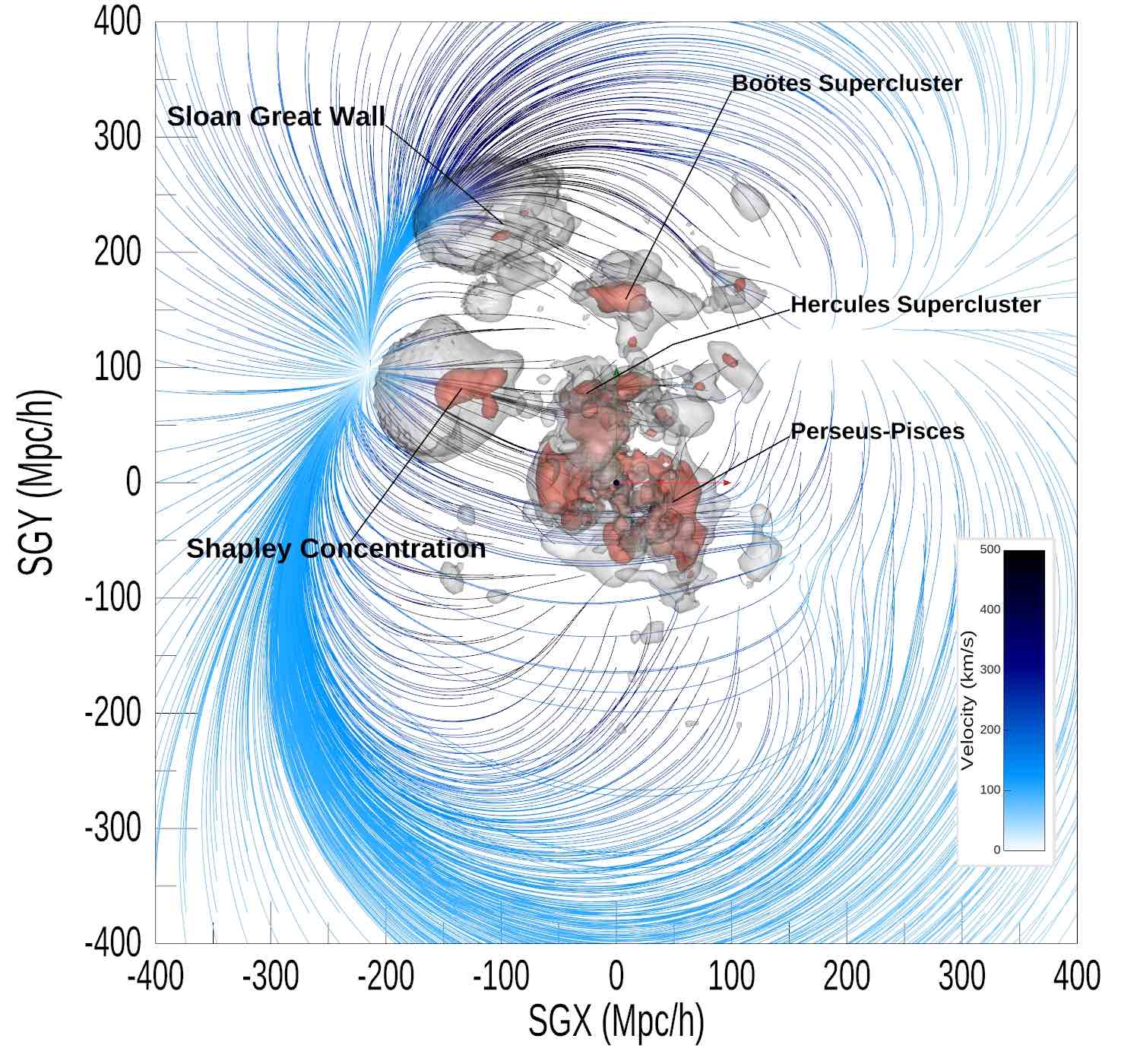}
\includegraphics[width=0.5\linewidth]{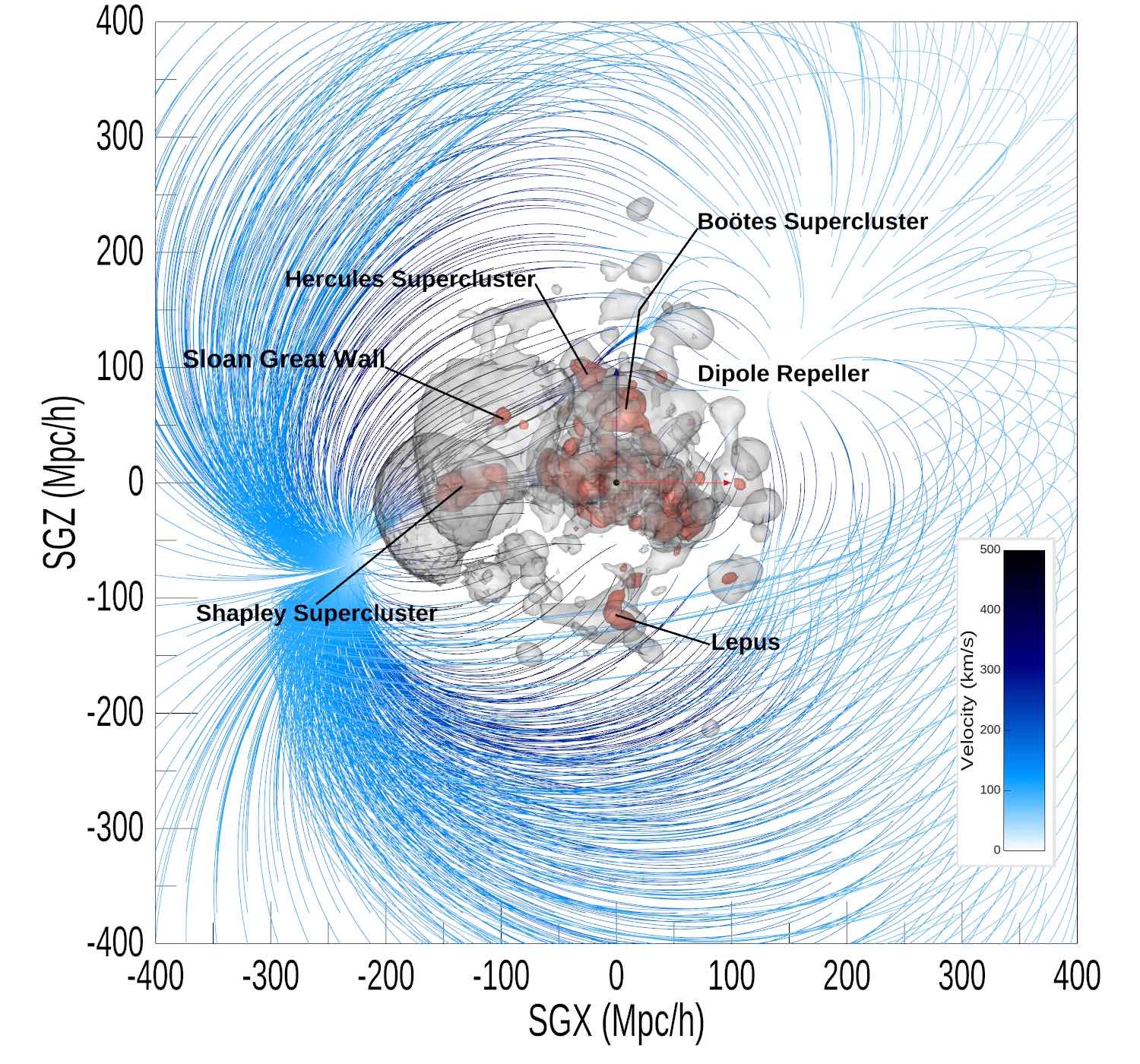}
}
\centerline{
\includegraphics[width=0.5\linewidth]{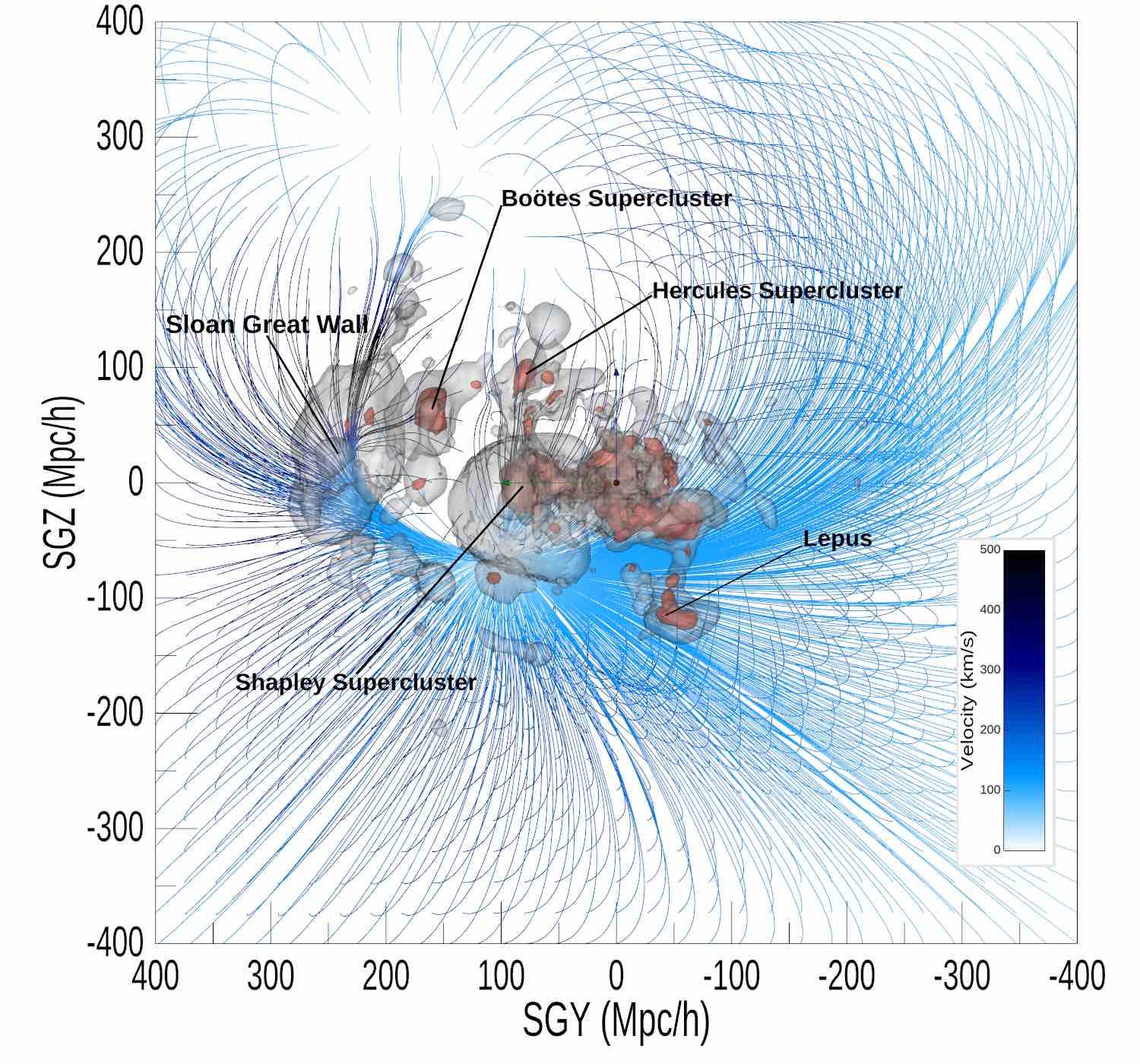}
}
\caption{ The Wiener filter reconstruction of the velocity field from the BGc corrected Cosmicflows-4 data, assuming $H_0=74.6\,{\rm km\,s^{-1}Mpc^{-1}}$. The field is represented by streamlines, with colors representing the amplitude of velocity field within the plane. Density isosurfaces correspond  to $\delta_L=0.2$ (grey) and $0.5$  (red). Labels denote the prominent nearby structures. }
\label{fig:3D}
        \label{fig:streamlines}
\end{figure*}

\begin{figure*}
\centerline{
\includegraphics[width=1.\linewidth]{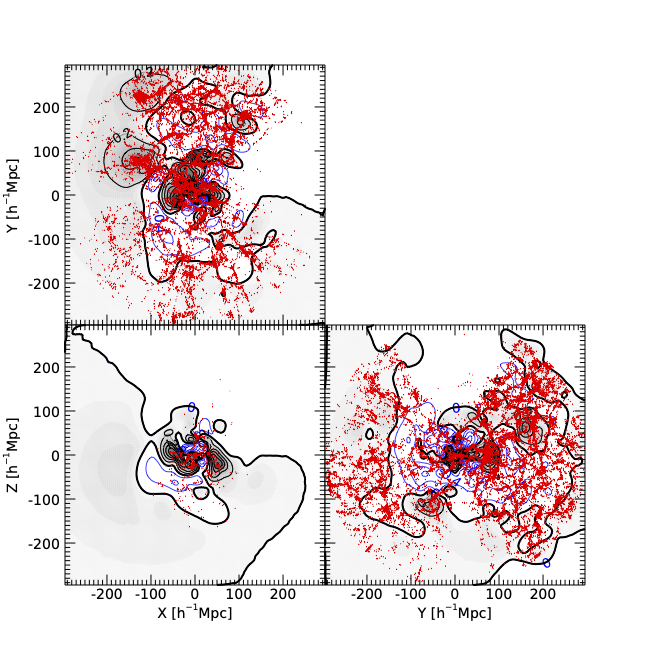}
}
\caption{The Wiener filter reconstruction of the linear density field ($\delta_L$) from the BGc corrected Cosmicflows-4 data, assuming $H_0=74.6\,{\rm km\,s^{-1}Mpc^{-1}}$. The field is represented by means of contour/color maps on the three principal planes of the Supergalactic Cartesian coordinates. The LEDA galaxies brighter than $M=-20$, within slabs of $\pm\,5\,\hmpc$ onto the principal planes are presented as a gauge of the quality of the reconstruction.  }
\label{fig:LSS}
\end{figure*}

 The main focus of the paper is the quantitative analysis of the LSS reconstructed from the  CF4 data, within the framework of the \LCDM\ serving as the Bayesian prior. Yet, we start here with a qualitative visual overview of the recovered density and velocity fields (Fig. \ref{fig:streamlines}). 
An insight into the issue of the validation of the reconstructed LSS is provided by 
Fig. \ref{fig:LSS} which presents the large scale WF reconstructed over-density field, $\delta=(\rho-\bar{\rho})/\bar{\rho}$ (where $\bar{\rho}$ is the mean density of the Universe) 
The distribution of the LEDA  galaxies \citep{2014A&A...570A..13M} is overlaid on the over-density  color/contour map, for the sake of orientation and visual inspection of the quality of the reconstruction. 
The large scale velocity field is shown by means of flow lines (Fig. \ref{fig:streamlines}).

A detailed analysis of the reconstruction of the LSS of our local patch of the Universe is beyond the scope of the current paper. Yet, we note here that the LSS that emerges here is in very good agreement with the one inferred from the CF2 WF reconstruction  (see Appendix \ref{appdx:CF4_CF2}).  A series of papers on the linear WF/CRs out of the CF2 data has been published and the interested readers are referred to these \citep[e.g. ][]{2014Natur.513...71T,2015MNRAS.449.4494H,2017NatAs...1E..36H}   

\subsection{Cosmic vs. constrained variance}
\label{sec:variance}

\begin{figure*}
\centering
\includegraphics[width=0.45\linewidth]{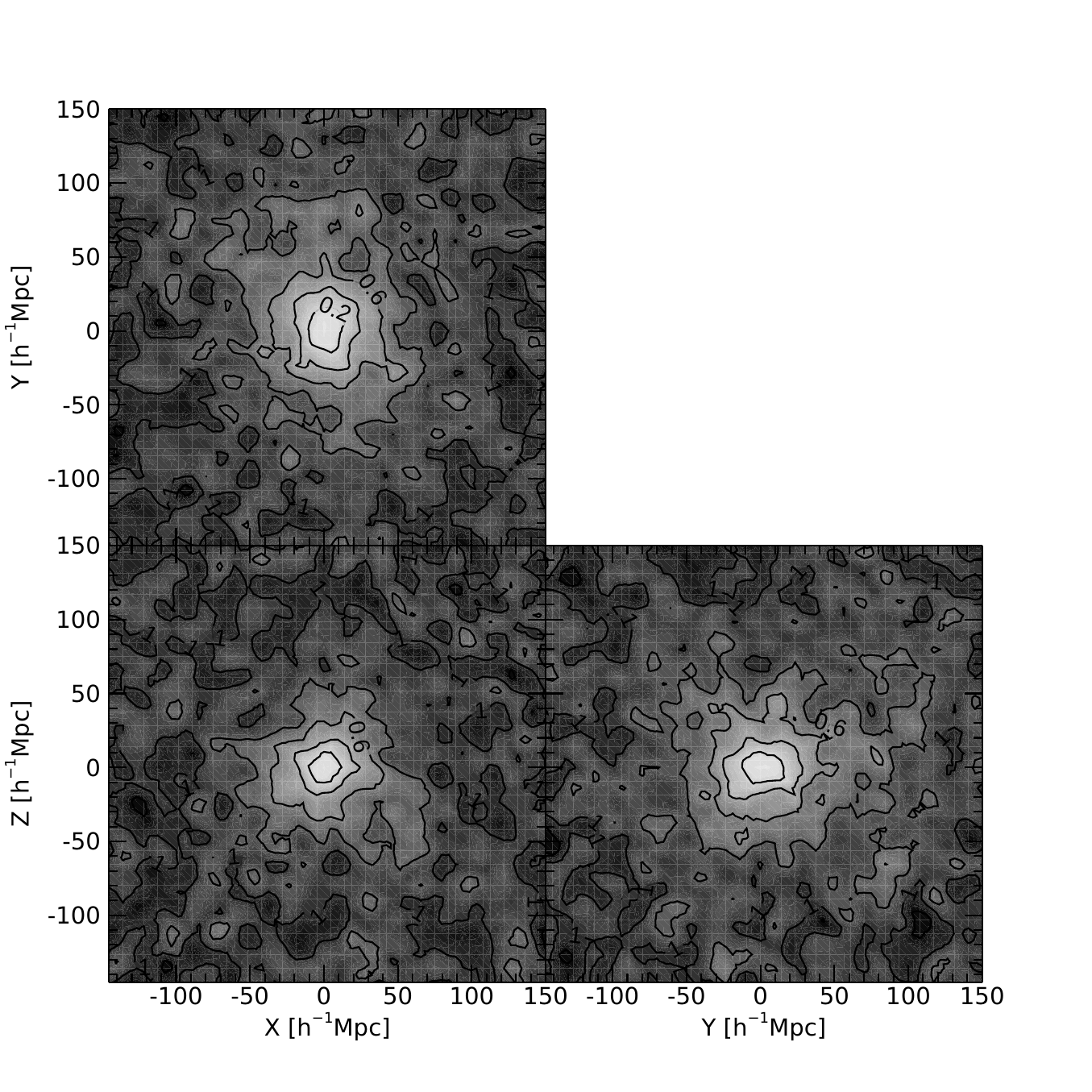}
\includegraphics[width=0.45\linewidth]{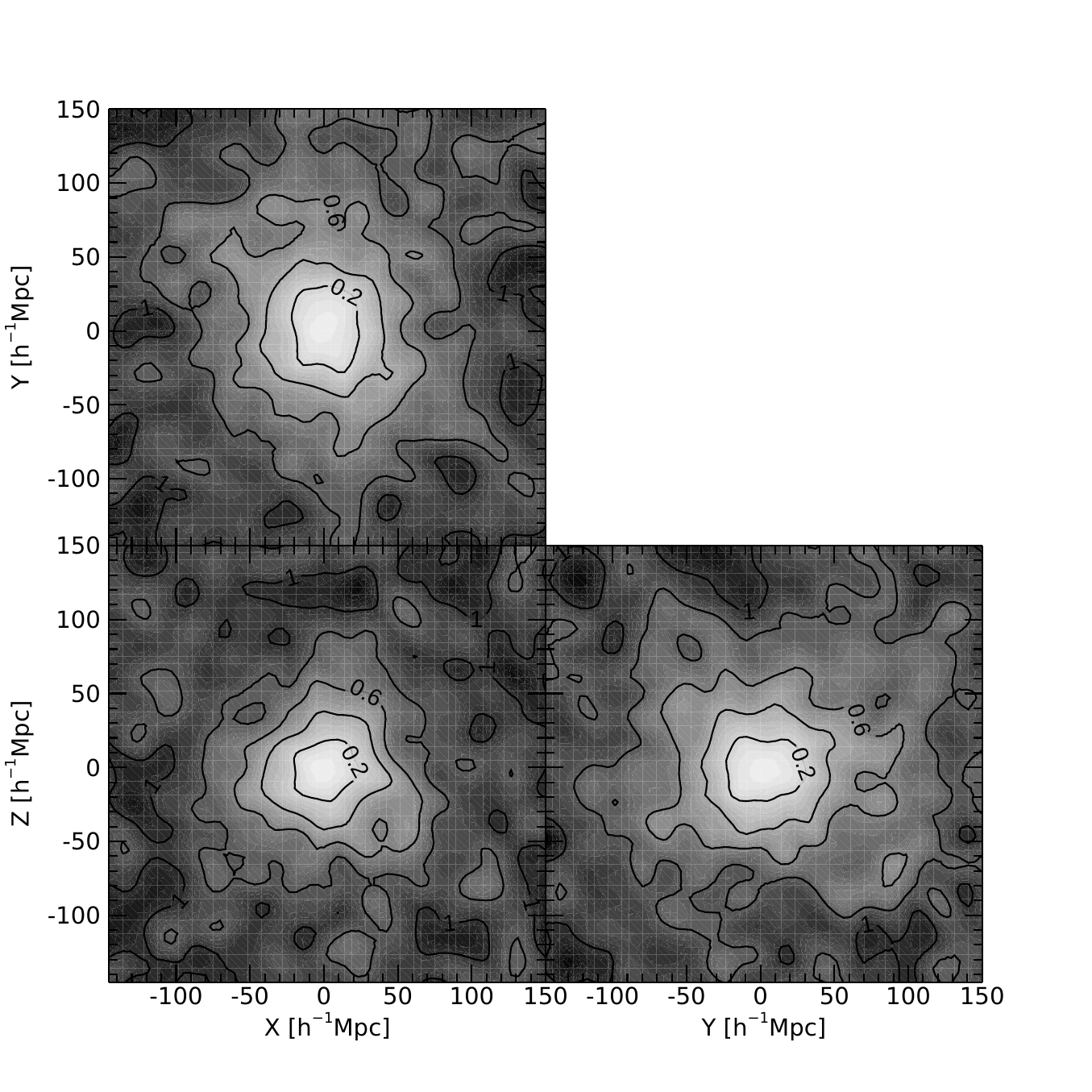}
\caption{Contour map of the constrained variance, normalized by the cosmic variance, taken over an ensemble of 50 CRs for the un-smoothed density field (left panel) and with Gaussian smoothing with $R_s=10.0\,\hmpc$ (right panel). The CRs are evaluated on a cubic grid of $N=128^3$ and a box of side length of $L=600\,\hmpc$.  }
\label{fig:variance}
\end{figure*}

Fig. \ref{fig:variance} depicts the behaviour of the constrained variance manifested by an ensemble of CRs constrained by the CF4 data. The constrained variance is calculated as the variance of the residual of the CRs from their mean value, namely the WF field. Two different resolutions are used - the case of no smoothing on a grid of $N=128^3$ and a box of side $L=600\,\hmpc$ (left panel) and Gaussian smoothing with a kernel of $Rs=10.0\,\hmpc$ (right panel). The contour/color maps presents the cosmic variance normalized by the (resolution dependent) cosmic variance. The no smoothing case is dominated by the virtually unconstrained short waves, hence the constrained variance is virtually identical to the cosmic one. For the $Rs=10.0\,\hmpc$ smoothed CRs one finds an inner region of a radius of $\sim25\ (70)\,\hmpc$ with the constrained variance smaller than 20 (50) \% of the cosmic variance. It follows that the Virgo cluster (at a distance of $\sim10\,\hmpc$) is strongly constrained by the data and the Coma cluster is less so. 

\subsection{Monopole and dipole moments}
\label{sec:multipoles}

\begin{figure}
  \centering
\includegraphics[width=1.\linewidth]{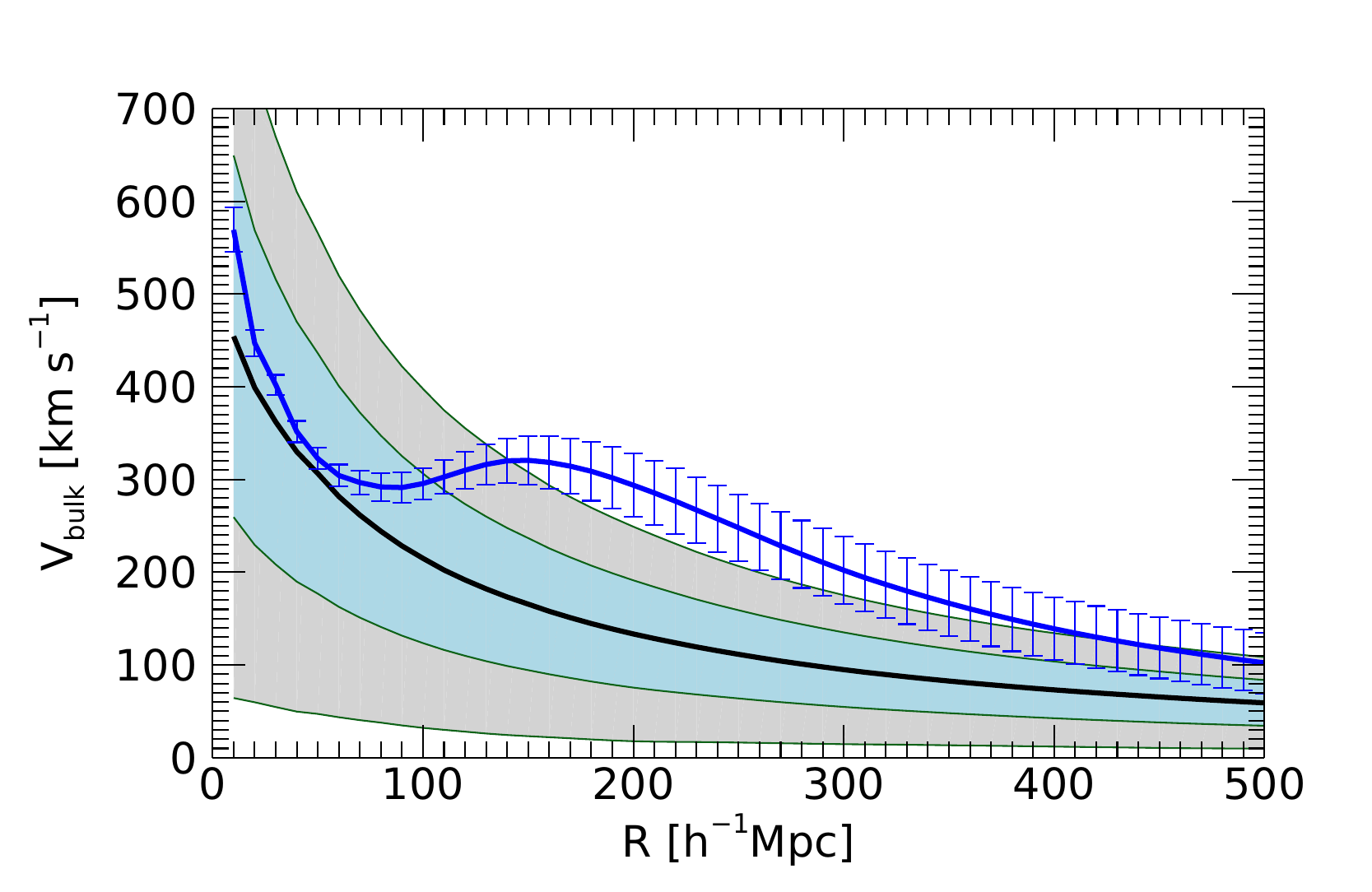}
\includegraphics[width=1.\linewidth]{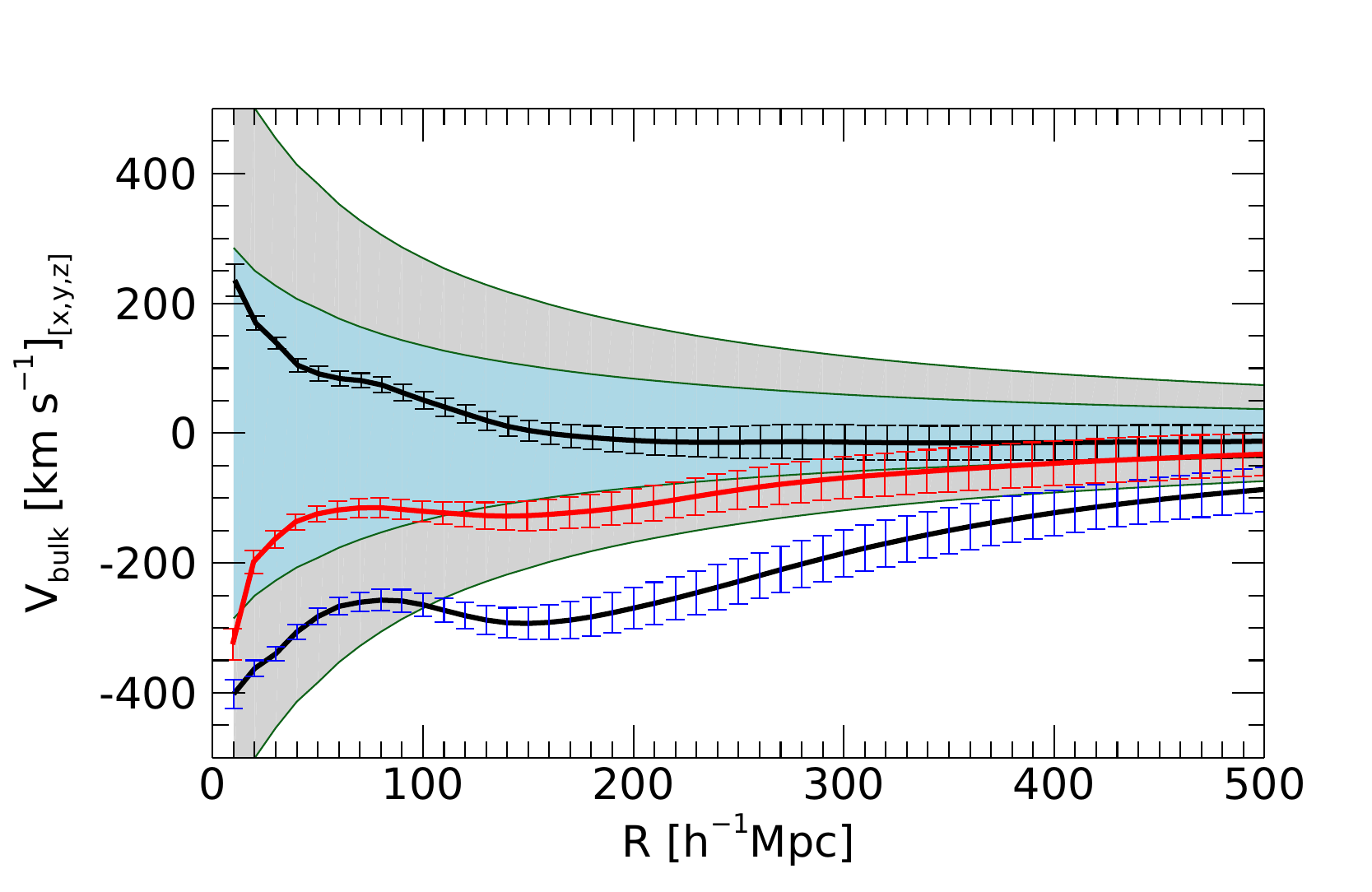}
\caption{  The amplitude of the bulk velocity (upper panel): The mean and scatter taken over the  ensembles   of the  constrained   (blue solid line, error bars) and of  the  random realizations (black solid lines, 1   sigma uncertainty (light blue) and 2 sigma (light grey) shaded regions).   The bulk velocity is calculated within spheres of radius $R$.  The three Supergalactic Cartesian components of the bulk velocity (lower panel): The mean and scatter of SGX (blue), SGY (black) and SGZ (red) components of the bulk velocity. The shaded regions (1   sigma, light blue, and 2 sigma, light grey) show the cosmic variance of one Cartesian component of the bulk velocity. The cosmic mean value of each of the individual components is zero and therefore is not shown. 
}
\label{fig:Vbulk}
\end{figure}

\begin{figure}
\centering
\includegraphics[width=1.\linewidth]{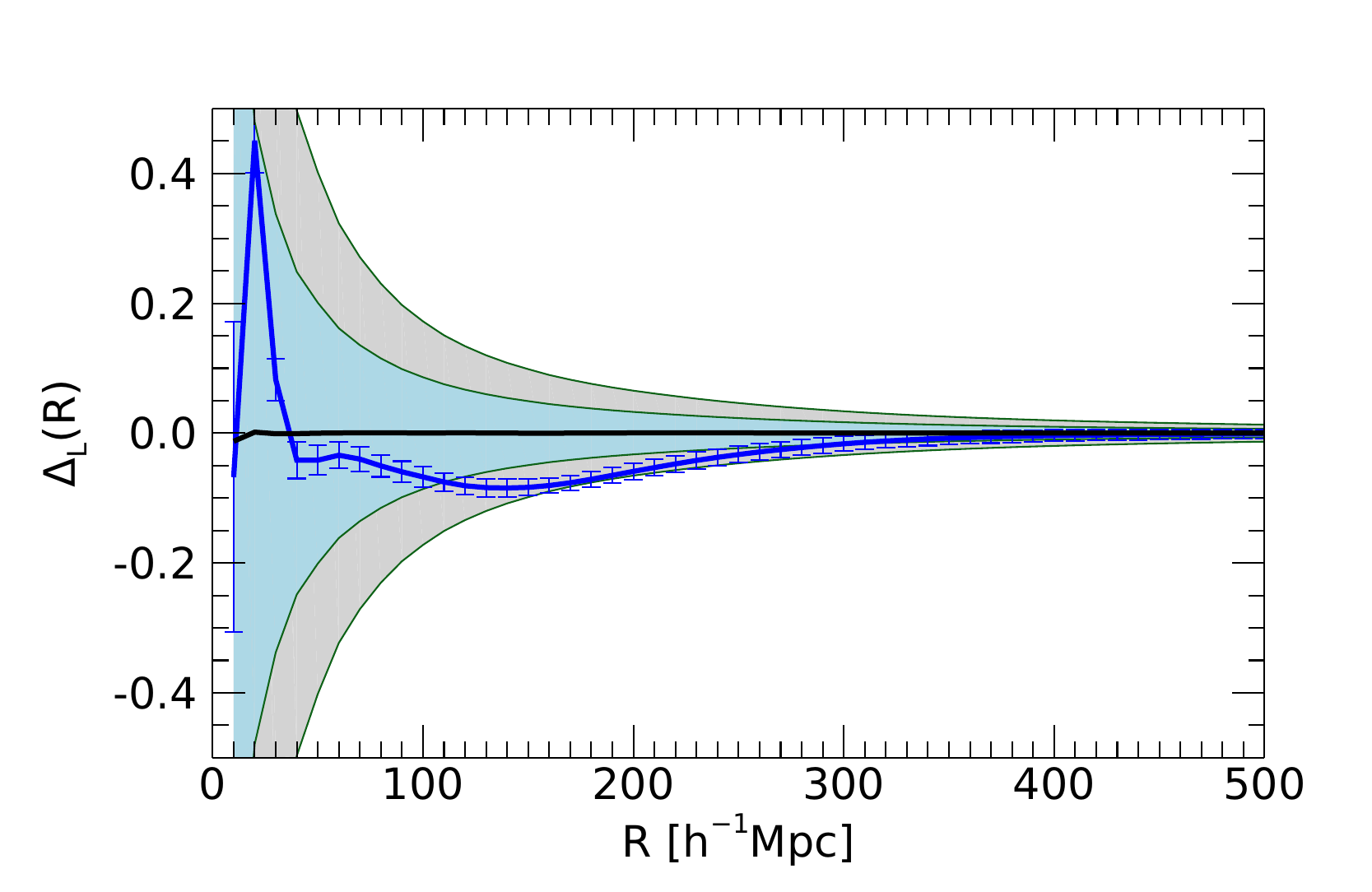}
\caption{The estimated cumulative  linear density field in spheres of radius $R$ (\DeltaLR) (conventions of the different lines are identical to those employed in the upper panel of Fig. \ref{fig:Vbulk}).  
}
\label{fig:monopole}
\end{figure}

The monopole and dipole moments, namely \DeltaLR\ and \VbulkR, have been calculated for the  ensembles of the CRs   and of the  random realizations of box sized of $L=1000\,\hmpc$, for which the mean profiles and the scatter (namely standard deviation) around the mean have been calculated. Fig. \ref{fig:Vbulk} presents the mean and variance of the constrained and random realizations of the norm of the bulk velocity (upper panel) and its three Supergalactic Cartesian components.   Fig. \ref{fig:monopole} shows the mean and scatter of the ensemble of the constrained  and random realizations of the monopole moment.      Table \ref{table:Vbulk}  shows the mean and scatter of the  ensemble of CRs of the alignment of ${\bf V}_{\mathrm bulk} (R)$ with CMB dipole velocity ( ${\bf V}_{\mathrm CMB} = [-410, 353, -324]\,\kms$),
$\mu_{\mathrm CMB}(R) = \hat{\bf V}_{\mathrm bulk}(R) \cdot \hat{\bf V}_{\mathrm CMB} $.  
The statistical significance  of the estimated alignment, $\mu_{\mathrm CMB}(R)$ is tested against the alignment calculated for an ensemble of 3,000 random realizations. As these realizations  do not have a counterpart to ${\bf V}_{\mathrm CMB}$, the alignment is calculated between the bulk velocity at a radius $R$ and its value at a minimal radius, taken to represent the zero lag, $\mu_{\mathrm self}(R) = \hat{\bf V}_{\mathrm bulk}(R_{\mathrm min}) \cdot \hat{\bf V}_{\mathrm bulk}(R) $. Here $R_{\mathrm min}=10\,\hmpc$ is chosen.
In practice this is done by constructing 3,000 random realization, evaluating the bulk velocity profile for each realization and calculating the resulting $\mu_{\mathrm self}(R)$ profiles. This is done also for 60 CRs of the actual CF4 data. Given the distribution of the  $\mu_{\mathrm self}(R)$ profiles of the random realizations the likelihood of a given  $\mu_{\mathrm self}(R)$ profile of a given CR at a given radius $R$ is gauged as follows. 
The mean and scatter, over the ensemble of CRs, of the fraction of the random realization for which $\mu_{\mathrm self}(R)$ is smaller than the corresponding value of a given CR is  $0.8 \pm 0.05$, 
with little variation as a function of R. Namely, the alignment of the bulk velocity with itself at zero lag is consistent with \LCDM\ to within roughly $\sim1.5\sigma$.

The cosmological mean profile of \VbulkR\  is a monotonically decreasing function of depth ($R$). The constrained mean radial profile of \Vbulk\ starts very close to CMB dipole velocity, in magnitude and direction, at the first radial bin considered here ($R=10\,\hmpc$), and then decays out to $R\sim70\,\hmpc$, follows by a hump that peaks at $R\sim160\,\hmpc$.  
Fig. \ref{fig:wf_CF4_CF3_CF2_dipole_divv} helps one  to trace the contribution of the three major sub-samples of the CF4 data to monopole (lower panel)  and dipole (upper panel)  moments.   It  shows the moments of the WF reconstructed velocity field from the full CF4 data, the CF4 without the 6dFGS and SDSS components (dubbed here as the 'others'), of the CF4 without the 6dFGS component and the combined `other' and SDSS data. One should note here that the 'others' data constitutes an updated version of the CF2 data, and the 'others' plus the 6dFGS data are the current equivalent  of the CF3 data.
The lower panel of Fig.  \ref{fig:Vbulk}      shows that the SGY and SGZ components of \Vbulk\ lies within or close to the 1 sigma of the cosmic variance as opposed to the SGX component - the one that deviates from the 2 sigma cosmic variance outside of $R\sim100\,\hmpc$.                                                                                                                                          
Fig. \ref{fig:wf_CF4_CF3_CF2_dipole_divv} depicts the radial profiles of the WF calculated bulk velocity (upper panel) and the monopole moment (lower panel). The plots further decompose the contributions to these moments by the main components of the CF4 data. Fig. \ref{fig:wf_CF4_CF3_CF2_dipole_divv} clearly shows that the hump in the \Vbulk\ profile is contributed by the 6dFGS sub-sample. The CF4 data without the 6dFGS does not show that rise in the \Vbulk\ profile. Fig. \ref{fig:Vbulk}  shows that it is the SGX component of the \Vbulk\ that is a responsible for the excess power. 

\begin{table*}
\begin{tabular}{||c c c c c c c c c c ||}
\hline
R [$\hmpc$]    &  $V_{\mathrm bulk}^{CRs}$  & $V_{\mathrm bulk}^{RANs}$   & $V_{\mathrm bulk,\,x}$ & $V_{\mathrm bulk,\,y}$ & $V_{\mathrm bulk,\,z}\ [km\ s^{-1}]$ & \DeltaLR\ & $\mu_{\mathrm CMB}$ & $\chi^2$ & $P(>\chi^2)$\\
\hline
 10  & 557 $\pm$ 11 & 441 $\pm$ 187 & -404 $\pm$ 8 & 245 $\pm$ 10& -293 $\pm$ 13 & 0.33 $\pm$     0.11  &  0.99 $\pm$  0.002 &  4.29 &  2.3e-01\\
 40  & 354 $\pm$ 10 &  320 $\pm$ 133 &     -308  $\pm$  10 &  103  $\pm$  9   & -137  $\pm$  14  &              -0.04  $\pm$     0.03       &       0.93  $\pm$  0.009  &   3.51  & 3.2e-01\\
 50  & 320 $\pm$ 13 & 295 $\pm$ 123 &     -280  $\pm$  13 &  90  $\pm$  10   & -125  $\pm$  13  &              -0.04  $\pm$     0.02       &       0.93  $\pm$  0.014  &   3.29  & 3.5e-01\\
100    &         295   $\pm$    18      &       211    $\pm$    84  &   -264 $\pm$ 19    &    46 $\pm$ 11   &   -121 $\pm$ 16           &       -0.07$\pm$    0.01        &        0.88 $\pm$ 0.021 &  5.34  & 1.5e-01 \\
150    &         318   $\pm$    27      &       162    $\pm$     65  &  -290 $\pm$  26    &    4 $\pm$  15   &  -127 $\pm$  21         &        -0.06 $\pm$     0.01        &       0.81 $\pm$  0.031 &  10.03  & 1.8e-02 \\
200    &         292   $\pm$    30      &       131    $\pm$     51  &  -267 $\pm$  31    &    -8 $\pm$  19   &  -111 $\pm$  27      &           -0.06 $\pm$     0.01     &          0.78 $\pm$  0.042  &  12.14 &  6.9e-03 \\
250    &         247   $\pm$   123      &       109    $\pm$     43  &   -227 $\pm$  32   & -14 $\pm$  23   &   -87 $\pm$  29        &         -0.03 $\pm$     0.01     &          0.75 $\pm$  0.059 & 11.43 &  9.6e-03 \\
300    &         210   $\pm$   36       &       94    $\pm$     40  &   -185$\pm$  32     & -13 $\pm$  26   &   -72 $\pm$  31        &         -0.02 $\pm$     0.01     &          0.73 $\pm$  0.097 & 11.14 &  1.1e-02 \\
500    &         102   $\pm$   32       &       59    $\pm$     24  &   -86$\pm$  34      & -12 $\pm$  24   &   -32 $\pm$  32        &         -0.005 $\pm$     0.01     &          0.65 $\pm$  0.195 & 6.36 &  9.5e-02 \\
\hline
\end{tabular}
\caption{
The mean and standard deviation taken over and ensemble of 60 constrained realizations (CRs) of the velocity field constrained by the CF4 data. The bulk velocity of an ensemble of 60 random realizations (RANs), with the same phases as the corresponding CRs, is shown for reference. 
The different columns present the mean and scatter of the amplitude of the bulk velocity  taken over an ensemble of constrained  ($V_{\mathrm bulk}^{CRs}$) and random ($V_{\mathrm bulk}^{RANs}$) realizations, of the Supergalactic Cartesian components of the bulk velocity ($V_{\mathrm bulk,\,x}$, $V_{\mathrm bulk,\,y}$, $V_{\mathrm bulk,\,z}$), the spherical mean linear over-density (\DeltaLR), the alignment of the bulk velocity with the CMB dipole  ($\mu_{\mathrm CMB}$), the  $\chi^2$ value of \Vbulk\ and the  probability of having such a $\chi^2$ of 3 degrees of freedom  larger than the estimated value ($P(>\chi^2)$). 
}
\label{table:Vbulk}
\end{table*}

\begin{figure}
\centering
\includegraphics[width=1.\linewidth]{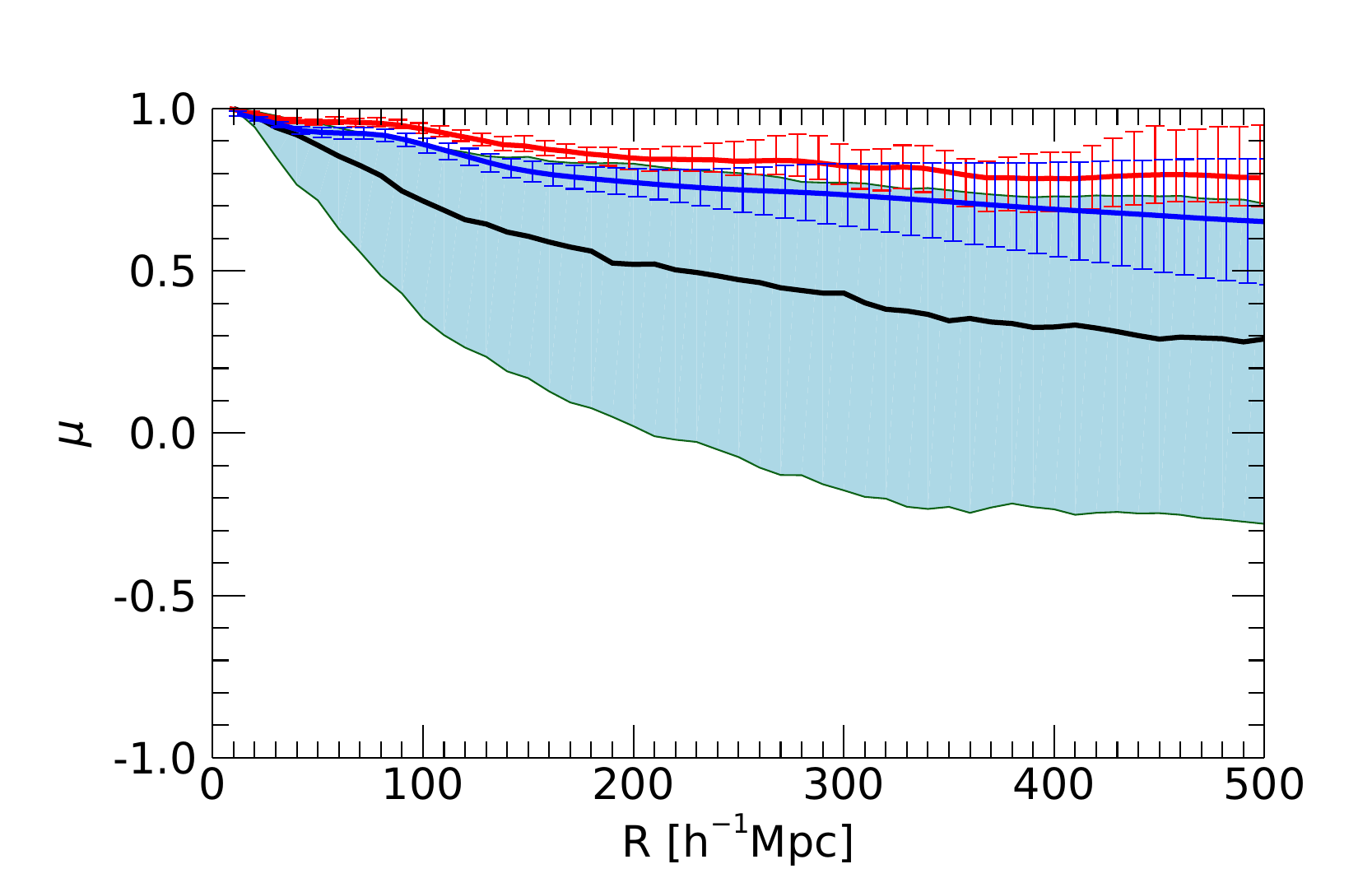}
\caption{The alignment of the bulk velocity of a sphere of radius R with the CMB dipole velocity ($\mu_{\mathrm CMB}$) is presented by mean and variance taken over an ensemble 60 CRs (blue)   To enable a comparison with the random realizations,  the alignment with the CMB dipole is 
closely approximated by plotting  the alignment of the bulk velocity at $R$ with itself at zero lag, 
$\mu_{\mathrm self}(R) = \hat{\bf V}_{\mathrm bulk}(R_{\mathrm min}) \cdot \hat{\bf V}_{\mathrm bulk}(R) $. 
The plot shows the median  and the 1st and 3rd quartiles  of the distribution of the ensembles of 60 CRs (red) and 3000 random realizations (median in black solid line  and the uncertainty in shaded light blue). Here $R_{\mathrm min}=10\,\hmpc$.
}
\label{fig:mu}
\end{figure}

\begin{figure}
    \centering
    \includegraphics[width=1.\linewidth]{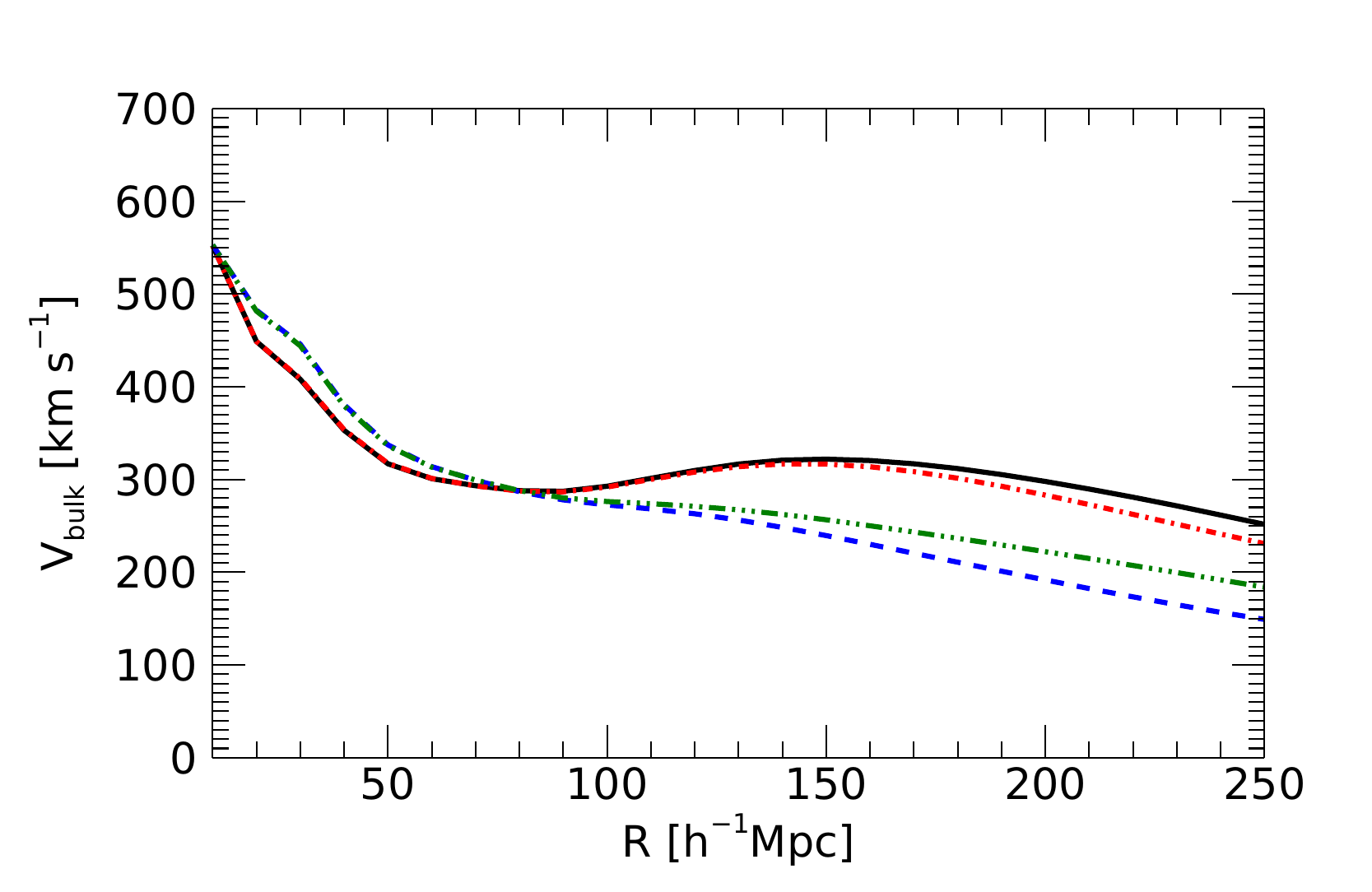}
    \includegraphics[width=1.\linewidth]{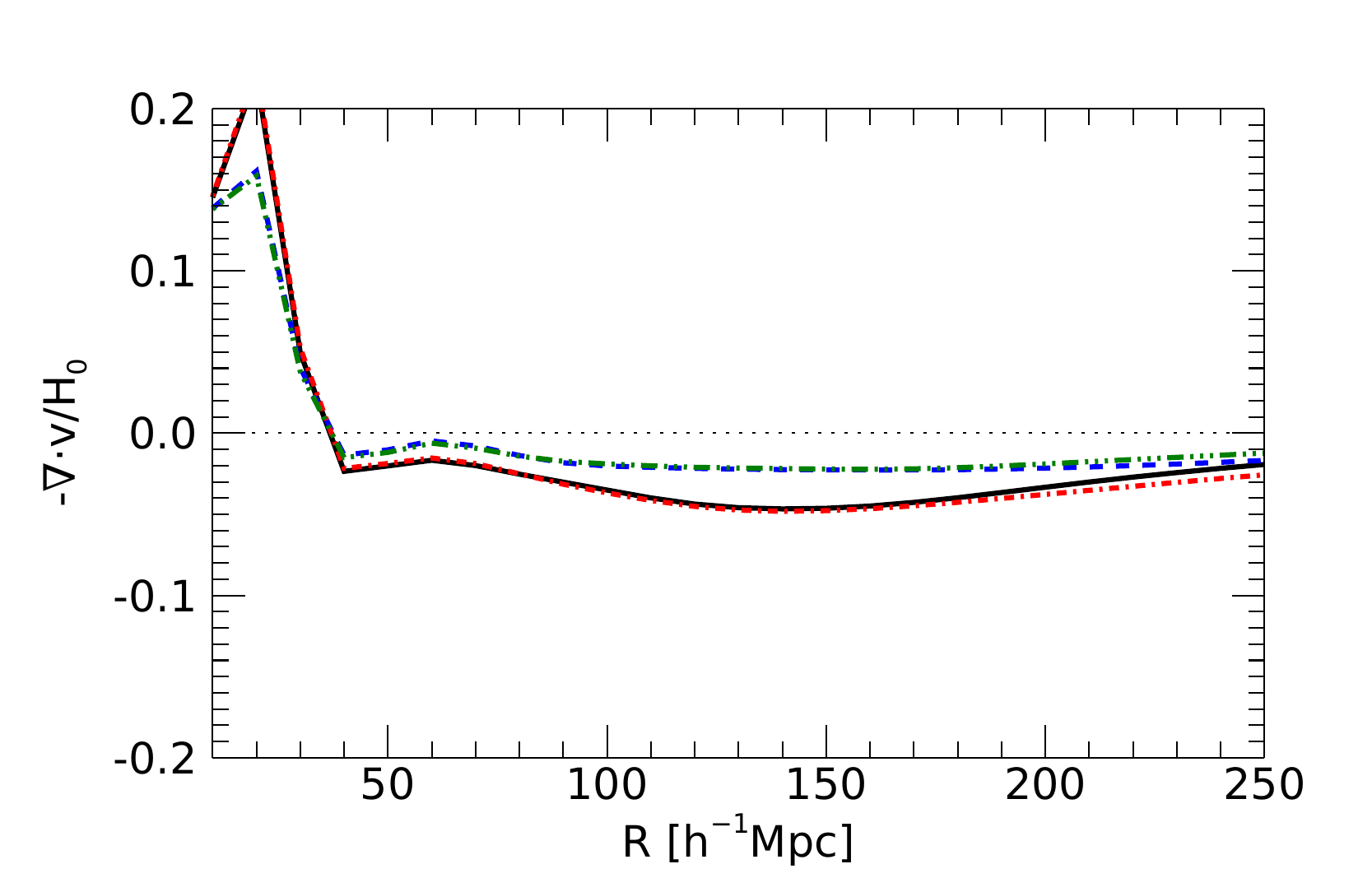}
    \caption{ 
    The amplitude of the WF reconstructed bulk velocity (upper panel) and of the monopole moment (lower panel) calculated from  the entire CF4 data (black - solid line), the CF2-like data, (i.e. the CF4 without the SDSS and without the 6dFGS data; blue - dashed), the CF3-like data (i.e. CF4 without the SDSS data; red - dot-dash) and the CF4 without the 6dFGS data (green - dot-dot-dot-dash). 
    }
    \label{fig:wf_CF4_CF3_CF2_dipole_divv}
\end{figure}

\subsection{Dipole velocity: $\chi^2$ statistics }
\label{sec:chisq}

\begin{figure}
    \centering
    \includegraphics[width=1.\linewidth]{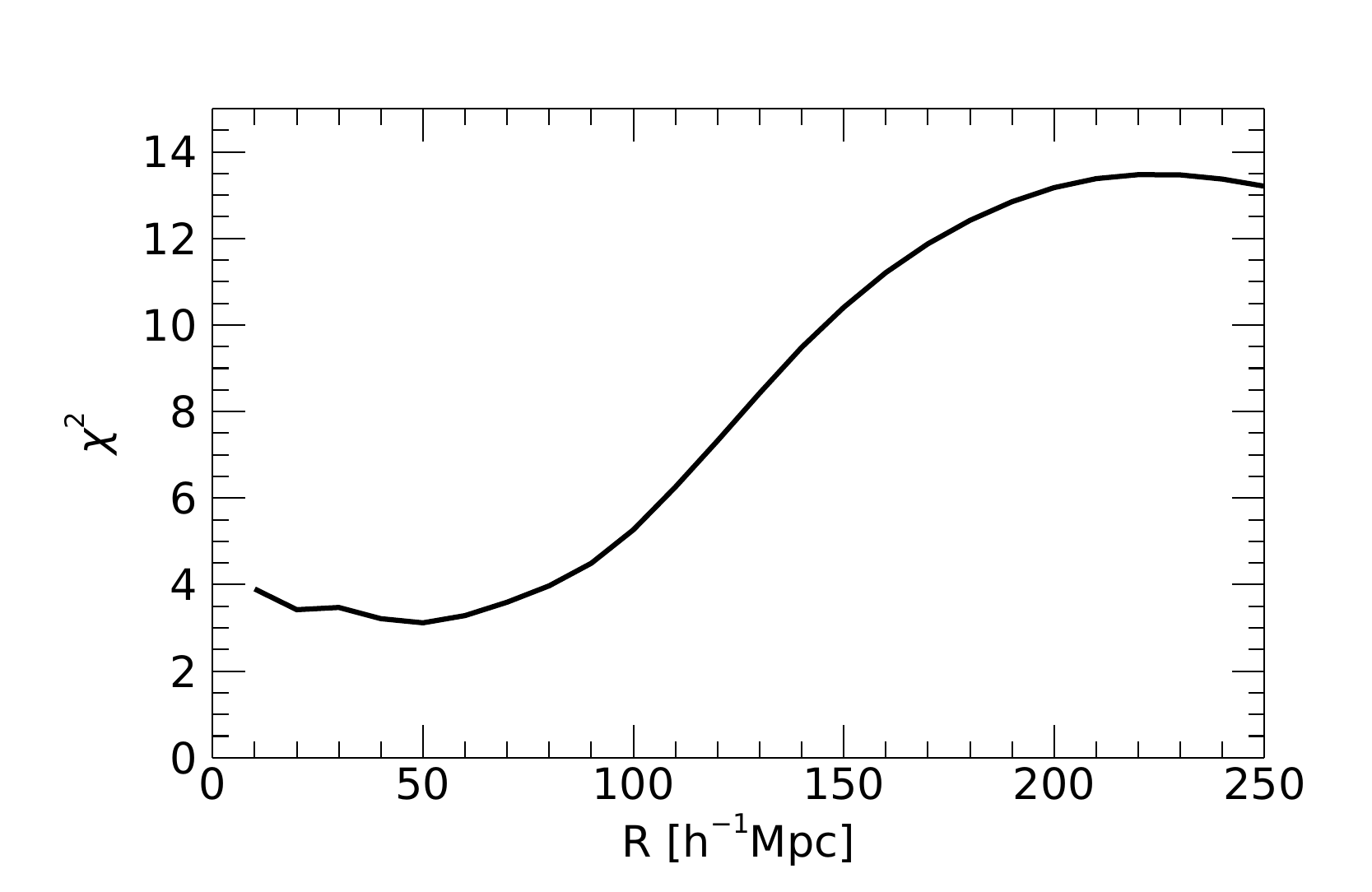}
    \includegraphics[width=1.\linewidth]{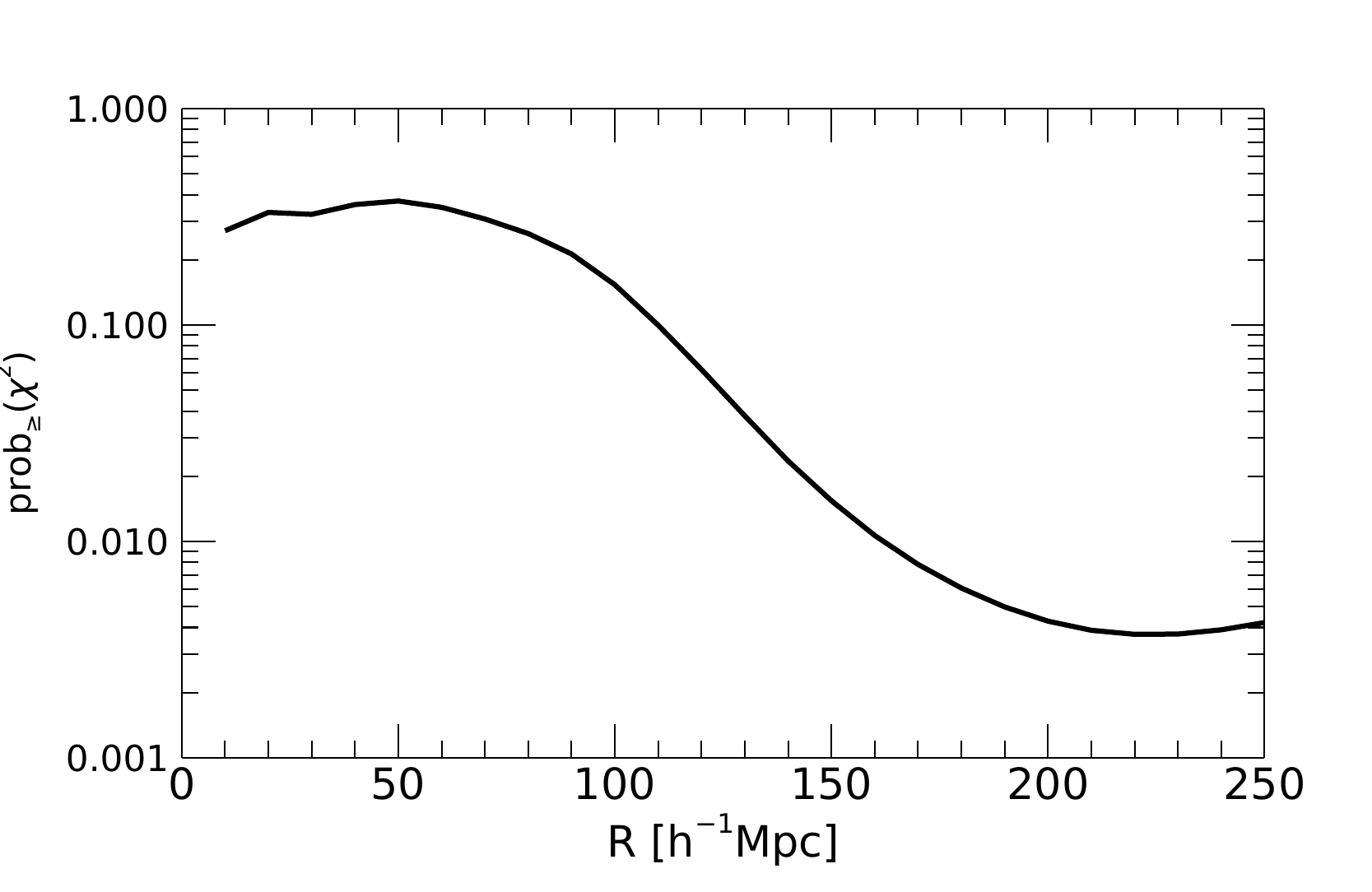}
    \caption{ 
    The $\chi^2$ (of 3 degrees of freedom)  statistics of the amplitude of the bulk velocity: The upper panel shows the $\chi^2$ profile as a function of $R$ - the radius within which \VbulkR\ is calculated. The lower panel presents the probability of obtaining a $\chi^2$ equal or larger than the one of the estimated \VbulkR.
    }
    \label{fig:chisq}
\end{figure}

The top panel of Fig.~\ref{fig:chisq} shows $\chi^2(R)$ plotted out to $R=250\,\hmpc$. The lower panel shows the probability of obtaining a $\chi^2$ equal or larger than the one calculated here, recalling a $\chi^2$ distribution with 3 degrees of freedom (d.o.f.). Table \ref{table:Vbulk} presents these numbers for a few cases of $R$.  In particular we find here  at $R=150\ (200)\,\hmpc$ a $\chi^2=9.99\ (12.48)$ and $P(>\chi^2)=1.9\times10^{-2}\ (5.9\times10^{-3})$. These results  are further discussed in \S\ref{sec:comparison}.

\subsection{Dipole velocity: tidal decomposition}
\label{sec:tidal}

\begin{figure}
  \centering
  \includegraphics[width=1.\linewidth]{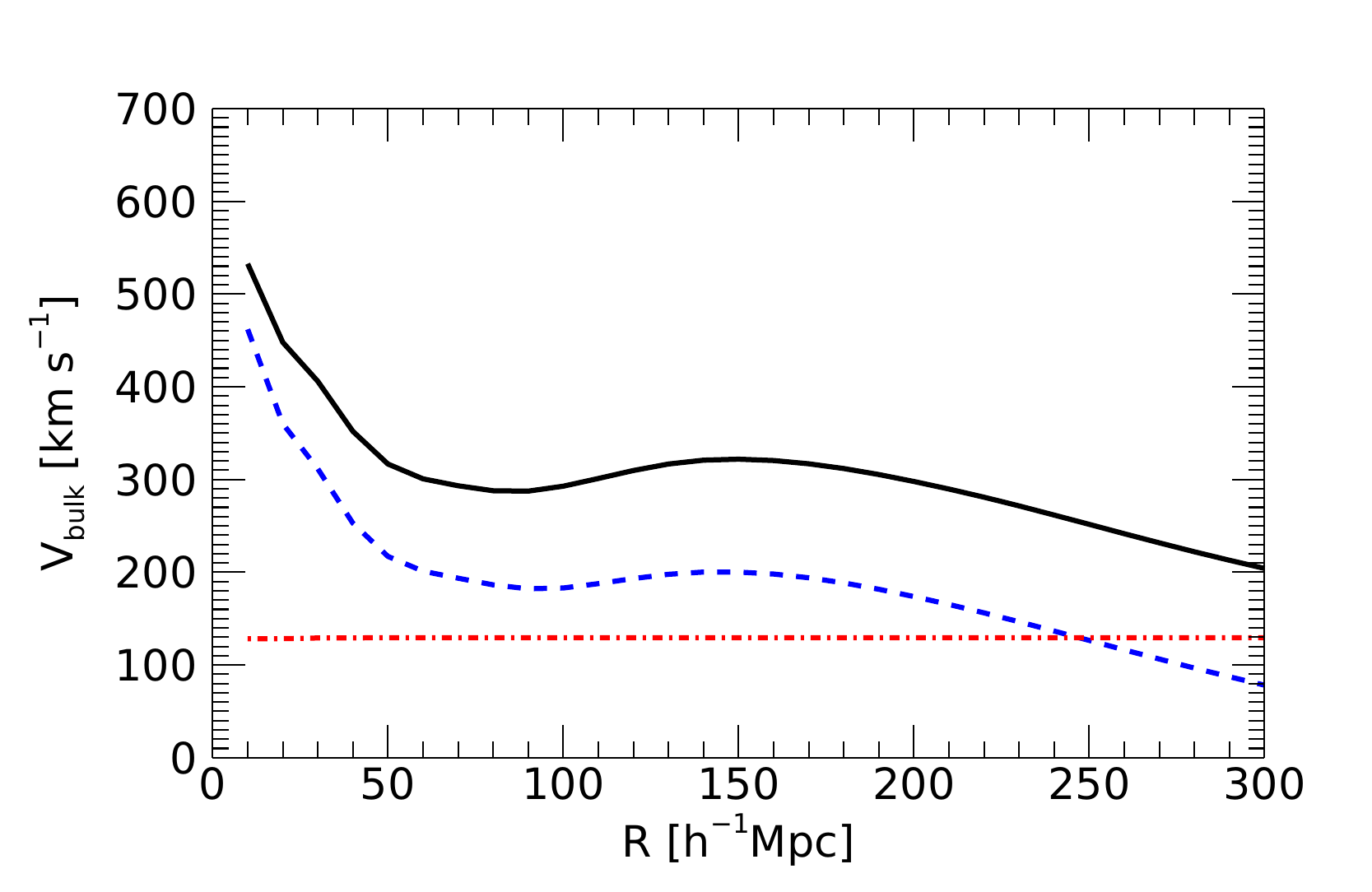}
\caption{Decomposition of the  velocity field into its tidal and local components with respect to sphere of radius $R=300\,\hmpc$ (see text in \S~\ref{sec:tidal}): The bulk velocity of the full velocity field (solid, black), of the local component (dashed, blue) and of the tidal component (dot-dashed, red). The error bars of the case of the full flow (shown in Fig. \ref{fig:Vbulk}) serve as upper limits on the bulk velocities of the two components.
}
\label{fig:tidal}
\end{figure}

 Fig. \ref{fig:tidal} shows the tidal decomposition of the WF reconstructed velocity field from the CF4 data, with respect to a sphere of a radius  $R=300\,\hmpc$.  The main feature is that the tidal component of the bulk velocity is essentially constant at $V_{\mathrm{bulk}}^{\mathrm tidal}\sim 129\,\kms$. This means that $\sim20\%$ of the CMB dipole velocity, i.e. the motion of the LG in the CMB frame of reference, is induce by structures in the universe beyond the edge of the CF4 data.

There are two reasons for dwelling here on the issue of the tidal decomposition. One is the issue of the convergence of the bulk flow. Namely, the bulk velocity converges to zero as the radius $R$ increases. For the CF4 data, with its effective depth of $\sim300\,\hmpc$, and within the \LCDM\ model, the tidal residual flow,  induced by structures beyond that depth, is roughly $63\,\kms$. 
The other motivation for the performance of such decomposition is for the sake of comparison with other reconstruction methods of the large scale velocity field.

Commonly, reconstructions of the local flow from  galaxy redshift surveys are based on the estimation of the local density, from which the  velocity field is recovered by using the linear theory (Eq. \ref{eq:v_k}).
That reconstructed `local'  velocity field needs to be augmented by an estimated tidal component, which is often done by requiring that the sum of the divergent and tidal fields equals the CMB dipole velocity field at the Local Group. Our evaluation of the divergent component enables a direct comparison of the WF reconstructed velocity field with that calculated from the galaxy distribution, and thereby bypassing the need to evaluate a tidal external field. A similar problem is encountered by Bayesian methods applied  to peculiar velocities where the target of the reconstruction is the density field in its Fourier representation - such as in the Markov Chain Monte Carlo \citep[MCMC;][]{2016MNRAS.457..172L,2019MNRAS.488.5438G} 
and the Hamiltonian Monte Carlo 
\citep[HMC;][]{2022MNRAS.517.4529B,2022MNRAS.513.5148V,2023MNRAS.519.2981V}. The periodic boundary conditions employed by the FFT suppresses the tidal component of the velocity field. The comparison of our work with those FFT based reconstructions needs to be done with respect to the divergent velocity field only.

\subsection{Dependence on Hubble's constant}
\label{sec:H0}

At its core the Cosmicflows is a database of the observed galaxy  redshifts and distance moduli, and an $H_0$ needs to be assumed in order to recast it as a distances and radial peculiar velocities catalog. Obviously, the most appropriate value of $H_0$ to use is the one obtained by fitting to the data. Random mock data catalogs have been used to assess the uncertainties in the determination of $H_0$ in the \LCDM\ standard model and the Cosmicflows data in the BGc framework \citep{2021MNRAS.505.3380H}. The mock data sampled the cosmic variance, namely the variation with the choice of the random observers,  and the errors variance, i.e. the variation due the observational errors. The BGc analysis of the CF3 data yielded $H_0 = 75.8 \pm 0.1 \pm 1.0 \pm 0.4\,\kmsMpc$, where first, second and third uncertainties are due to fitting formal error, the cosmic variance and the errors variance, respectively \citep{2021MNRAS.505.3380H}.
Repeating that analysis and assuming its assessed uncertainties we find here $H_0 = 74.6 \pm 1.1\,\kmsMpc$, where the uncertainty includes the cosmic variance. Without the cosmic variance the uncertainty drops to $0.4\,\kmsMpc$. This stands in excellent agreement with the  $H_0 = 74.6 \pm   0.8\,\kmsMpc$ of \cite{2023ApJ...944...94T}. The latter result refers to the actual CF4 data, observed by humans on earth, hence it does not account for the cosmic variance. It should be stressed here that we do not account here to the possibility of systematic errors - the CF4 data is taken here at  face value. In particular the zero-point calibration of the sub-samples of the CF4 data is taken here as is from the Cosmicflows analysis.  

\begin{figure}
\centerline{
\includegraphics[width=1.\linewidth]{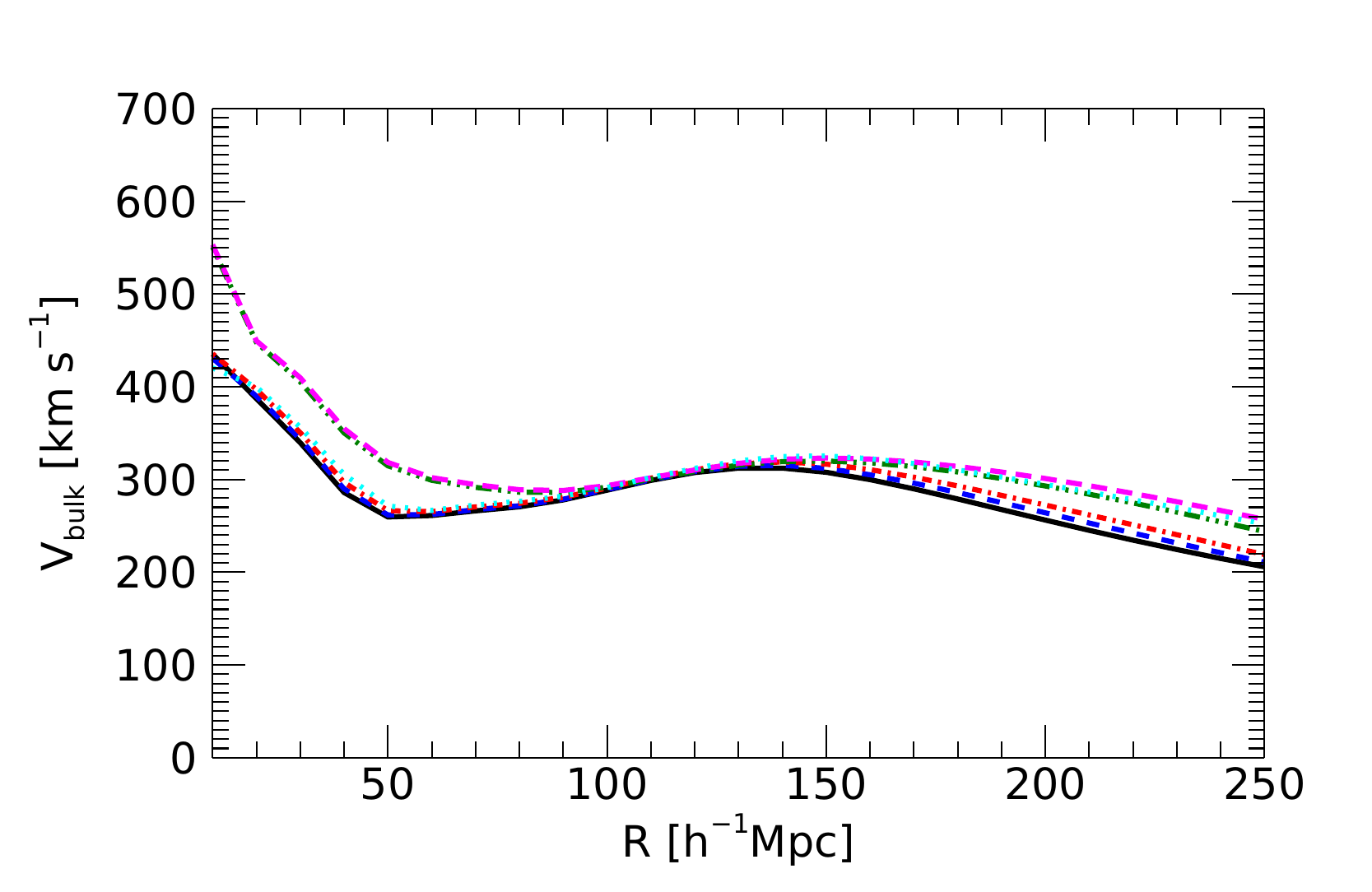}
}
\centerline{
\includegraphics[width=1.\linewidth]{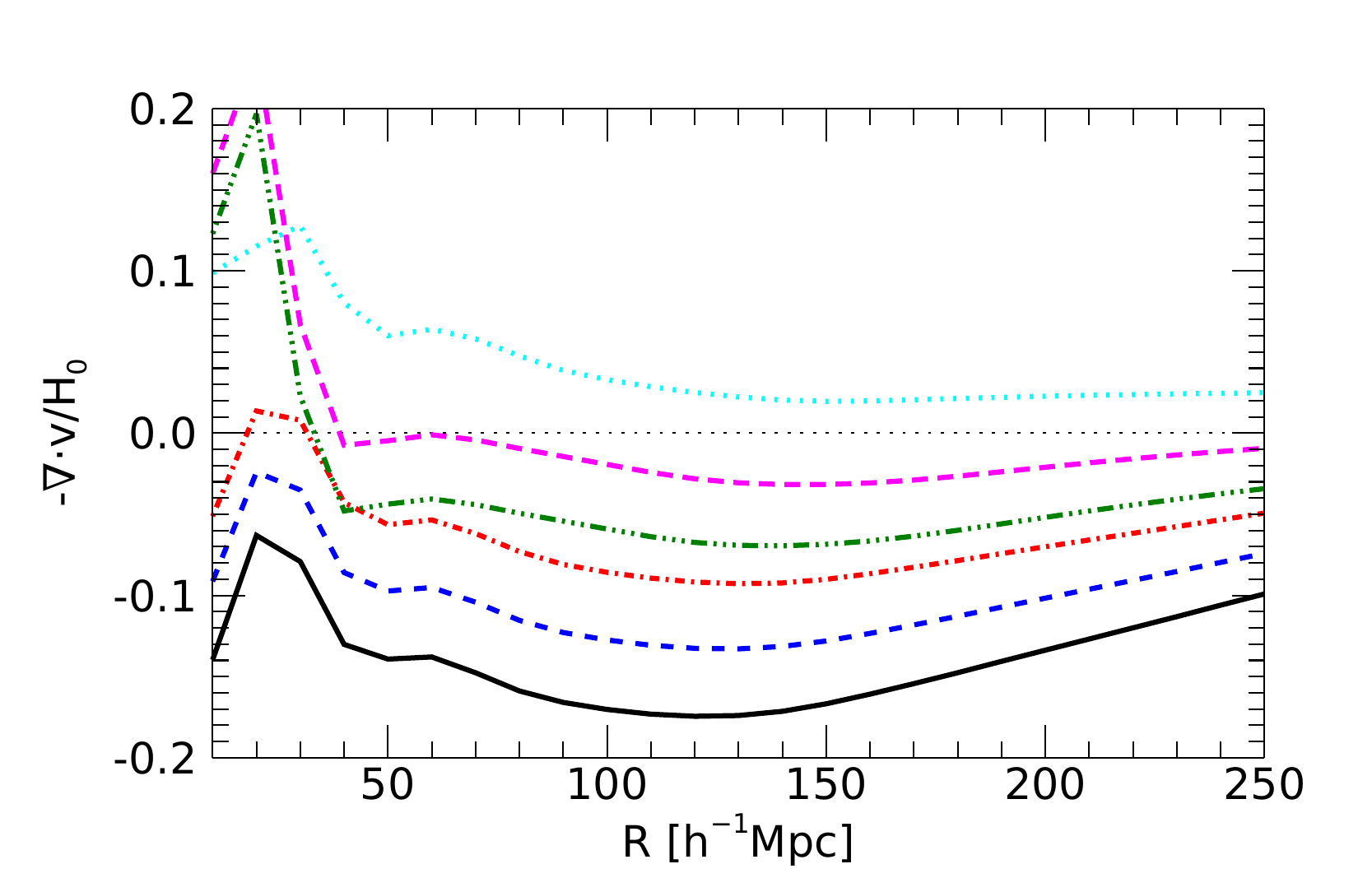}
}
\caption{The dependence on the monopole (lower panel)  and dipole (upper panel) moments  moments on Hubble's constant:  $H_0=71.0$,  (black, solid), $72.0$ (blue, dashed), $73.0$ (red, dot-dashed), $74.0$ (green, dot-dot-dashed), $75.0$ (magenta, dashed), and $76.0\,\hmpc$ (cyan, dotted). 
}
\label{fig:H0}
\end{figure}

Given the $H_0$ controversy 
the  dependence of the estimated monopole and dipole moments on $H_0$ is of interest and is examined here. The monopole moment corresponds to the local deviation  from the global value of $H_0$ - hence a strong dependence on the assumed $H_0$ is expected, as indeed depicted by  Fig. \ref{fig:H0}. Considering the bulk velocity then such a dependence is possible only in the case of an anisotropic distribution of the data points. For perfectly isotropically distributed data the monopole and dipole moments are mutually orthogonal, hence the bulk flow cannot be affected by the choice of $H_0$. The CF4 data is extremely anisotropic and inhomogeneous   (Fig. \ref{fig:aitoffl}) hence a dependence of \Vbulk\ on $H_0$ is expected.      Fig. \ref{fig:H0} indeed confirms that expectation, yet \Vbulk\ hardly changes for $H_0$ in the range of  $(40\ -\  60)\,\hmpc$.                                                                                                                                                                                                                                                                                                                                                                                                                                                                                                                                 

\section{Interpretation and comparison}
\label{sec:compare}

\subsection{Validation and assessment}
\label{sec:validation}

The quality of the combined BGc and WF/CRs algorithm is tested here against ensembles of random mock CF3 data (Appendix \ref{appdx:random_mocks}) and of constrained mock CF3 data (Appendix \ref{appdx:constrained_mocks}). The details of mock data sets, their constructions and their analysis are given in the two appendices.

We use here the monopole and dipole moments, namely the mean of the linear overdensity (up to the scaling by the growth factor) and the bulk velocity in  a sphere of radius $R$, as tracers of the large scale velocity field. Detailed analysis of reconstruction of these from  random and constrained  mock data is presented in Appendices \ref{appdx:random_mocks} and \ref{appdx:constrained_mocks}. A brief summary is given here. The main conclusion that follows from that analysis is that the residual of the WF reconstructed monopole and dipole moments from the ones of the target velocity field of the simulation is always smaller, and often much smaller, than the cosmic variance of these moments. Over the majority of ranges of $R$ considered here the WF reconstruction is within 1 to 2 sigma of  the target simulation, where sigma is the constrained variance of the ensemble of CRs. 

The conclusion we draw here is that the WF reconstruction provides a good estimation of the underlying velocity to within 1 to 2 times the constrained variance that is materialized by the scatter of the CRs.

\subsection{Comparison with other studies: bulk velocity}
\label{sec:comparison}

The bulk velocity is arguably the most basic and common characteristic of the large scale velocity field and therefore we compare here our estimation of it with some of its recent estimations from   galaxy peculiar velocities    and of galaxy redshift surveys. We start by a brief description of these studies:  
\vskip -0.5cm
\begin{enumerate}
\vskip -1.0cm
\item \cite{2015MNRAS.450..317C}:  The velocity field is calculated by means of the linear theory from the 2M++ galaxy redshift compilation and the external tidal contribution has been fitted so as to minimize the residual from the observed redshifts. The bulk velocity was estimated by Gaussian weighting.  
\item \cite{2015MNRAS.449.4494H}: The LSS was reconstructed from the CF2 data by means of the Bayesian WF/CRs methodology, same as the one employed here, but with a different bias correction scheme. The bulk velocity is defined by the volume weighting in a sphere.
\item  \cite{2021MNRAS.507.1557L}: The velocity field is estimated from the 2MRS galaxy redshift survey by means of linear CRs, assuming the \LCDM\ standard model as a prior. The external tidal component is estimating by fitting to the CF3 data. A Gaussian window function is used to calculate the bulk flow.  
\item \cite{2016MNRAS.455..386S}: Minimal variance estimation of the bulk flow from the 6dFGS velocities survey.
\item \cite{2019MNRAS.482.1920Q}: Bulk flow is estimated by means of a maximum likelihood analysis from the 2MTF velocity survey. 
\item  \cite{2022MNRAS.517.4529B}: The large scale velocity field is reconstructed by a linear Bayesian estimator - the Hamiltonian Monte Carlo algorithm - from the combined SFI++ and 2MTF velocity surveys. The bulk flow is calculated by means of a Gaussian filter.  
\item  \cite{2023MNRAS.524.1885W}: A minimal variance estimation of the bulk velocity of a spherical volume from the CF4 data. The authors use a novel bias correction algorithm of the observed radial peculiar velocities.  
\item \cite{2023A&A...670L..15C}: A Bayesian reconstruction of the LSS from the grouped CF4 data by means of the Hamiltonian Monte Carlo algorithm. The bulk velocity is the volume weighted average within spheres of radius $R$. 
\item   \cite{2023arXiv230611269W}: An estimation of the bulk flow from the CF4 data by means of the maximum likelihood   and by  the minimum variance estimators.
\item  Current work: The present work estimations of the bulk velocity for an effective radii of $40$ and $50\,\hmpc$ are expressed in terms of the Gaussian weighting, so as to enable a direct comparison with the other studies. The Gaussian weighting calculation is based on a small  ensemble of 60 random realizations, hence the uncertainty of the calculated $\chi^2$   is    $\sim 0.2$.  
\end{enumerate} 
\vskip -0.5cm
One should recall that all the HMC-based reconstructions \citep[and Valade et al., in prep.]{2022MNRAS.517.4529B,2023A&A...670L..15C} are done in Fourier space employing the periodic boundary conditions, hence recovering the local component of the velocity field.

The fact that all the estimations of the direction of the bulk velocity quoted here agree to within their formal errors is gratifying. The angular positions  of the data points are very precisely determined, hence in spite of the sparseness of the data and the very incomplete sky coverage of some of the databases, all the methods and databases considered here provide consistent results about the direction of the bulk velocity. This is a sanity check that all studies have passed successfully. This is not the case with the amplitude of the bulk velocity. We focus here mostly on the comparison of our results with the other ones quoted here.   
Our best agreement is found with the $V_{\mathrm{bulk}}$ profile of Fig. 11 of \cite{2021MNRAS.507.1557L}. We are also in good agreement with \cite{2019MNRAS.482.1920Q} both in terms of the bulk velocity itself, at a single effective distance, and also with respect to its $\chi^2$
 significance with respect to the \LCDM\ model. 

All the Bayesian reconstructions considered here - the present WF/CRs and the HMC ones - reconstruct the full LSS, densities and velocities, out of which the bulk velocity is calculated by means of a volume weighted average within spherical volumes. 
This is the case with the bulk flow presented in Fig. 10 of  \cite{2022MNRAS.517.4529B}, given that it is the local component of the velocity field that is calculated by the HMC algorithm, it needs to be compared with the dashed-blue curve of Fig. \ref{fig:tidal}. Visual inspection of the two figures finds an agreement out to $R\approx50\,\hmpc$, beyond which the results diverge. The bulk velocity of \cite{2022MNRAS.517.4529B} decreases monotonically with $R$ over the entire plotted range, compared with our results that show an upturn in the \Vbulk\ profile with a `hump' that peaks at $R\approx150\,\hmpc$. A similar upturn is exhibited by the \Vbulk\ profile of \cite{2021MNRAS.507.1557L}. 
Yet, Fig. \ref{fig:wf_CF4_CF3_CF2_dipole_divv}, which shows the \Vbulk\ profiles for the three components of the CF4 data,  provides the explanation of the discrepancy. 
The \Vbulk\ profile of the 'other' component, 
after correcting for its local component, is similar in shape and consistent in amplitude with the profile of \cite{2022MNRAS.517.4529B}. 
We note here that the effective depths  of the 'others' component
(Fig. \ref{fig:CDF}) and of the combined SFI++ and 2MTF \citep[used by][]{2022MNRAS.517.4529B} are similar, $cz\sim10^4\kms$ and both have quite a uniform angular distribution outside the ZOA. We conclude here that for similar data, the \cite{2022MNRAS.517.4529B} and our results are in good agreement.

The HMC analysis of the CF4 grouped data by   \cite{2023A&A...670L..15C} stands in formal agreement with the present results. Namely, given the much larger uncertainty of the estimated \Vbulk\ of these authors the present results are within the  1 sigma uncertainty of that work. Yet, an inspection of Fig. 3 of \cite{2023A&A...670L..15C} finds  a  radial profile of the bulk velocity that is very different than the one found here (Fig. \ref{fig:Vbulk}). The present work does not find any hint of the second hump, that peaks at $\sim275\,\hmpc$, of \cite{2023A&A...670L..15C}. 
A closer look at  Table 1 of that work finds a marked difference not only in the amplitude of \Vbulk\ but also in its direction. The entry for the case of $R=300\,\hmpc$ (for the grouped CF4 data) shows $V_{\mathrm{bulk}}=230\pm136\,\kms$  (${\bf V}_{\mathrm bulk}=(1, 229, -29)\,\kms$) (no error bars on the Supergalactic Cartesian components are given), compared with our $V_{\mathrm{bulk}}=202\pm36\,\kms$  (${\bf V}_{\mathrm bulk}=(-185 \pm 35,      -13 \pm 26,  -69 \pm 32)\,\kms$). The directions of the two estimated bulk flows are very different. Comparable disagreements are found for other values of $R$.

Our work stands in disagreement with  \cite{2023MNRAS.524.1885W}. That study uses the same CF4 data that is used here, yet the bias correction scheme is very different from the BGc used here. Also, a different weighting scheme for \Vbulk\ is used there. On the face of it one cannot compared the minimal variance and the WF/CRs estimations at their quoted depth, $R$, yet the fact that cosmic variance calculated by the two methods are essentially the same argues that such a comparison is meaningful. There is a strong discrepancy, of a few sigma of one method from the other. Moreover, the statistical significance of the results, as gauged by the $\chi^2$ statistics with respect to the standard \LCDM\ model, varies dramatically between to the two works. The probability found by \cite{2023MNRAS.524.1885W} of finding a $\chi^2$ as large or larger than the estimated one is $2.3\times10^{-4}$ and $2.1\times10^{-6}$ for $R=150$ and $200\hmpc$, respectively. Here we find $6.9\times10^{-3}$ and $9.6\times10^{-3}$ for the same radii (Table \ref{table:Vbulk}). \cite{2023MNRAS.524.1885W} results constitute a $\sim3.5$ and $\sim4.5$ sigma deviation from the standard \LCDM\ model, compared with the  $\sim2.6$ of the current work. Given that the two studies use the same data it seems that the differences between the bulk velocities   stem from the different bias correction schemes of the two studies.

The minimal variance analysis of the 6dFGS survey \citep{2016MNRAS.455..386S} finds at a depth of $R=50\,\hmpc$ an amplitude of \Vbulk\ in close agreement with the present finding, yet the direction of the vector deviates from the one calculated here by $\sim40^\circ$, which constitutes a more than $1\sigma$ discrepancy. One might argue that such  a discrepancy is not surprising given the very anisotropic distribution over the sky of the 6dFGS data point (see Fig. \ref{fig:aitoffl}).

There is also tension between the present work and the minimum variance estimation of \cite{2023arXiv230611269W}. These authors estimate \Vbulk\ at an effective depth of $R=173\,\hmpc$ to be $428\pm108\,\kms$ (${\bf V}_{\mathrm bulk}=(-391\pm104, -119\pm93, -126\pm122)\,\kms$). We find here  
$V_{\mathrm{bulk}}=311\pm28\,\kms$  (${\bf V}_{\mathrm bulk}=(-285 \pm 28,      -1 \pm 17,  -122 \pm 24)\,\kms$) at   $R=170\,\hmpc$. The \LCDM\ cosmic variance at that depth is $194\pm 86$ for the minimum variance estimator and $144\pm 60\,\kms$ for the current WF/CRs estimator. One can fit an effective depth,    of the WF/CRs algorithm, that corresponds to the effective depth of the minimum variance estimation by equating their corresponding \LCDM\ predictions. This yields $R\sim110\,\hmpc$ for the $173\,\hmpc$ of \cite{2023arXiv230611269W}.  For that depth we find  $V_{\mathrm{bulk}}=303\pm 20\,\kms$ (${\bf V}_{\mathrm bulk}=(-273 \pm 21,      35 \pm  12, -124 \pm 16)\,\kms$) - still some tension exists between the current results and   \cite{2023arXiv230611269W}, for the amplitude and the direction of the bulk velocity.

\subsection{Comparison with other studies: density}
\label{sec:comparison_density}

The WF estimated linear density field   provides a good proxy to the actual density field on scales, i.e resolution, of roughly $20\,\hmpc$ or larger. It follows that the WF predicted cumulative overdensity profile (\DeltaLR) constitutes a good approximation to the  real density field for $R\,\gtrsim\,20\,\hmpc$.

The prediction of the linear density field profile, \DeltaLR, with the density inferred from surveys of galaxies or galaxy clusters is hampered by issues of selections and bias. We are tempted here to compare the WF predicted \DeltaLR\ with the corresponding density profile of the compilation of CLASSIX galaxy clusters   \citep{2020A&A...633A..19B}. Fig. 11 of these authors shows the cumulative  profile of the clusters number   density and using a simple bias model also that of the total matter density. Detailed analysis is to be presented elsewhere and only a qualitative comparison shows that   both the measured  (from clusters) and the predicted (from the CF4 data) local underdensity extends over the range of $30\,\lesssim\,R\,\lesssim\, 250\,\hmpc$. Note that range of the local underdensity quite strongly depends on the value of $H_0$ ({\it cf.}  Fig. \ref{fig:H0}). The agreement quoted here is obtained from $H_0=74.6\,\kmsMpc$.

The maximal statistical departure of  \DeltaLR\ occurs at $R\sim190\,\hmpc$, where it attains a value of roughly $-1.9$ times the \LCDM\ cosmic variance.

\begin{table*}
    \begin{tabular}{l l l c c  c c c c }
       \hline
       \hline
      source & data & weighting  & radius [$\hmpc$] & $V_{\mathrm{bulk}}\ [\kms]$ &  \LCDM\   [$\kms$] & $l$ [deg]  & $b$ [deg] & $\chi^2$ \\
       \hline
       \cite{2015MNRAS.450..317C}  & 2M++ (redshift)   & Gaussian & $50$     & $230 \pm 30$  & & $293\pm 8$   &  $14 \pm 10 $  & \\
       \cite{2015MNRAS.449.4494H}  & CF2               & volume   & $50$     & $258 \pm 21$  &  $292 \pm 147$ & 280  &  $18 $  & \\
       \cite{2016MNRAS.455..386S} & 6dFGS    & minimum variance &  $50$      & $248 \pm 58$ &  & $318 \pm 20$ & $40 \pm 13 $  &  \\
       \cite{2019MNRAS.482.1920Q}  & 2MTF   &  max. likelihood & $37$  & $259 \pm 15$    & $231^{+118}_{-101}$ & $300 \pm 4$  & $23 \pm 3 $  & 2.76 \\
       \cite{2021MNRAS.507.1557L}  &  2MRS (redshift) &  Gaussian   &  $50$     & $274 \pm 50$  & & $287\pm 9$   &  $11 \pm 10 $  & \\
     \cite{2022MNRAS.517.4529B}   &  SFI++ \& 2MTF &  Gaussian  &  $40$     & $220 \pm 21$  & & $295\pm 6$   &  $21\pm5 $     & \\
     \cite{2023MNRAS.524.1885W}    &  CF4   &   min. variance    &  $150$ & $387 \pm 28$ & $139$  & $297 \pm 4$  & $-6 \pm 3 $  & 19.34 \\
        &  CF4   &   min. variance    &  $200$ & $419 \pm 36$ & $120$ & $298 \pm 5$  & $ -8 \pm 4 $  & 29.13 \\
    Courtois et al. (2023) &  CF4   &  volume & $50$     & $255  \pm 49$ &   & 296  &  15   &   \\
                           &  CF4   &  volume & $150$    & $272  \pm 105$ &   & 295  &  14   &   \\
                           &  CF4   &  volume & $200$    & $205  \pm 123$ &   & 279  &   60  &   \\
      \cite{2023arXiv230611269W}    &  CF4   &   max. likelihood    &  $49$ & $408 \pm 165$ & $196\pm82$  & $301$ &    $ -18 $  & $6.2$ \\
                                    &  CF4   &   min. variance    &  $173$ & $420 \pm 108$ & $194\pm86$  & $297$ &    $ 5 $  & $16.0$ \\
    \hline
    current work  & CF4  &  Gaussian & $40$  & $303 (196) \pm 14$ & $229 \pm 129$ & $291 \pm 2$  & $ 9 \pm 1 $  & 4.20 \\    
                  &  CF4 &  Gaussian & $50$  & $303 (191) \pm 16$ & $200 \pm 94$  & $292 \pm 2$  & $ 5 \pm 2 $  & 5.06 \\
       &  CF4   &  volume   & $50$  & $320 (217) \pm 13$ & $291 \pm 914$ & $291 \pm 2$  & $13 \pm 1 $  & 3.12 \\
       &  CF4   &  volume   & $150$ & $318 (200) \pm 26$ & $158 \pm 65$  & $293 \pm 3$  & $-1 \pm 2 $  & 10.41 \\
       &  CF4   &  volume   & $200$ & $292 (174) \pm 30$ & $127 \pm 53$  & $294 \pm 5$  & $-4 \pm 3 $  & 13.17 \\
       &  CF4   &  volume   & $300$ & $202 (78) \pm 36$ & $94 \pm 40$  & $297 \pm 9$  & $-5 \pm 7 $  & 11.14 \\
    \hline
    \hline
    \end{tabular}
    \caption{ Comparison with other studies (see text for details). The radius given here is the effective radius whose definition depends on the weighting scheme.  The \LCDM\ column provides the cosmic variance of the bulk velocity. The numbers that appear in the parentheses in the $V_{\mathrm{bulk}}$ column for the current work correspond  to the bulk velocity of the local flow  induced within a sphere of $R=300\,\hmpc$ (see \S \ref{sec:tidal}). }
\end{table*}

\section{Summary and Discussion}
\label{sec:summary}

The large scale structure of the Universe out to a distance of $\sim300\,\hmpc$ is reconstructed here from the Cosmicflows-4 (CF4) grouped data of redshifts and distance moduli. An unbiased transformation of the input data to distances and peculiar velocities is performed by the Bias Gaussianization correction (BGc) algorithm.  The  reconstruction is performed by means of the Wiener filter (WF) and  constrained realizations (CRs) assuming the linear theory of the \LCDM\ standard cosmological model as the Bayesian prior. The combined BGc and WF/CRs algorithm has been tested against  a constrained and a set of random mock CF3 like data sets.

The main conclusion that follows from the testing of the BGc and WF/CRs algorithm against the mock data is that the WF reconstruction of the velocity field recovers the actual underlying density field to within 1 to 2 times the scatter exhibited by the ensemble of the CRs, namely the constrained variance. Furthermore, for the CF4 data and within the \LCDM\ prior the constrained variance is significantly smaller than the cosmic one.  

The WF/CRs predicted underdensity over the range of $30\,\lesssim\,R\,\lesssim\, 250\,\hmpc$ stands in qualitative agreement with the density inferred  from the compilation of the  CLASSIX galaxy clusters \citep{2020A&A...633A..19B}. That  range depends quite strongly on $H_0$, hence the agreement with the distribution of clusters provides an extra support to the claim that local probes  give rise to higher values of $H_0$ and thereby intensifies the so-called Hubble constant tension.

The WF/CRs approach is a conservative one. Namely, where the data is `weak' the Bayesian prior dominates the reconstruction and nothing new is learnt from the reconstruction. In the case of the standard cosmological model, in regions in configuration or Fourier spaces dominated by `weak' data the WF reconstructed density and velocity fields converges towards  the null fields and the CRs become random realizations. Yet in spite of this reservation it is of interest to check in what sense and to what degree our  local `patch' of the Universe is typical, or maybe  atypical.  Our analysis shows that on distances exceeding $\sim150\,\hmpc$ and all the way to the edge of the data  the local Universe is somewhat an outlier, on the 2 - 3 sigma level.  This is manifested by the amplitude of the \Vbulk\ and the monopole moment profiles. This is also indicated by the direct $\chi^2$  analysis of the bulk velocity vector. Also, the coherent nature of the direction of the bulk velocity vector is somewhat atypical.    
Our local patch of the Universe is also a `bit' unusual, at the $\sim2$ sigma level,  in the sense of the coherence of the direction of the bulk velocity vector. This is also the maximal level of discrepancy of the density profile, \DeltaLR.   

How meaningful is such discrepancy? It is clear that being at the roughly 2 sigma edge of the cosmic variance is not enough to declare a possible tension with the \LCDM\ model. Yet, it is enough to recognize our local cosmological neighborhood is somewhat atypical. It is interesting to see whether such a roughly 2 sigma discrepancy can be corroborated by other reconstructions of the local Universe, in particular from redshift surveys of galaxies.

\section*{Data availability}
The estimated density and velocity fields are available upon reasonable request to the authors.

\section*{Acknowledgements}
 
This paper is dedicated to the memory of Nick Kaiser, whose seminal paper \citep{1988MNRAS.231..149K} set the stage and provided the framework for subsequent studies of  the cosmological velocity surveys. 
This work has been done within the framework of the Constrained Local UniversE Simulations (CLUES) simulations.
YH has been partially supported by the Israel Science Foundation grant ISF 1358/18.
NIL \& AV  acknowledge financial support of the Project IDEXLYON at the University of Lyon under the Investments for the Future Program (ANR-16-IDEX-0005). JS acknowledges support from the ANR LOCALIZATION project, grant ANR-21-CE31-0019 of the French Agence Nationale de la Recherche. SP acknowledges financial support from the Deutsche Forschungs Gemeinschaft joint Polish-German research project  LI 2015/7-1 (LUSTRE)

\bibliographystyle{mn2e}
\bibliography{CF4_23}

\appendix
\section{WF/BGc reconstruction: Random Mock CF3-like data  }
\label{appdx:random_mocks}
\numberwithin{equation}{section}
\setcounter{equation}{0}
\numberwithin{figure}{section}
\setcounter{figure}{0}

\cite{2021MNRAS.505.3380H} present a detailed description of the mock catalogs used here. A brief description of these follows. Mock catalogs of the grouped CF3 data were   constructed so as to test the BGc bias correction scheme.  The mock catalogs are drawn from  the publicly available Multi-Dark-2 simulation\footnote{see https://www.cosmosim.org} (Klypin et al 2016).  
This is a DM-only $N$-body simulation  with $N=3840^3$ in a periodic box of side length  $1.0 \hgpc$ particles,  assuming a Planck cosmology ($H_0=67.7\kmsMpc$, $\Omega_{\Lambda}=0.69$, $\Omega_{\rm b}=0.04$, $\Omega_{\rm m}=0.31$ $\sigma_{8}=0.82$)  \citep{Planck:2013}.
A FOF algorithm is run on the $z=0$ particle distribution and all groups larger than 20 particles are retained. Ten different mock observers have been selected at random within the computational box, subject to a single constraint - that they `reside' within $\approx10^{12}M_\odot$ halos. The distribution of the mock data points was designed to reproduce the spatial distribution of the CF3 data and the measurement uncertainties are inherited from the actual CF3 data \citep[for details]{2021MNRAS.505.3380H}. Ten different random realizations of the observational errors were constructed for each mock observer, resulting in a total of 100 mock catalogs - representing the cosmic variance (10 mock observers) and the error variance (10 errors realizations). 

\begin{figure*}
\centerline{
\includegraphics[width=0.33\linewidth]{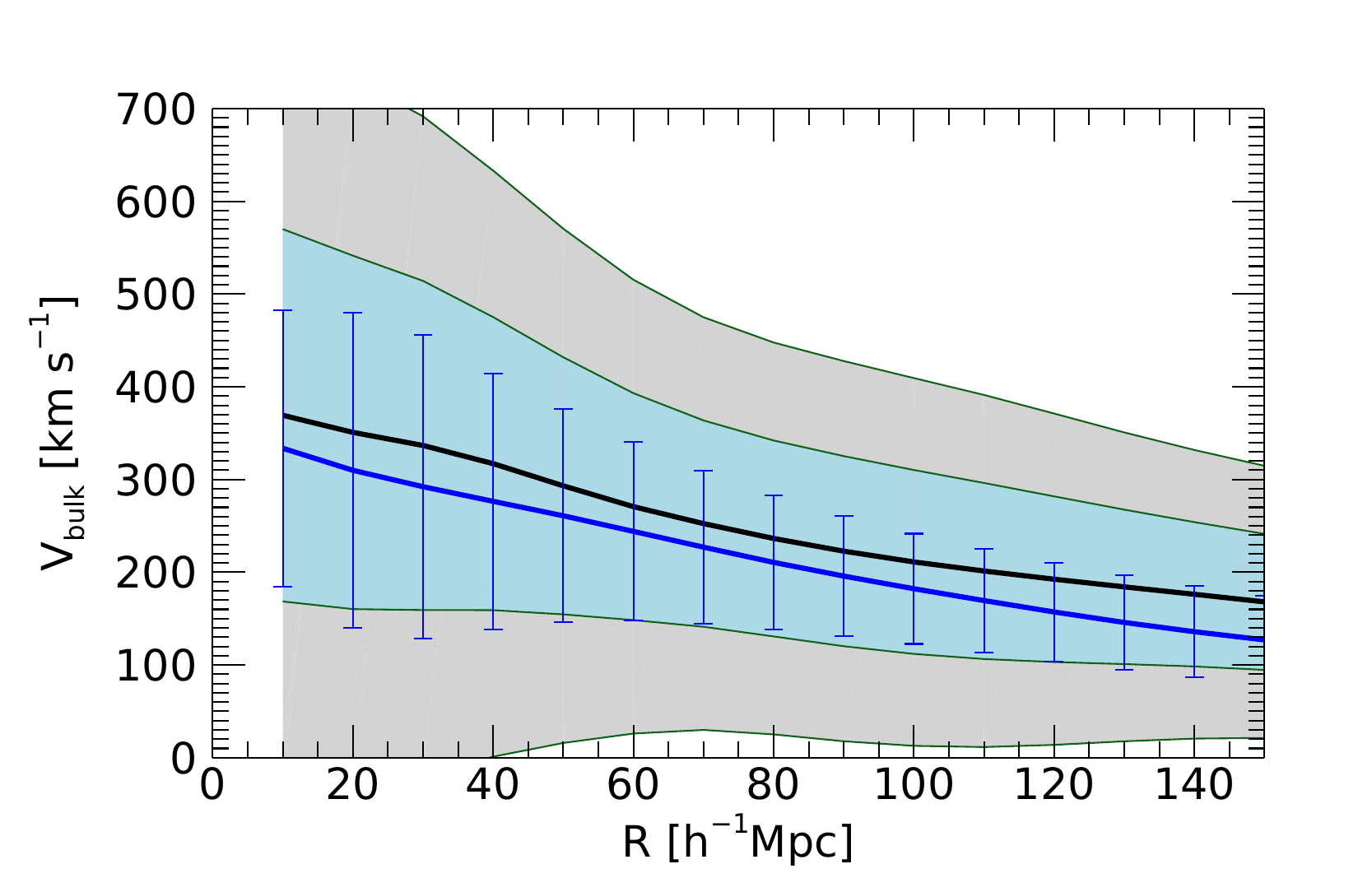}
\includegraphics[width=0.33\linewidth]{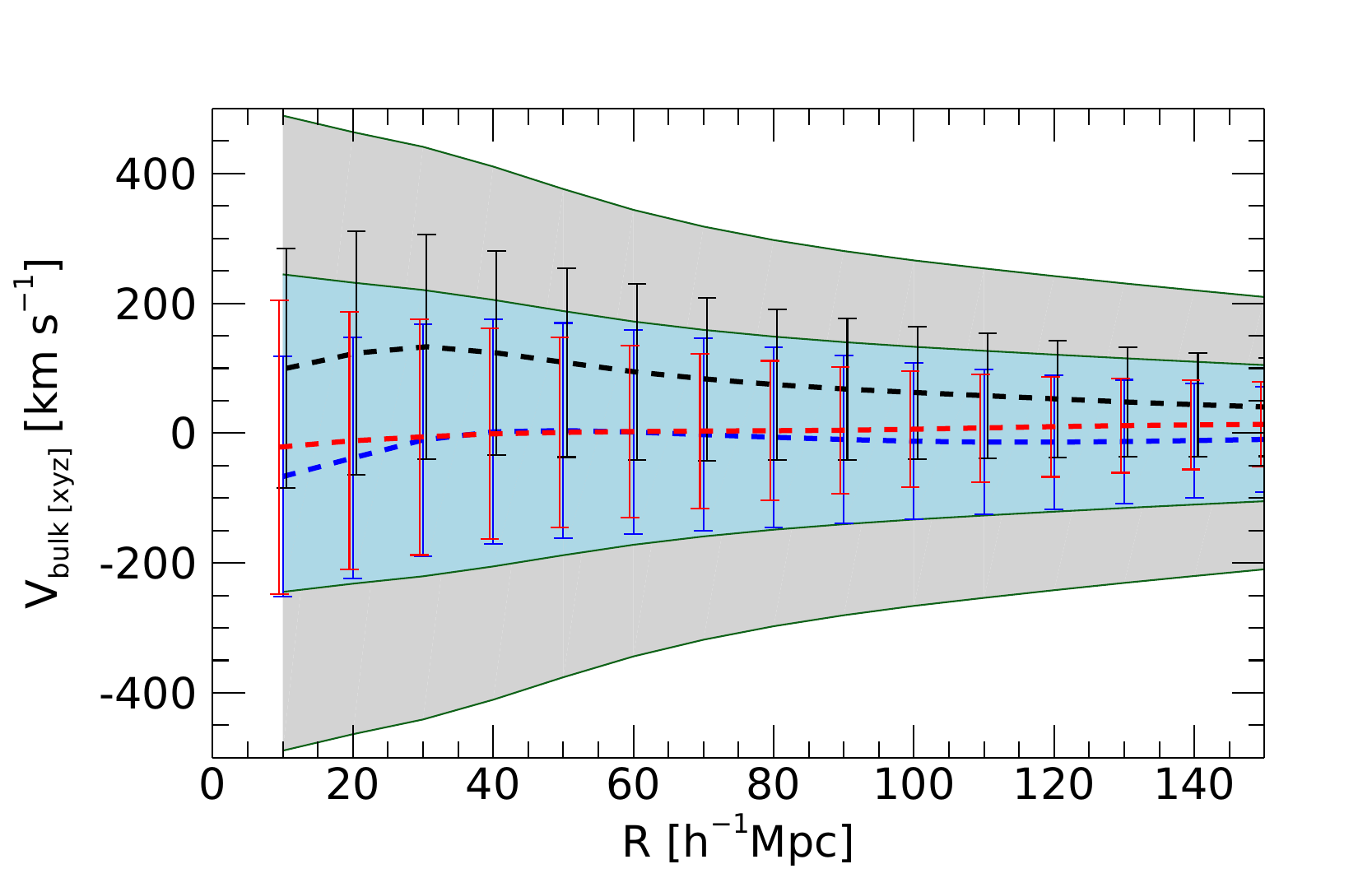}
\includegraphics[width=0.33\linewidth]{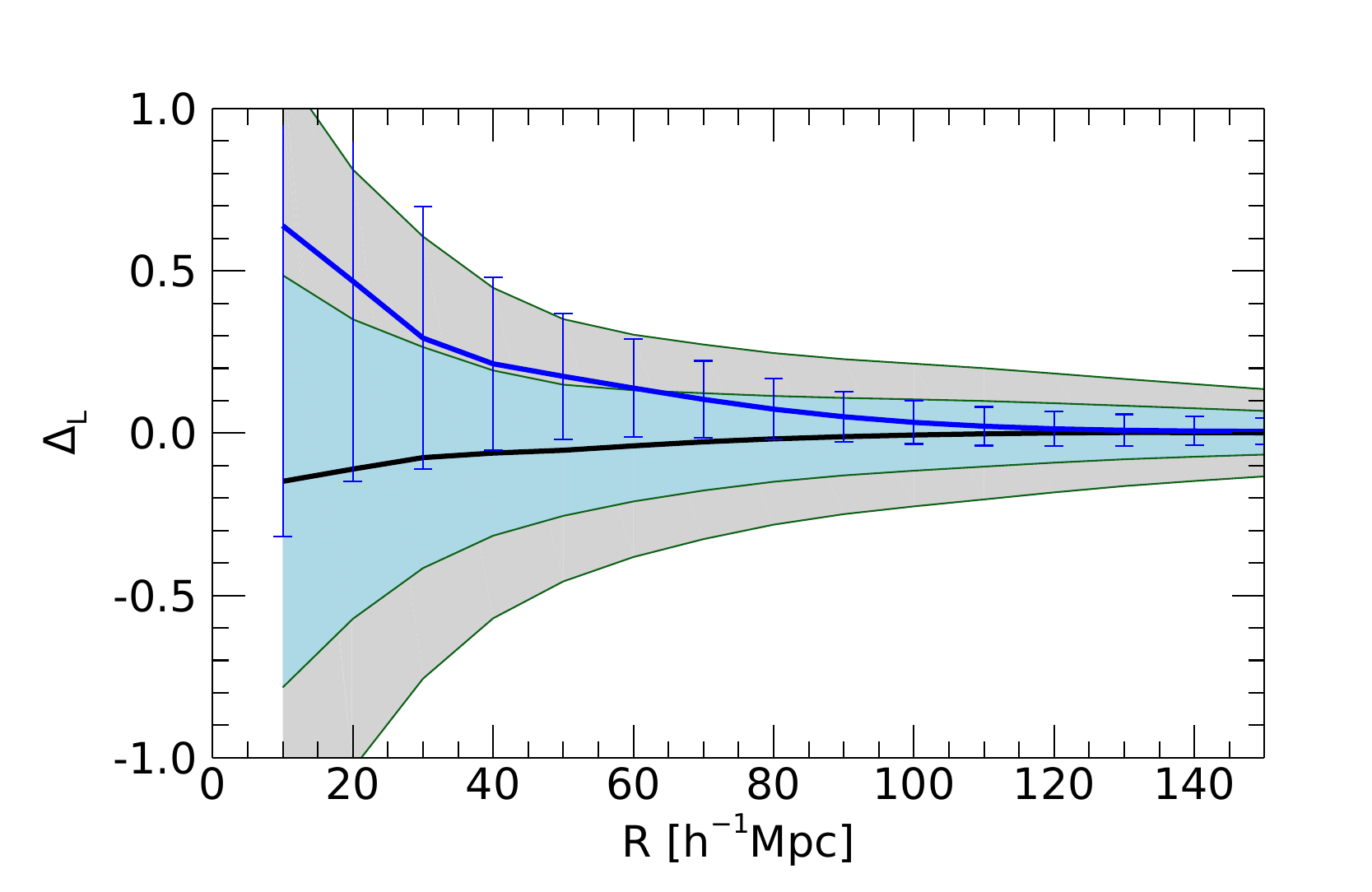}
}
\caption{The combined cosmic and error mean and variance  of an ensemble of     WF reconstructions of 100 mock data realizations, of the 10 random observers and of the 10 errors realizations for each observer. The shaded uncertainties correspond to an ensemble of 200 random realizations and are given for reference.
Color and line style conventions are the ones used in Figs. \ref{fig:Vbulk} and \ref{fig:monopole}.   }
\label{fig:WF-mocks}
\end{figure*}

The  BGc algorithm has been applied to the mock catalogs, so as to correct the lognormal bias. The LSS has been reconstructed, within the linear regime, by means of the WF/CRs algorithm. Fig. \ref{fig:WF-mocks} presents the monopole and dipole moments of the WF reconstructed from the 100 mock catalogs. The plots of the figures present the mean and scatter of the ensemble of the 10 errors realizations of the mock data of each observer. The WF reconstruction is done on a cubic grid of $N=128^3$ spanning a box of side length $L=500\,\hmpc$. The resolution of the first two bins of the profiles ($R=10$ and $20\,\hmpc$) is rather poor but we opted to show them nevertheless.

The mean and variance of the amplitude  and  the Supergalactic coordinates   of the bulk velocity and of the cumulative linear density  profiles of the ensemble of   WF reconstructions from an ensemble of 100 mock data sets are presented in Fig.~\ref{fig:WF-mocks}. The mean and variance of an ensemble of random realizations  is presented in shaded confidence intervals for reference. The bulk velocity plots show that the WF reconstruction recovers the 'true' underlying profiles to better than one standard deviation of the distribution of the WF profiles. The figure further shows the expected underestimation of the WF reconstruction. The mean spherical density profiles deviate from the expected random realizations. This is a reflection of the fact that the random observers are selected to coincides with $\approx10^{12}\hmsun$ halos. Such halos tends to reside in overdense environments, and hence the bias.

\begin{figure*}
\centerline{
\includegraphics[width=0.33\linewidth]{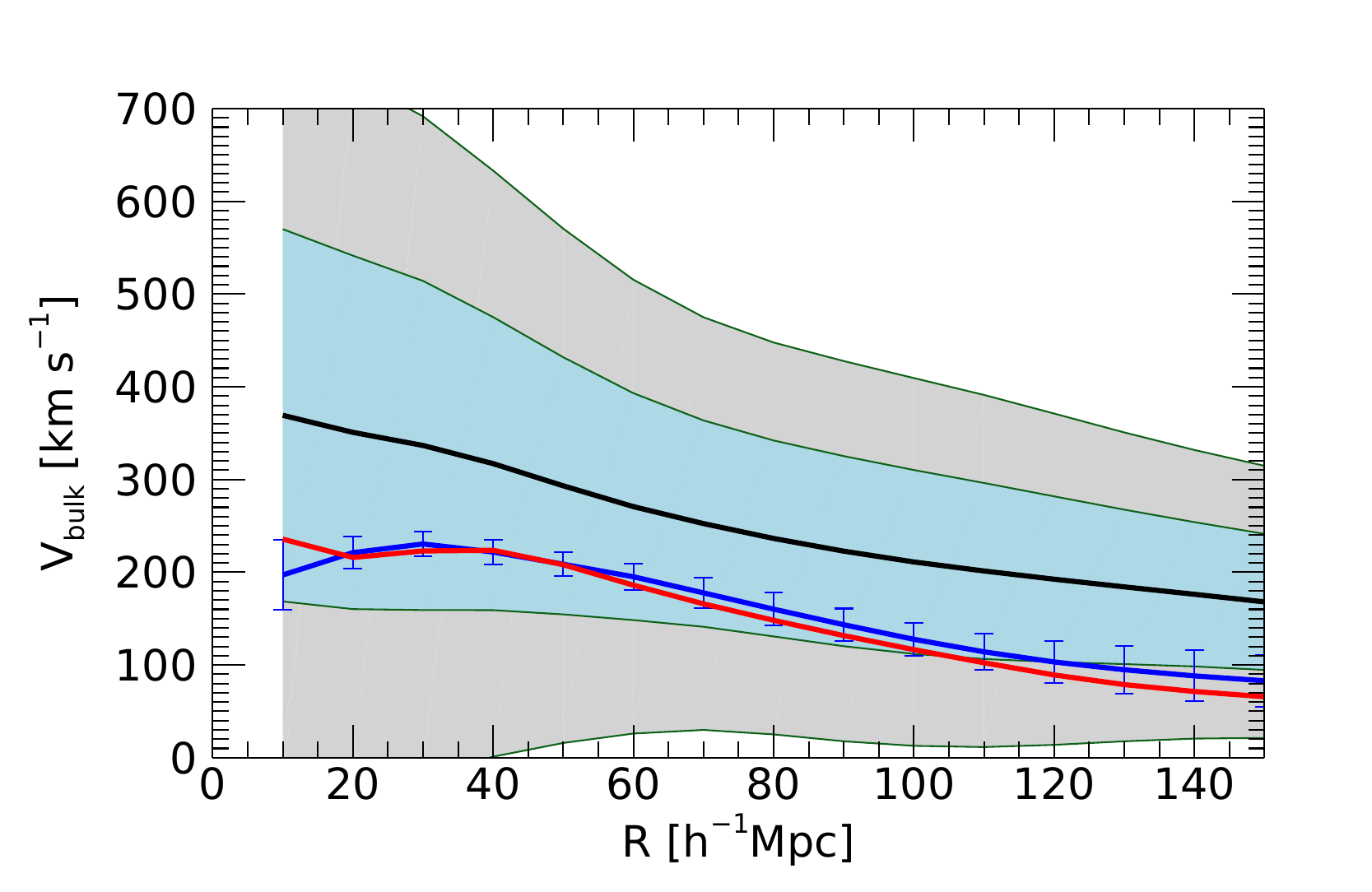}
\includegraphics[width=0.33\linewidth]{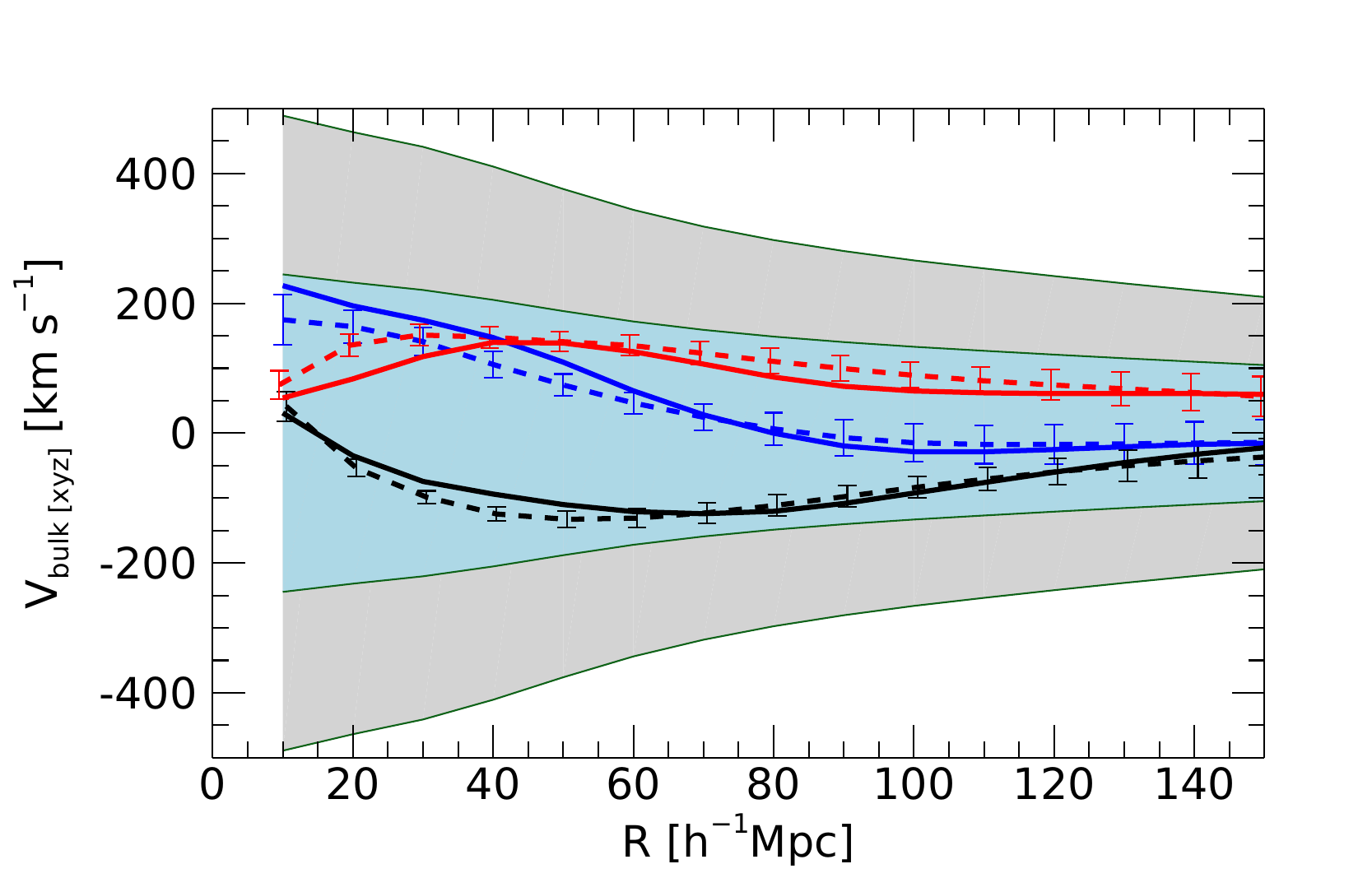}
\includegraphics[width=0.33\linewidth]{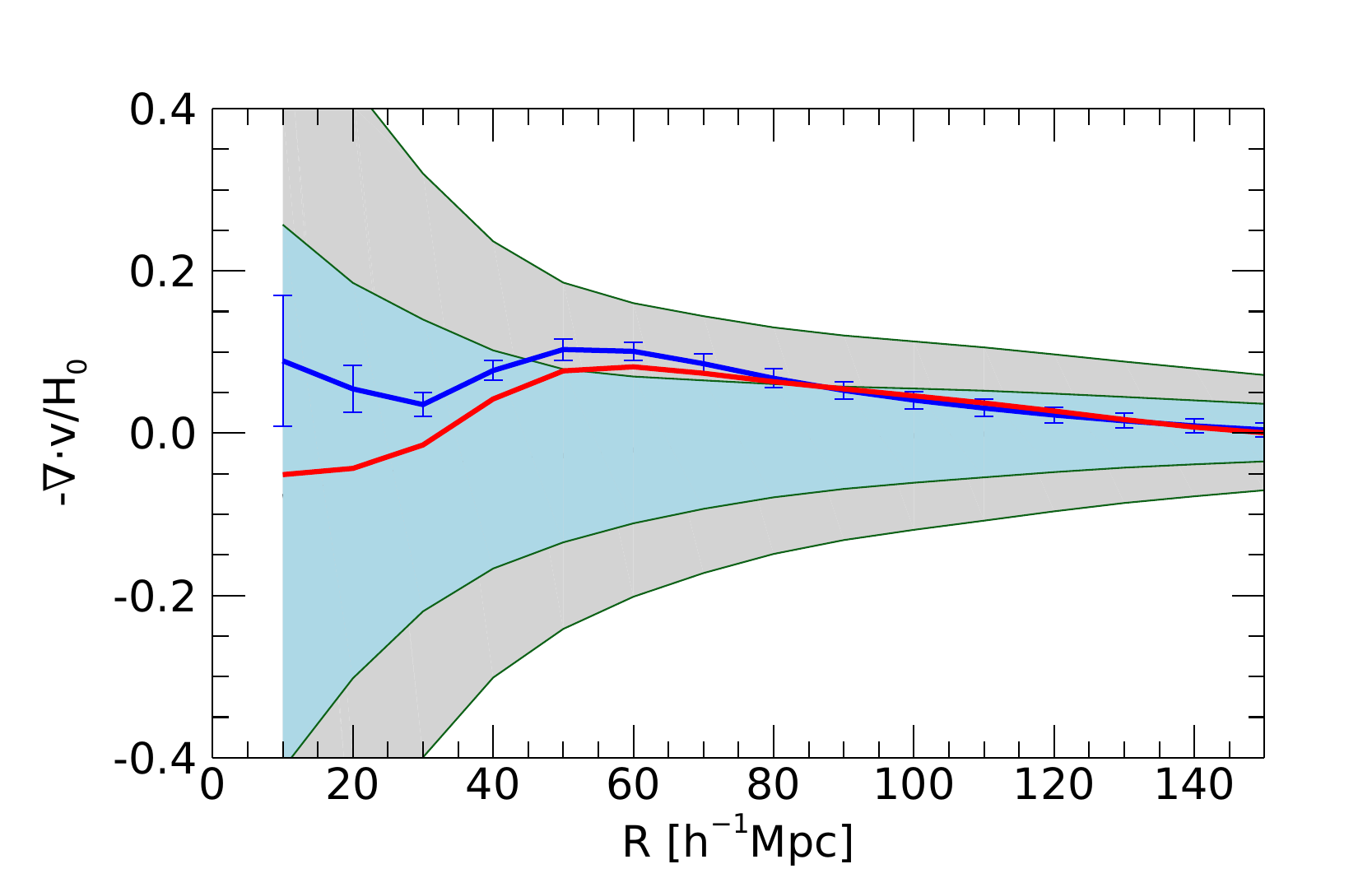}
}
\caption{The variance of an ensemble of 30 CRs of a single errors realization of one of the mock observers. 
Color and line style conventions are the ones used in Figs. \ref{fig:Vbulk} and \ref{fig:monopole}.  In addition  the red solid line represents the corresponding profiles of that particular  target simulation.}
\label{fig:CRs-mocks}
\end{figure*}

Next,  the more relevant analysis  of an ensemble of 30 CRs using one errors realization of one particular observer, namely one of the 100 cosmic and errors realization, is considered here. Such an analysis is actually employed in the case of the actual CF4 data, drawn from our actual Universe. Fig. \ref{fig:CRs-mocks} presents the mean and cosmic variance of the amplitude and the values of the 3 Cartesian components of the bulk velocity and of the monopole moment. 
The plots show the  constrained   mean and variance, calculated over an ensemble of 30 constrained realizations, and the corresponding `true' profiles of the underling velocity field of the target simulation. The mean and variance of an ensemble of 30 random realizations are shown as well as a manifestation of the cosmic variance,
The plots show that the constrained variance is much smaller that the cosmic one and that the residual of the mean of the CRs, which is effectively equal to the WF estimated profiles,  is always much smaller than the cosmic variance. It should be recalled here that the scatter of residual of the CRs from the actual field depends only on the measurement  uncertainties and not the actual values of the underlying field \citep{1991ApJ...380L...5H}, hence all statements made here about the residual from the `truth' apply to all 10 different mock observers.

\section{WF/BGc reconstruction: Constrained CF3-like Mock data  }
\label{appdx:constrained_mocks}

Doumler et al. (2013a,b,c)
presented a new and alternative way for constructing mock galaxy velocity catalogs. This was done by drawing the mock data from a constrained simulation, one that was constrained by actual observed data \citep{2003ApJ...596...19K, 2023arXiv230101305S}. 
The main merit of such constrained mock data is that the choices of the mock data points are correlated with the underlying density field of the actual Universe. Namely, the constrained simulation reproduce, to a certain degree, the neighborhood  of prominent structures such as the Virgo   cluster, say. The cluster itself is represented by one data point only, in  
the grouped data catalog, of a reduced error, and is surrounded by many nearby data points.  Such a data distribution enhances  the constraining power of the data and makes the Virgo cluster and its neighborhood more constrained than other nearby neighborhoods. The constrained mock data captures that feature of the actual data.

\begin{figure*}
\centerline{
\includegraphics[width=0.5\linewidth]{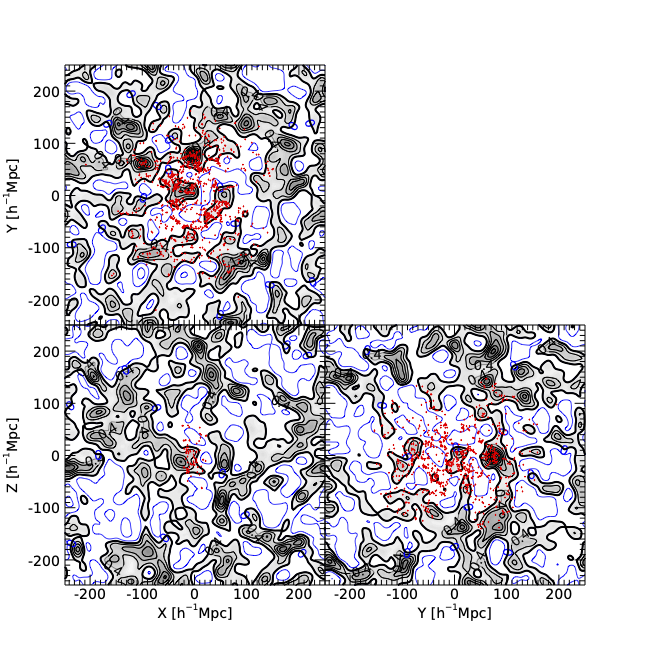}
\includegraphics[width=0.5\linewidth]{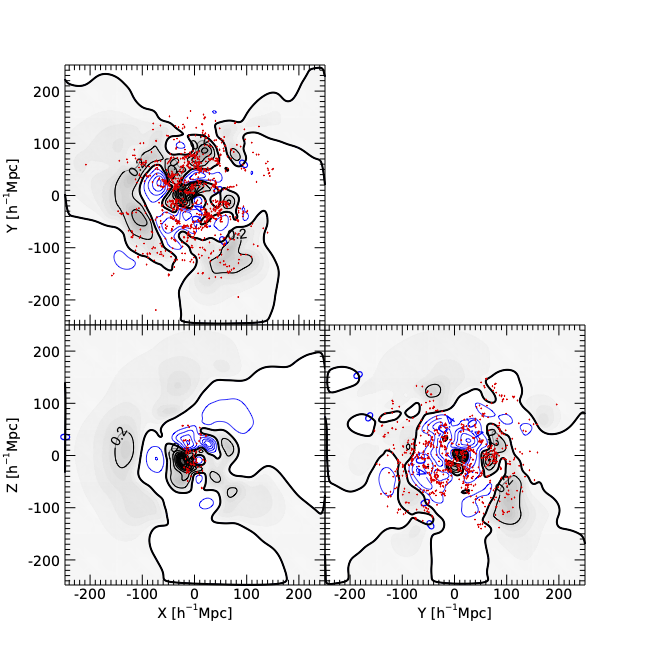}
}
\caption{A comparison of the target CF2-constrained simulation with the WF reconstruction: The    CIC interpolated linear density field ($\delta_L$ of the target simulation (left panel) is compared with the WF reconstructed linear density field (right panel). Both fields are Gaussian smoothed with a kernel of $R_s=5\,\hmpc$. The constrained CF3-like mock data points (within a slab of $\pm 5\,\hmpc$) are superimposed on both images (red dots).
}
\label{fig:Jmock_dens}
\end{figure*}

\begin{figure*}
\centerline{
\includegraphics[width=0.5\linewidth]{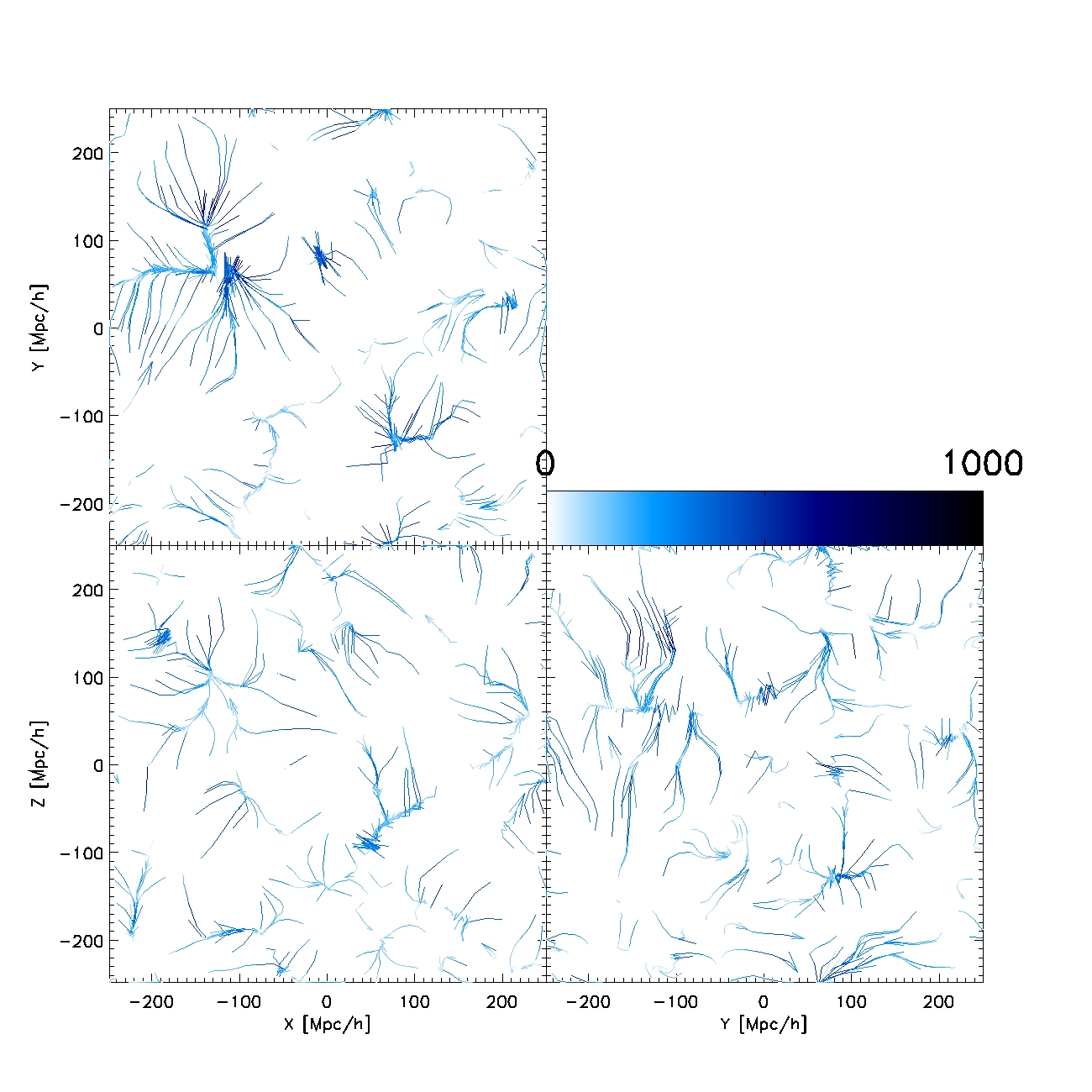}
\includegraphics[width=0.5\linewidth]{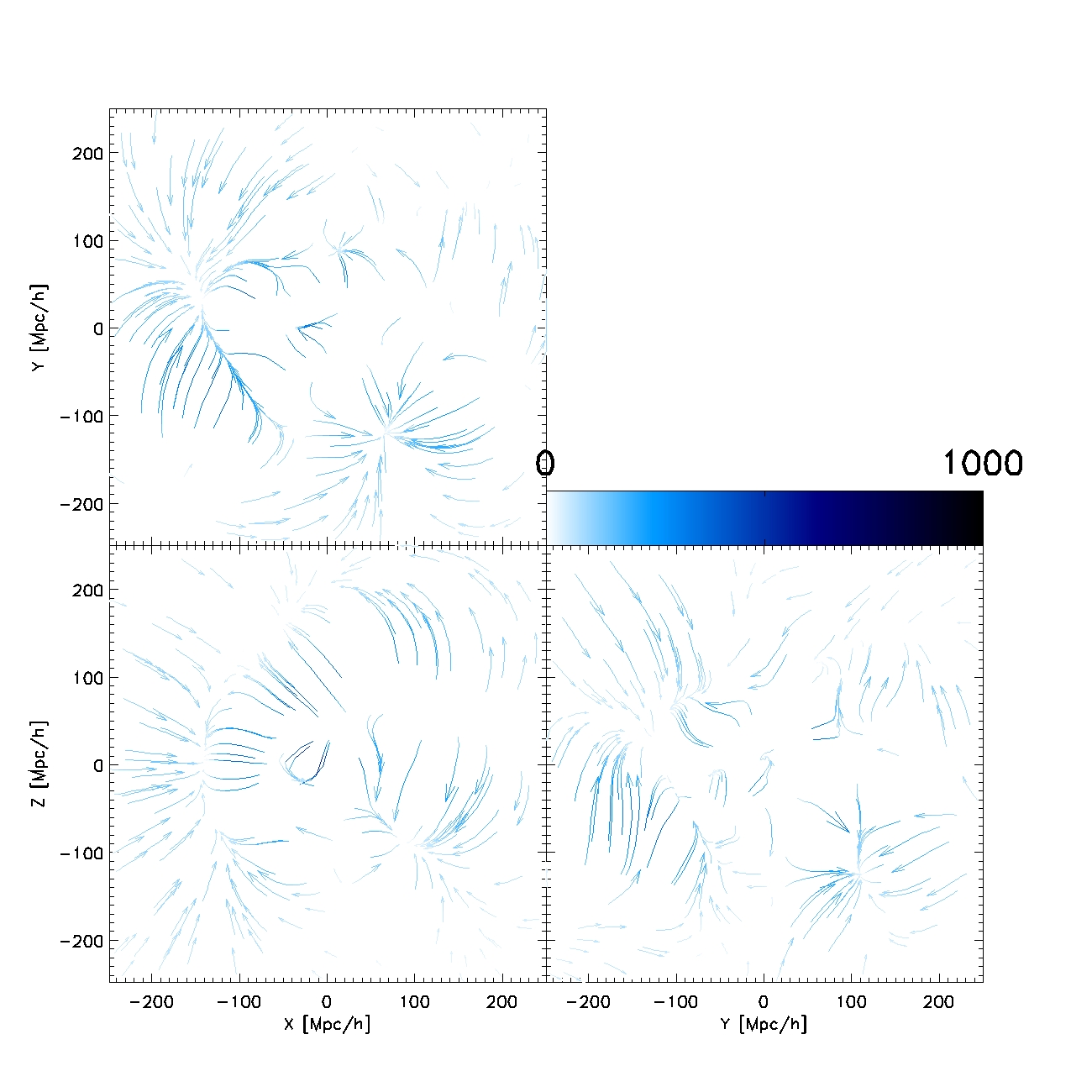}
}
\caption{A comparison of the target CF2-constrained simulation with the WF reconstruction: The    CIC interpolated velocity field of the target simulation (left panel) is compared with the WF reconstructed velocity field (right panel). Both fields are Gaussian smoothed with a kernel of $R_s=5\,\hmpc$ and the field is presented by streamlines. 
}
\label{fig:Jmock_vel}
\end{figure*}

A CF3-like constrained mock grouped data, drawn from a \LCDM\ simulation constrained by the CF2 data has been recently constructed \citep{2023arXiv230101305S}. The BGc bias correction scheme has been applied to that data and the WF/CRs algorithm has been applied to it. The quality of the BGc and the WF/CRs algorithms is gauged by the comparison with the target constrained simulation.  Fig. \ref{fig:Jmock_dens} presents a comparison of the $\delta_L$ field (Eq. \ref{eq:delta_L}; Gaussian smoothed with a kernel of $5\,\hmpc$) with the WF reconstructed $\delta$ field. The constrained CF3-like mock data points (within a slab of $\pm5\,\hmpc$)  are superimposed for the sake of orientation and comparison. The effective depth of the CF3 data is $\sim160\,\hmpc$ and  beyond it the quality of the reconstruction degrades to the null field.  As expected, the reconstructed velocity field recovers the LSS of the target simulation better than the density field (Fig. \ref{fig:Jmock_vel}).

\begin{figure}
  \centering
\includegraphics[width=1.\linewidth]{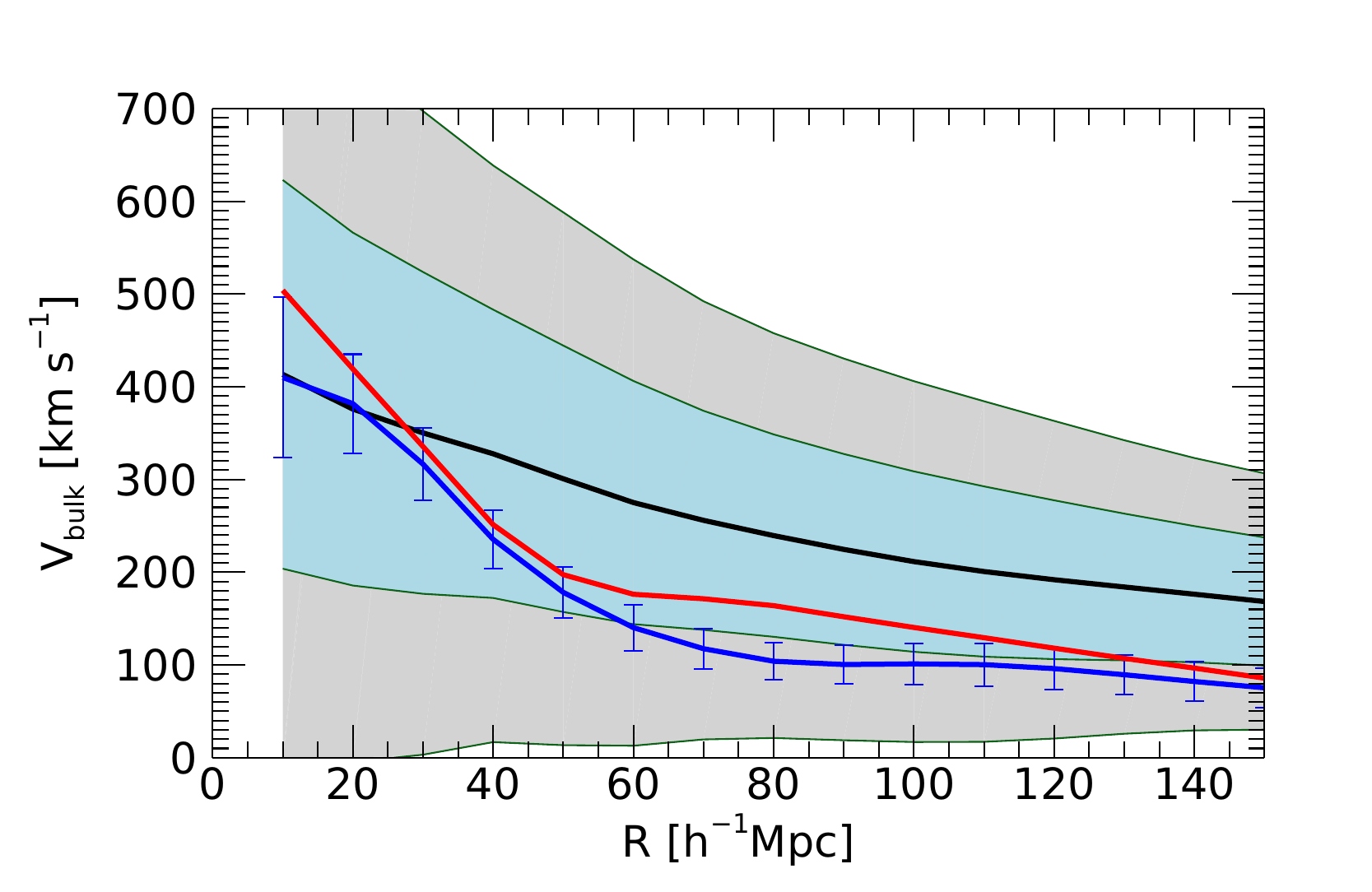}
\includegraphics[width=1.\linewidth]{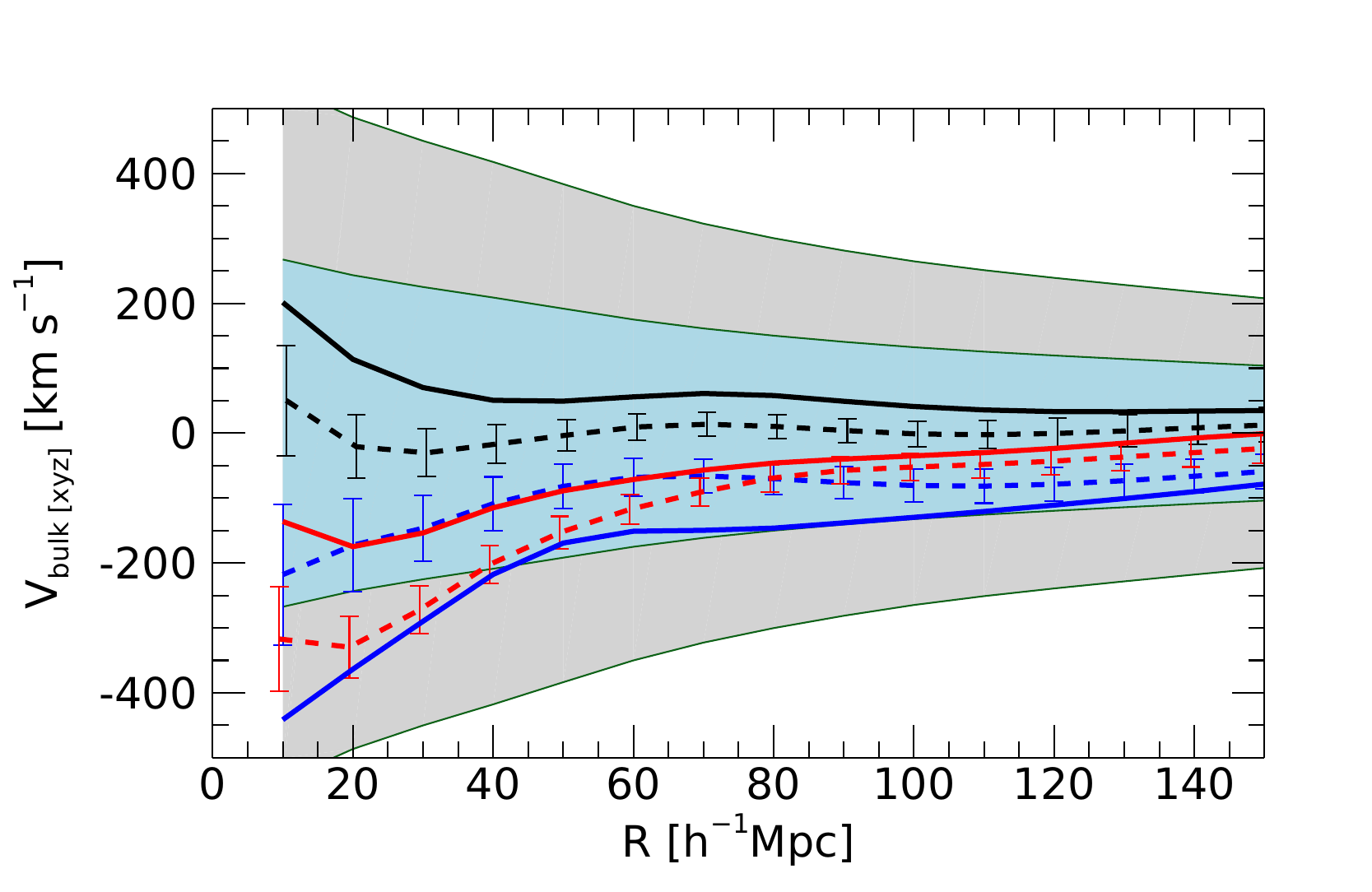}
\caption{
A comparison of the target CF2-constrained simulation with the WF/CRs reconstruction from a CF3-like mock data drawn from that simulation. The mean and scatter are calculated for ensembles of 60 constrained and 60 random realizations
The amplitude of the bulk velocity (upper panel): The mean and scatter taken over the  ensembles   of the  constrained   (blue solid line, error bars) and of  the  random realizations (black solid lines, 1   sigma uncertainty (light blue) and 2 sigma (light grey) shaded regions).   The bulk velocity is calculated within spheres of radius $R$. The red solid line presents the target bulk velocity profile. The three Supergalactic Cartesian components of the bulk velocity (lower panel): The mean and scatter of SGX (blue), SGY (black) and SGZ (red) components of the bulk velocity. The dashed  lines with error bars correspond to the WF/Crs reconstructions and the solid lines to the target profiles. The shaded regions (1   sigma, light blue, and 2 sigma, light grey) show the cosmic variance of one Cartesian component of the bulk velocity.  The cosmic mean value of each of the individual components is zero and therefore is not shown.
}
\label{fig:Jmock-Vbulk}
\end{figure}

Fig. \ref{fig:Jmock-Vbulk} presents the amplitude and the 3 Cartesian coordinates of the bulk velocity out to a distance of $R=150\,\hmpc$. Over most of that range the  \Vbulk\ of the  target simulation lies within the scatter around the mean of the ensemble of CRs. The discrepancy is larger than that, within the 2 sigma level, within the range of $60 < R < 100\,\hmpc$.   As expected, the discrepancy of the individual Cartesian components is somewhat larger, yet it is considerably smaller than the cosmic variance. 
The monopole moment of the target simulation is very well recovered by WF/CRs reconstruction (Fig. \ref{fig:Jmock-mu}). The constrained variance around the mean profile of the CRs, hence of the WF, is much smaller than the cosmic variance. The slight deviation from zero of the mean of the CRs is due to the very finite sampling at small radii and the relatively small number of CRs.
The alignment of \Vbulk\ with itself (at zero lag) is considerably larger than that of random realization, with the constrained variance of the alignment being much smaller than the cosmic variance (Fig. \ref{fig:Jmock-mu}).                                                                                                                                                                                                                                                                                                                                                                                                                                                                                                                                                                                                                                                                                                                                                                                                                                                                                                                                                                                                                                                                                                                                                                                                             

\begin{figure}
\centering
\includegraphics[width=1.\linewidth]{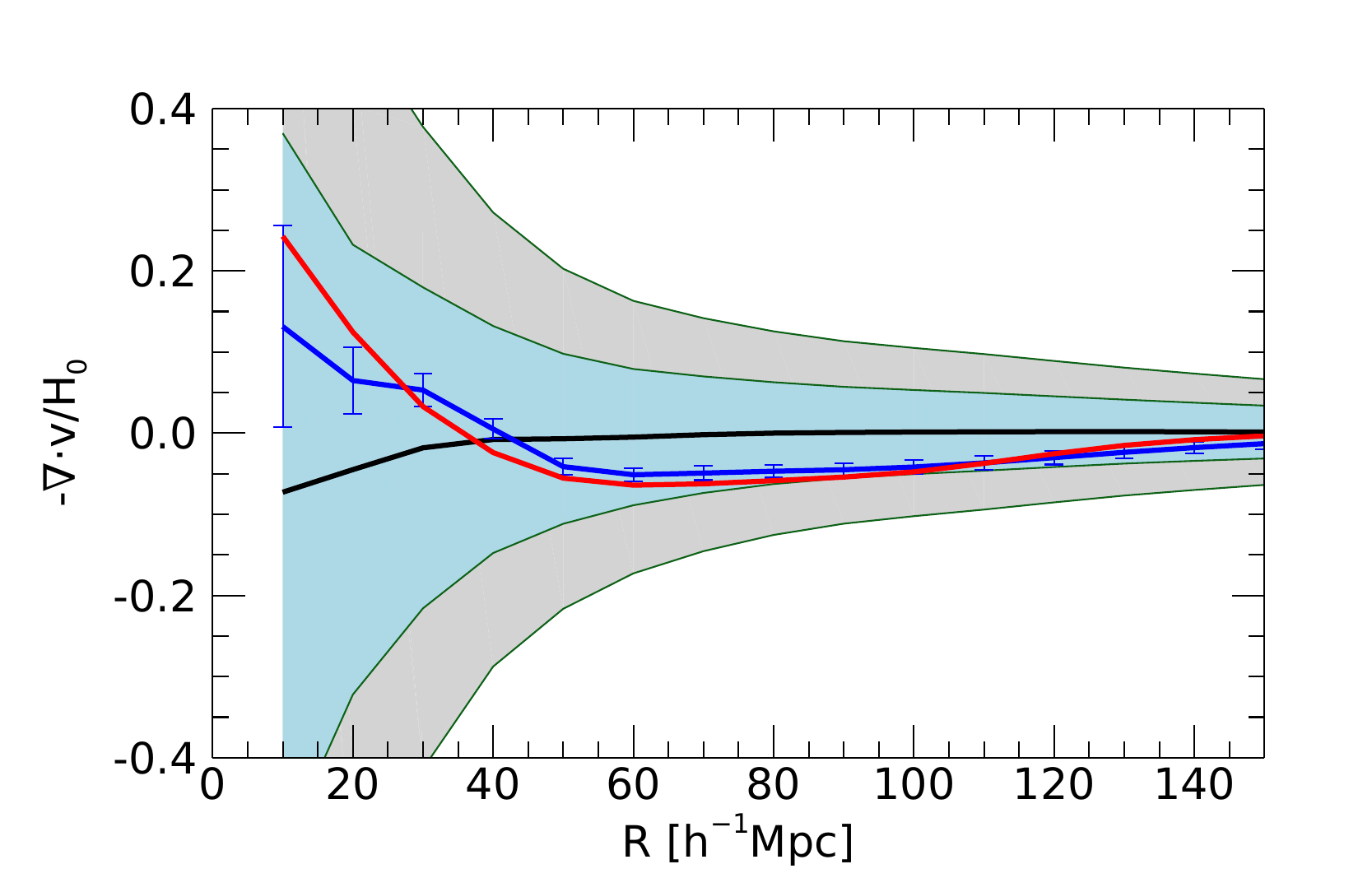}
\caption{
The monopole moment of the velocity field ($ -\nabla\cdot\bf{v}/H_0 $), 
calculated in spheres of radius $R$ (conventions of the different lines are identical to those employed in the upper panel of Fig. \ref{fig:Vbulk}. The minus sign and the $H_0$ scaling are introduced so as to make it proportional to the  linear density.
}
\label{fig:Jmock-monopole}
\end{figure}

\begin{figure}
\centering
\includegraphics[width=1.\linewidth]{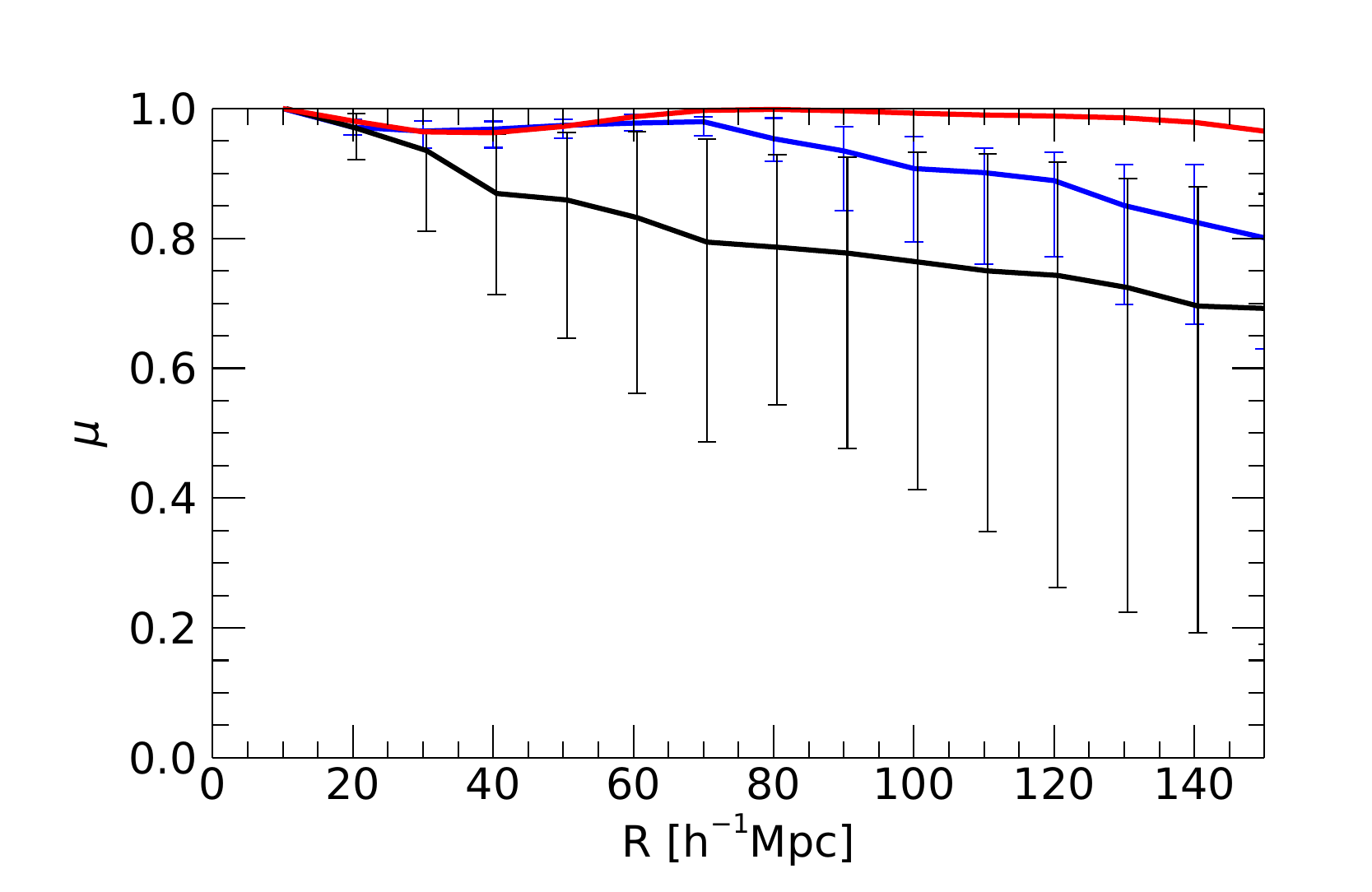}
\caption{The alignment of the bulk velocity of a sphere of radius R  with itself at zero lag, 
$\mu_{\mathrm self}(R) = \hat{\bf V}_{\mathrm bulk}(R_{\mathrm min}) \cdot \hat{\bf V}_{\mathrm bulk}(R) $. The zero lag is defined by $R_{\mathrm min} = 10\,\hmpc$.
The plot shows the median and the lower and upper quartiles  of the distribution the ensembles of 50 CRs (blue) and the corresponding 50 random realizations (black). The alignment of the bulk velocity of the target simulation with itself at zero lag is show as well (red).
}
\label{fig:Jmock-mu}
\end{figure}

\section{WF reconstruction: Cosmicflows-4 vs. Cosmicflows-2  }
\label{appdx:CF4_CF2}

\begin{figure*}
\centering
\includegraphics[width=0.48\linewidth]{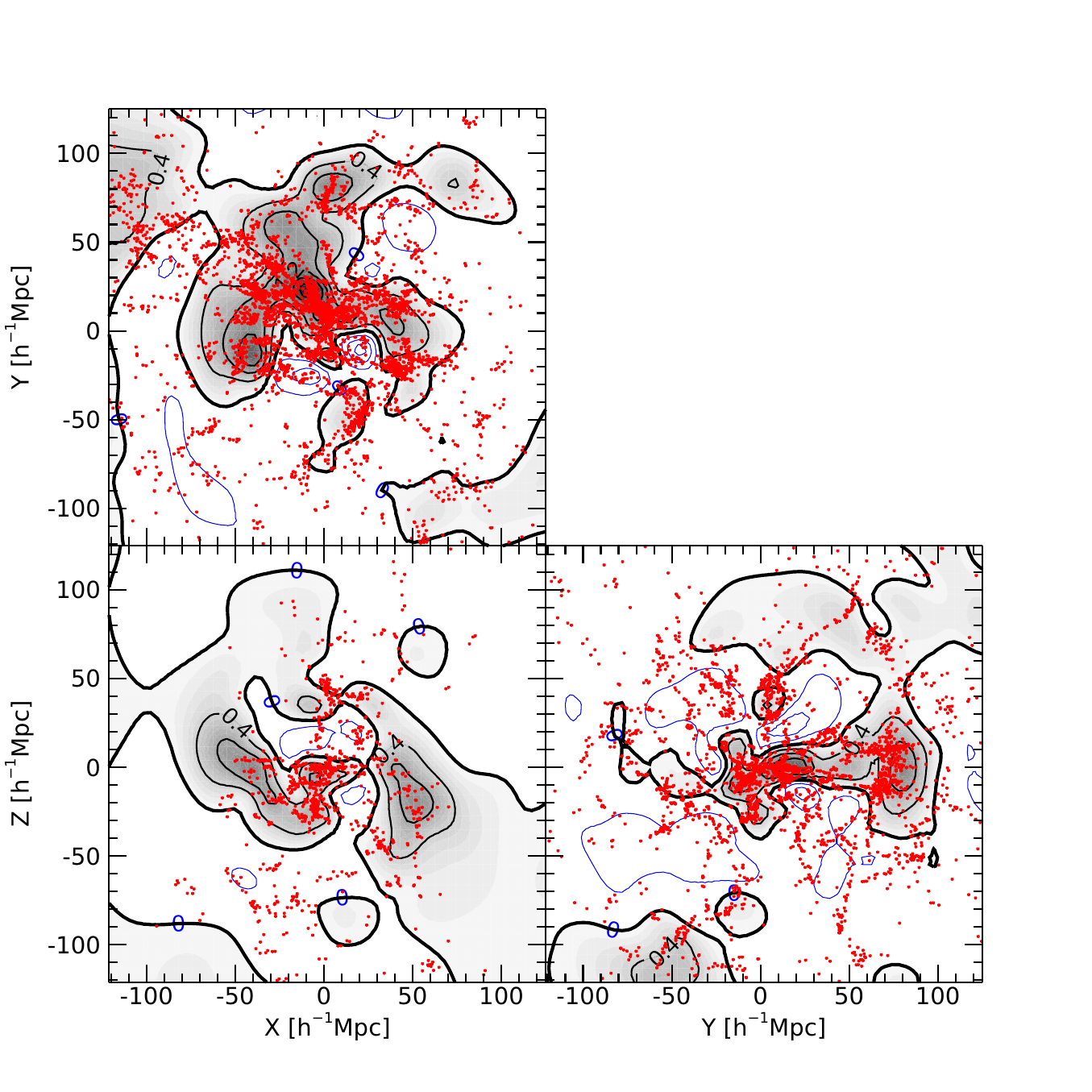}
\includegraphics[width=0.48\linewidth]{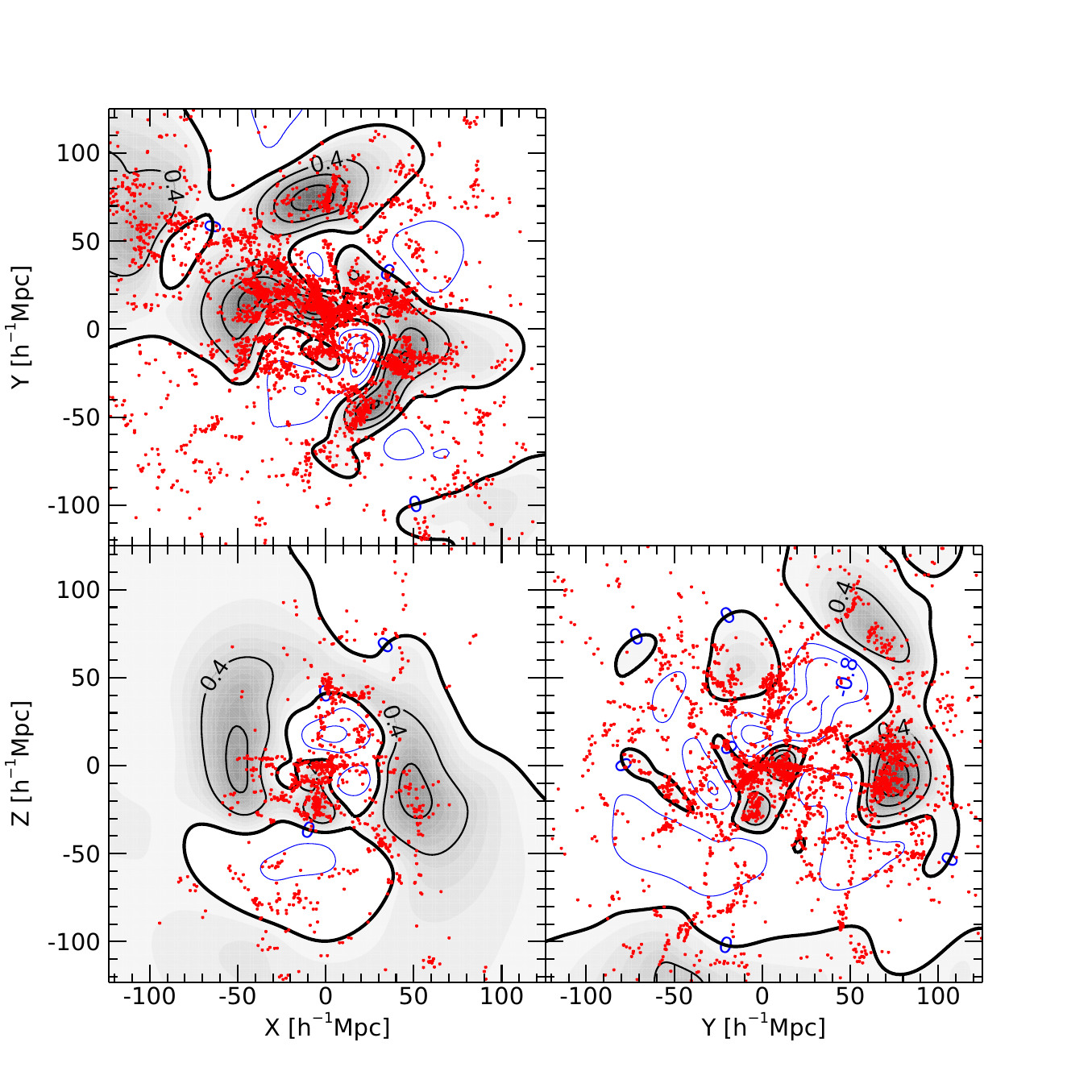}
\caption{The WF reconstructed  over-density field of the CF4 data (left panel is compared with that of the CF2 one (right panel). The density fields are smoothed with a $5.0\,\hmpc$ Gaussian kernel.The 2MRS galaxies, within $\pm5.0,\hmpc$ of the principal Supergalactic planes, are superimposed as a mean for validation of the reconstruction and for the sake of orientation. 
}
\label{fig:CF4-CF2}
\end{figure*}

A comparison of the large scale density field reconstructed from the CF4 data with the one from the CF2 is depicted by Fig. \ref{fig:CF4-CF2}. The 2MRS galaxies \citep{2012ApJS..199...26H} are superimposed on the density maps as a means of validation of the reconstruction and for the sake of orientation.

The CF2 grouped data consists of 4,814 data points within an effective depth of $\sim100\,\hmpc$, compared with the 38,060  data points within $\sim300\,\hmpc$ of the grouped CF4 catalog. Also, the lognormal bias correction of the CF4 data is much better than the one used for the CF2 data. Yet, the density field of the CF4 reconstruction and of the CF2 data are in very agreement within the CF2 data zone. 

\bsp
\label{lastpage}
\end{document}